\newcounter{myctr}
\def\myitem{\refstepcounter{myctr}\bibfont\noindent\ifnum\themyctr>9\else\phantom{0}\fi\hangindent17pt\themyctr.\enskip}

\documentclass{ws-ijqi}
\usepackage{hyperref}
\usepackage[super,sort,compress]{cite}

\usepackage{tikz}
\usepackage{dsfont}
\usepackage{adjustbox}

\begin{document}

	\markboth{M. F. Araujo de Resende et al.}{Quantum Double Models coupled to matter fields: a detailed review for a dualization procedure}

%
\catchline{}{}{}{}{}
%

	\title{QUANTUM DOUBLE MODELS COUPLED TO MATTER FIELDS: A DETAILED REVIEW FOR A  DUALIZATION PROCEDURE}

	\author{M. F. ARAUJO DE RESENDE}

	\address{Instituto de F\'{\i}sica, Universidade de S\~{a}o Paulo, 05508-090 S\~{a}o Paulo SP, Brasil \\ \email{resende@if.usp.br}}
	
	\author{J. P. IBIETA JIMENEZ}
	
	\address{Instituto de F\'{\i}sica, Universidade de S\~{a}o Paulo, 05508-090 S\~{a}o Paulo SP, Brasil \\\email{pibieta@if.usp.br}} 
	
	\author{J. LORCA ESPIRO}

	\address{Departamento de Ciencias F\'{\i}sicas, Facultad de Ingenier\'{\i}a, Ciencias y Administraci\'{o}n,\\ Universidad de La Frontera, Avda. Francisco Salazar 01145, Casilla 54-D Temuco, Chile \\ \email{javier.lorca@ufrontera.cl}}
	
	\maketitle


	\begin{abstract}
		In this paper, we investigate how it is possible to define a new class of lattice gauge models based on a dualization procedure of a previous generalization of the Kitaev Quantum Double Models. In the case of this previous generalization that will be used as a basis, it was defined by adding new qudits (which were denoted as matter fields in reference to some works) to the lattice vertices with the intention of, for instance, interpreting its models as Kitaev Quantum Double Models coupled with Potts ones. Now, with regard to the generalization that we investigate here, which we want to define as the dual of this previous one, these new qudits were added to the lattice faces. And as the coupling between gauge and matter qudits of the previous generalization was performed by a gauge group action, we show that the dual behaviour of these two generalizations was achieved by coupling these same qudits in the second one through a gauge group co-action homomorphism. One of the most striking dual aspects of these two generalizations is that, in both, part of the quasiparticles that were inherited from the Kitaev Quantum Double Models become confined when these action and co-action are non-trivial. But the big news here is that, in addition to the group homomorphism (that defines this gauge group co-action) allows us to classify all the different models of this second generalization, this same group homomorphism also suggests that all these models can be interpreted as two-dimensional restrictions of the $ 2 $-lattice gauge theories. 
	\end{abstract}

	\section{Introduction}
	
		One of the current issues of interdisciplinary research explores some theoretical models and technologies that try to support some kind of quantum computing [\citen{benioff,deutsch,loss,cirac,akama}]. And since the original purpose of Quantum Computation is to be interpreted as a generalization of Classical Computation [\citen{nielsen,bellac}], some of these theoretical models are defined by assigning \emph{quantum bits} (qubits) [\citen{schumacher}] to the edges of some oriented lattice $ \mathcal{L} _{2} $. In general, it is not wrong to say that, in order to avoid any problems with reading the data encoded by these qubits, $ \mathcal{L} _{2} $ is usually chosen to be the one that discretizes some two-dimensional compact orientable manifold $ \mathcal{M} _{2} $. Nevertheless, a crucial advantage of using these lattices, which discretize these two-dimensional compact orientable manifolds, is the possibility of evaluating/implementing theoretical models that, because they have a topological order [\citen{wen}], can perform some fault-tolerant quantum computation [\citen{shor,dennis,kitaev-1}]. These are precisely the cases of
		\begin{itemize}
			\item the \emph{Toric Code} ($ D \left( \mathds{Z} _{2} \right) $), which gets its name from the fact that $ \mathcal{M} _{2} $ is homeomorphic to a two-dimensional torus [\citen{kitaev-1,mf-pedagogical}], and
			\item its natural generalization, so-called \emph{Quantum Double Models} ($ D \left( G \right) $), which are defined by using (i) a group $ G $ that is not necessarily Abelian and (ii) an $ \mathcal{L} _{2} $ that discretizes an $ \mathcal{M} _{2} $ not necessarily homeomorphic to a two-dimensional torus, but that assigns \emph{quantum dits} (qudits) instead of qubits to its edges [\citen{kitaev-1,pachos,naaij}].
		\end{itemize}
		
		Given that the $ D \left( G \right) $ models do not associate any qudits with the faces or/and vertices of $ \mathcal{L} _{2} $, one paper was published a few years ago [\citen{miguel-1}] to understand what happens, for instance, when these models are coupled to new qudits assigned to the lattice vertices ($ D_{M} \left( G \right) $). After all, as the $ D \left( G \right) $ models can be understood in terms of pure lattice gauge theories [\citen{pachos,mf-lattice-gauge}], it was desirable that there was another class of lattice models, which would be able to mimic some more general gauge theories, where some kind of matter was also present [\citen{aza,seiler}]. Therefore, since these new $ D_{M} \left( G \right) $ models needed to be interpreted as $ D \left( G \right) $ generalizations, which also needed to be understood in the light of these more general gauge theories, these new qudits were purposely denoted as \emph{matter fields} similarly to what was done in Ref. [\citen{fradkin}], where lattice gauge theories were coupled to fixed-length scalar (Higgs) fields allocated on the lattice vertices. 
		
		In this Ref. [\citen{miguel-1}] it became clear that, as the $ D_{M} \left( G \right) $ magnetic quasiparticles increase the energy of the system when they are transported through the lattice, the ground state of these models does not necessarily depend on the first homotopy group $ \pi _{1} \left( \mathcal{M} _{2} \right) $. As a consequence of this result, which points out that these magnetic quasiparticles should be interpreted as \emph{confined}\footnote{We will explain it all (what these $ D_{M} \left( G \right) $ magnetic quasiparticles are, why they increase the energy of the system when they are transported, why they were interpreted as confined etc.) later on, in Subsubsection \ref{confined-tcm}.} in the $ D_{M} \left( G \right) $ models, another paper was published shortly thereafter, presenting a new class of theoretical lattice models ($ H_{M} / \mathds{C} \left( \mathds{Z} _{N} \right) $) where this increase no longer happens [\citen{pramod-wrong}]. This class of (Abelian) $ H_{M} / \mathds{C} \left( \mathds{Z} _{N} \right) $ models is a subclass of the $ D_{M} \left( G \right) $ models in which
		\begin{itemize}
			\item the (gauge) group $ G $ is the cyclic Abelian group $ \mathds{Z} _{N} $, and
			\item the operators, which detect the magnetic quasiparticles in the $ D_{M} \left( G \right) $ models, were excluded from their Hamiltonian\footnote{At this point, we need to make an addendum: after all, although Ref. [\citen{miguel-1}] refers to its models by using the same notation as Ref. [\citen{pramod-wrong}], here we prefer to use the \textquotedblleft $ D_{M} \left( G \right) $\textquotedblright \hspace*{0.01cm} notation not only to differentiate the models of these two works, but also to highlight the fact that the models of Ref. [\citen{miguel-1}] satisfy the same \emph{Drinfeld's quantum double algebra} [\citen{drinfeld}] of the $ D \left( G \right) $ models [\citen{kitaev-1}].}.
		\end{itemize}
		And as strange as it may seem to build this subclass (which have the same gauge group dependence as the $ D \left( G \right) $ models) without the operators that detect these magnetic quasiparticles, the fact is that these $ H_{M} / \mathds{C} \left( \mathds{Z} _{N} \right) $ models have, at least, a very interesting property: some matter excitations (i.e., those energy excitations that can be produced by manipulating the matter fields) exhibit non-Abelian fusion rules [\citen{pramod-wrong,mf-errata}]. Therefore, since the only difference between the $ H_{M} / \mathds{C} \left( \mathds{Z} _{N} \right) $ and $ D_{M} \left( \mathds{Z} _{N} \right) $ models are the operators that measure magnetic quasiparticles, it is not difficult to conclude that these matter excitations, which exhibit non-Abelian fusion rules, are also present in the $ D_{M} \left( \mathds{Z} _{N} \right) $ models. 
		
		Nevertheless, and in contrast to the $ D \left( G \right) $ models [\citen{aguado}], the fact is that these two generalizations do not lead, for instance, to self-dual models. And as the confinement of the $ D_{M} \left( \mathds{Z} _{N} \right) $ magnetic quasiparticles has some similarity with that of quarks in Quantum Chromodynamics [\citen{quigg}] (which is precisely the gauge theory whose non-perturbative problems fostered the development of the lattice gauge theories [\citen{rothe,wilson-loops}]), a natural question that arises is: how to use these $ D_{M} \left( G \right) $ models as a kind of basis for defining a self-dual generalization of the $ D \left( G \right) $ ones where, for instance, qudits are assigned to all the faces and vertices of $ \mathcal{L} _{2} $? By the way, can a generalization of the $ D \left( G \right) $ models, intentionally defined by using the dual framework of the $ D_{M} \left( G \right) $ models, show us if it is possible to construct this self-dual generalization? Thus, in order to answer these questions, this paper is rightly devoted to the analysis of a new class of models ($ D^{K} \left( G \right) $), which is intentionally defined by using the dual framework of these $ D_{M} \left( G \right) $ models. That is, this new generalization of the $ D \left( G \right) $ models, whose construction/definition will be detailed in Section \ref{QDMf-construction}, has
		\begin{itemize}
			\item the same gauge structure as them, but
			\item the matter qudits attached only with the centre of all the faces of $ \mathcal{L} _{2} $, since all these centres can be interpreted as the vertices of a dual lattice $ \mathcal{L} ^{\ast } _{2} $ [\citen{wenninger}].
		\end{itemize}
		However, as we need to do this construction/definition (and, consequently, analyse it) based on what we already know about the $ D_{M} \left( G \right) $ models, we will deliberately use the next Section to do a detailed and judicious review of these $ D_{M} \left( G \right) $ models, by analysing their algebraic and topological properties. And just for the sake of simplicity, we will consider that $ \mathcal{L} _{2} $ and, consequently, $ \mathcal{L} ^{\ast } _{2} $ are square lattices, even though all the considerations that will be presented in this paper can be applied to general lattices. 
		
	\section{\label{QDM-review}A brief overview on the $ D_{M} \left( G \right) $ models}
	
		According to what we said in the Introduction, the $ D_{M} \left( G \right) $ models are straightforward generalizations of the $ D \left( G \right) $ ones. And in order to prove this statement, it is important to analyse what are the similarities and differences between these two classes of models. As a matter of fact, with regard to the issue of the similarities, it is worth noting that, due to the quantum computing proposal that permeates these $ D \left( G \right) $ and $ D_{M} \left( G \right) $ models, both are defined by assigning a $ \left\vert G \right\vert $-dimensional Hilbert space $ \mathfrak{H} _{\left\vert G \right\vert } $ to each of the edges of $ \mathcal{L} _{2} $. Observe that, since $ \mathfrak{H} _{\left\vert G \right\vert } $ is responsible for supporting the gauge qudits that need to be manipulated in these two classes of models, its single-qudit computational basis is $ \mathcal{B} _{g} = \bigl\{ \left\vert g \right\rangle : g \in G \bigr\} $.
		
		Now, with regard to the differences that exist between these two classes of models, it is correct to say that all these differences have their origin in the fact that, in the case of the $ D_{M} \left( G \right) $ models, there is also an $ M $-dimensional Hilbert space $ \mathfrak{H} _{M} $ assigned to each of the vertices of $ \mathcal{L} _{2} $. After all, by remembering that these $ D_{M} \left( G \right) $ models
		\begin{itemize}
			\item were intentionally defined to mimic some more general lattice gauge theories, and
			\item have additional qudits assigned to the lattice vertices in order to mimic matter fields similarly to what was done, for instance, in Ref. [\citen{fradkin}], where Higgs fields were allocated on the lattice vertices, 
		\end{itemize}
		an additional Hilbert space
		\begin{equation*}
			\underbrace{\mathfrak{H} _{M} \otimes \ldots \otimes \mathfrak{H} _{M}} _{N_{v} \ \textnormal{\tiny{times}}}
		\end{equation*}
		becomes necessary to support all the additional matter qudits that need to be manipulated in the $ D_{M} \left( G \right) $ models. That is, these $ D_{M} \left( G \right) $ models were defined by assigning an $ \mathfrak{H} _{M} $ to each of the vertices of $ \mathcal{L} _{2} $ because this Hilbert space is responsible for supporting these new matter qudits.
		
		By the way, since this Ref. [\citen{fradkin}] couples the Higgs fields to the lattice gauge fields by using a \emph{group action} [\citen{james}], it is also correct to say that these $ D_{M} \left( G \right) $ models were also defined by exploiting this fact: i.e., they were defined by coupling these matter qudits to the gauge ones by using a group action $ \mu : G \times S  \rightarrow S $, which defines how the gauge group $ G $ acts on the elements of the single-qudit computational basis $ \mathcal{B} _{m} = \bigl\{ \left\vert \alpha \right\rangle : \alpha \in S \bigr\} $ of $ \mathfrak{H} _{M} $. For now, $ S $ should only be interpreted as a set of indices (i.e., $ S = \left\{ 0 , 1 , 2 , \ldots , M-1 \right\} $). Nevertheless, since this group action $ \mu $ is one of the protagonists of the $ D_{M} \left( G \right) $ models (because it tells us how the matter and gauge qudits are coupled), it is interesting to note that it allows us to interpret $ \mathfrak{H} _{M} $ as a (left) $ \mathds{C} \left( G \right) $\emph{-module} [\citen{fulton}]. 
		
		\subsection{A few words about the $ D_{M} \left( G \right) $ Hamiltonian operator}
		
			But since we are talking about $ \mu $, it is important to take the opportunity to mentioning that its role is explicit in two of the three operators that make up the Hamiltonian
			\begin{equation}
				H_{D_{M} \left( G \right) } = - \mathcal{J} _{A} \sum _{v  \in \mathcal{L} _{2}} A_{v} - \mathcal{J} _{B} \sum _{f \in \mathcal{L} _{2}} \ B_{f} - \mathcal{J} _{C} \sum _{\ell \in \mathcal{L} _{2}} C_{\ell } \label{H-qdmv}
			\end{equation}
			of these $ D_{M} \left( G \right) $ models, where $ \mathcal{J} _{A} $, $ \mathcal{J} _{B} $ and $ \mathcal{J} _{C} $ are three positive parameters. Scilicet, $ \mu $ appears in the vertex and link operators\footnote{Although we are referring to $ \mathcal{L} _{2} $ in terms of its vertices, faces and \emph{edges}, it is worth noting that operators analogous to $ C_{\ell } $ are often denoted as \emph{link operators} in other lattice gauge models. And as the origin of this name lies in the fact that the purpose of such operators is to make clear what is the \emph{link} between what is contained in two neighbouring vertices, we will use this same denotation since, for example, this is also the purpose of $ C_{\ell } $.}
			\begin{equation}
				A_{v} = \frac{1}{\vert G \vert } \sum _{g \in G} A^{\left( g \right) } _{v} \quad \textnormal{and} \quad C_{\ell } = C^{\left( 0 \right) } _{\ell } \label{ac-qdmv-operators}
			\end{equation}
			respectively, whose components are shown in Figure \ref{QMDv-operators-components}
			\begin{figure}[!t]
				\begin{center}
					\begin{tikzpicture}[
						scale=0.3,
						equation/.style={thin},
						trans/.style={thin,shorten >=0.5pt,shorten <=0.5pt,>=stealth},
						flecha/.style={thin,->,shorten >=0.5pt,shorten <=0.5pt,>=stealth}
						]
						\draw[equation] (-7.8,0.15) -- (-7.8,0.15) node[midway,right] {$ A^{\left( g \right) } _{v} $};
						\draw[trans] (-4.8,2.3) -- (-4.8,-2.3) node[above=2pt,right=-1pt] {};
						\draw[trans] (1.9,2.3) -- (3.1,-0.06) node[above=2pt,right=-1pt] {};
						\draw[trans] (1.9,-2.3) -- (3.1,0.06) node[above=2pt,right=-1pt] {};
						\draw[flecha] (-3.1,0.0) -- (-2.2,0.0) node[above=2pt,right=-1pt] {};
						\draw[flecha] (-2.4,0.0) -- (0.7,0.0) node[above=2pt,right=-1pt] {};
						\draw[trans] (0.2,0.0) -- (1.1,0.0) node[above=2pt,right=-1pt] {};
						\draw[flecha] (-1.0,-1.7) -- (-1.0,-1.0) node[above=2pt,right=-1pt] {};
						\draw[flecha] (-1.0,-1.2) -- (-1.0,1.6) node[above=2pt,right=-1pt] {};
						\draw[trans] (-1.0,1.4) -- (-1.0,1.8) node[above=2pt,right=-1pt] {};
						\draw[trans,fill=white] (-1.0,0.0) circle (0.9);
						\draw[equation] (-1.0,0.0) -- (-1.0,0.0) node[midway] {$ \alpha $};
						\draw[equation] (-1.0,2.4) -- (-1.0,2.4) node[midway] {$ a $};
						\draw[equation] (2.6,0.0) -- (2.6,0.0) node[midway,left] {$ b $};
						\draw[equation] (-1.0,-2.4) -- (-1.0,-2.4) node[midway] {$ c $};
						\draw[equation] (-4.7,0.0) -- (-4.7,0.0) node[midway,right] {$ d $};
						\draw[equation] (4.5,-0.07) -- (4.5,-0.07) node[midway] {$ = $};
						\draw[equation] (6.6,0.15) -- (6.6,0.15) node[midway] {$ \sum $};
						\draw[equation] (6.6,-1.1) -- (6.6,-1.1) node[midway] {$ _{\gamma \in S} $};
						\draw[equation] (11.0,0.1) -- (11.0,0.1) node[midway] {$ \delta \left( \mu \left( g , \alpha \right) , \gamma \right) $};
						\draw[trans] (15.3,2.3) -- (15.3,-2.3) node[above=2pt,right=-1pt] {};
						\draw[trans] (24.6,2.3) -- (25.8,-0.06) node[above=2pt,right=-1pt] {};
						\draw[trans] (24.6,-2.3) -- (25.8,0.06) node[above=2pt,right=-1pt] {};
						\draw[flecha] (18.9,0.0) -- (19.8,0.0) node[above=2pt,right=-1pt] {};
						\draw[flecha] (19.6,0.0) -- (22.7,0.0) node[above=2pt,right=-1pt] {};
						\draw[trans] (22.2,0.0) -- (23.1,0.0) node[above=2pt,right=-1pt] {};
						\draw[flecha] (21.0,-1.8) -- (21.0,-1.0) node[above=2pt,right=-1pt] {};
						\draw[flecha] (21.0,-1.2) -- (21.0,1.6) node[above=2pt,right=-1pt] {};
						\draw[trans] (21.0,1.4) -- (21.0,1.8) node[above=2pt,right=-1pt] {};
						\draw[trans,fill=white] (20.95,0.0) circle (0.9);
						\draw[equation] (20.9,0.0) -- (20.9,0.0) node[midway] {$ \gamma $};
						\draw[equation] (21.1,2.3) -- (21.1,2.3) node[midway] {$ ga $};
						\draw[equation] (25.2,-0.2) -- (25.2,-0.2) node[midway,left] {$ gb $};
						\draw[equation] (21.1,-2.3) -- (21.1,-2.3) node[midway] {$ cg^{-1}  $};
						\draw[equation] (15.4,0.0) -- (15.4,0.0) node[midway,right] {$ dg^{-1}  $};		
					\end{tikzpicture} \\
					\begin{tikzpicture}[
						scale=0.3,
						equation/.style={thin},
						trans/.style={thin,shorten >=0.5pt,shorten <=0.5pt,>=stealth},
						flecha/.style={thin,->,shorten >=0.5pt,shorten <=0.5pt,>=stealth}
						]
						\draw[equation] (-8.8,0.0) -- (-8.8,0.0) node[midway,right] {$ B^{\left( h \right) } _{f} $};
						\draw[trans] (-5.6,2.3) -- (-5.6,-2.3) node[above=2pt,right=-1pt] {};
						\draw[trans] (-0.1,2.3) -- (1.1,-0.06) node[above=2pt,right=-1pt] {};
						\draw[trans] (-0.1,-2.3) -- (1.1,0.06) node[above=2pt,right=-1pt] {};
						\draw[flecha] (-4.7,1.2) -- (-2.1,1.2) node[above=2pt,right=-1pt] {};
						\draw[trans] (-2.3,1.2) -- (-0.1,1.2) node[above=2pt,right=-1pt] {};
						\draw[flecha] (-4.7,-1.2) -- (-2.1,-1.2) node[above=2pt,right=-1pt] {};
						\draw[trans] (-2.3,-1.2) -- (-0.1,-1.2) node[above=2pt,right=-1pt] {};
						\draw[flecha] (-3.8,-2.0) -- (-3.8,0.3) node[above=2pt,right=-1pt] {};
						\draw[trans] (-3.8,0.0) -- (-3.8,2.0) node[above=2pt,right=-1pt] {};
						\draw[flecha] (-1.0,-2.0) -- (-1.0,0.3) node[above=2pt,right=-1pt] {};
						\draw[trans] (-1.0,0.0) -- (-1.0,2.0) node[above=2pt,right=-1pt] {};
						\draw[equation] (-2.3,2.0) -- (-2.3,2.0) node[midway] {$ a $};				
						\draw[equation] (0.7,0.0) -- (0.7,0.0) node[midway,left] {$ d $};
						\draw[equation] (-2.3,-2.0) -- (-2.3,-2.0) node[midway] {$ c $};
						\draw[equation] (-5.5,0.0) -- (-5.5,0.0) node[midway,right] {$ b $};
						\draw[equation] (6.7,-0.07) -- (6.7,-0.07) node[midway] {$ = \ \delta \left( h , a^{-1} b^{-1} cd \right) $};
						\draw[trans] (12.2,2.3) -- (12.2,-2.3) node[above=2pt,right=-1pt] {};
						\draw[trans] (17.7,2.3) -- (18.8,-0.06) node[above=2pt,right=-1pt] {};
						\draw[trans] (17.7,-2.3) -- (18.8,0.06) node[above=2pt,right=-1pt] {};
						\draw[flecha] (13.0,1.2) -- (15.6,1.2) node[above=2pt,right=-1pt] {};
						\draw[trans] (15.4,1.2) -- (17.6,1.2) node[above=2pt,right=-1pt] {};
						\draw[flecha] (13.0,-1.2) -- (15.6,-1.2) node[above=2pt,right=-1pt] {};
						\draw[trans] (15.4,-1.2) -- (17.6,-1.2) node[above=2pt,right=-1pt] {};
						\draw[flecha] (13.9,-2.0) -- (13.9,0.3) node[above=2pt,right=-1pt] {};
						\draw[trans] (13.9,0.0) -- (13.9,2.0) node[above=2pt,right=-1pt] {};
						\draw[flecha] (16.7,-2.0) -- (16.7,0.3) node[above=2pt,right=-1pt] {};
						\draw[trans] (16.7,0.0) -- (16.7,2.0) node[above=2pt,right=-1pt] {};
						\draw[equation] (15.4,2.0) -- (15.4,2.0) node[midway] {$ a $};
						\draw[equation] (18.4,0.0) -- (18.4,0.0) node[midway,left] {$ d $};
						\draw[equation] (15.4,-2.0) -- (15.4,-2.0) node[midway] {$ c $};
						\draw[equation] (12.2,0.0) -- (12.2,0.0) node[midway,right] {$ b $};
					\end{tikzpicture} \\
					\begin{tikzpicture}[
						scale=0.3,
						equation/.style={thin},
						trans/.style={thin,shorten >=0.5pt,shorten <=0.5pt,>=stealth},
						flecha/.style={thin,->,shorten >=0.5pt,shorten <=0.5pt,>=stealth}
						]
						\draw[equation] (-8.7,0.0) -- (-8.7,0.0) node[midway,right] {$ C^{\left( \Lambda \right) } _{\ell } $};
						\draw[trans] (-5.6,2.3) -- (-5.6,-2.3) node[above=2pt,right=-1pt] {};
						\draw[trans] (0.1,2.3) -- (1.3,-0.06) node[above=2pt,right=-1pt] {};
						\draw[trans] (0.1,-2.3) -- (1.3,0.06) node[above=2pt,right=-1pt] {};
						\draw[flecha] (-4.4,0.0) -- (-2.1,0.0) node[above=2pt,right=-1pt] {};
						\draw[trans] (-2.3,0.0) -- (-0.2,0.0) node[above=2pt,right=-1pt] {};
						\draw[trans,fill=white] (-4.4,0.0) circle (0.9);
						\draw[trans,fill=white] (-0.2,0.0) circle (0.9);
						\draw[equation] (-2.3,-0.8) -- (-2.3,-0.8) node[midway] {$ a $};
						\draw[equation] (-4.4,0.0) -- (-4.4,0.0) node[midway] {$ \alpha $};
						\draw[equation] (-0.2,-0.02) -- (-0.2,-0.02) node[midway] {$ \beta $};
						\draw[equation] (2.6,-0.07) -- (2.6,-0.07) node[midway] {$ = $};
						\draw[equation] (10.6,0.0) -- (10.6,0.0) node[midway] {$ \delta \left( \mu \left( a , \alpha \right) , \left( \beta + \Lambda \right) \textnormal{mod} \ M \right) $};
						\draw[trans] (18.0,2.3) -- (18.0,-2.3) node[above=2pt,right=-1pt] {};
						\draw[trans] (23.7,2.3) -- (24.9,-0.06) node[above=2pt,right=-1pt] {};
						\draw[trans] (23.7,-2.3) -- (24.9,0.06) node[above=2pt,right=-1pt] {};
						\draw[flecha] (19.2,0.0) -- (21.5,0.0) node[above=2pt,right=-1pt] {};
						\draw[trans] (21.2,0.0) -- (23.4,0.0) node[above=2pt,right=-1pt] {};
						\draw[trans,fill=white] (19.2,0.0) circle (0.9);
						\draw[trans,fill=white] (23.4,0.0) circle (0.9);
						\draw[equation] (21.3,-0.8) -- (21.3,-0.8) node[midway] {$ a $};
						\draw[equation] (19.2,0.0) -- (19.2,0.0) node[midway] {$ \alpha $};
						\draw[equation] (23.4,-0.02) -- (23.4,-0.02) node[midway] {$ \beta $};
					\end{tikzpicture}
				\end{center}
				\caption{\label{QMDv-operators-components} Definition of the components $ A^{\left( g \right) } _{v} $, $ B^{\left( h \right) } _{f} $ and $ C^{\left( \Lambda \right) } _{\ell } $ whose effective action, on the sectors $ S_{v} $, $ S_{f} $ and $ S_{\ell } $ of $ \mathcal{L} _{2} $ respectively, can also be better understood by looking at Figure \ref{QMDv-rede}. Here, the group element $ a $ is indexing an $ \left\vert a \right\rangle $ basis element of $ \mathfrak{H} _{\left\vert G \right\vert } $, the symbol $ \alpha $ indexes an $ \left\vert \alpha \right\rangle $ basis element of $ \mathfrak{H} _{M} $, and $ \Lambda $ is a natural index such that $ 0 \leqslant \Lambda \leqslant M-1 $. Moreover, it is also worth noting that, here, $ \delta \left( x , y \right) $ should be interpreted as a Kronecker delta that was written differently for the sake of intelligibility (i.e., $ \delta \left( x , y \right) = \delta _{xy} $).}
			\end{figure}
			\begin{figure}[!t]
				\begin{center}
					\begin{tikzpicture}
						\draw[color=blue!20,fill=blue!20] (6,5) rectangle (8,7);
						\draw[color=red!20,fill=red!20] (1,2) rectangle (3,4);
						\draw[color=orange!20,fill=orange!20] (5.5,2.5) rectangle (8.5,3.5);
						\draw[->, color=gray, ultra thick, >=stealth] (0,0) -- (0,2.2);
						\draw[->, color=gray, ultra thick, >=stealth] (0,2) -- (0,4.2);
						\draw[->, color=gray, ultra thick, >=stealth] (0,4) -- (0,6.2);
						\draw[-, color=gray, ultra thick] (0,6) -- (0,8);
						\draw[->, color=gray, ultra thick, >=stealth] (2,0) -- (2,2.2);
						\draw[->, color=gray, ultra thick, >=stealth] (2,2) -- (2,4.2);
						\draw[->, color=gray, ultra thick, >=stealth] (2,4) -- (2,6.2);
						\draw[-, color=gray, ultra thick] (2,6) -- (2,8);
						\draw[->, color=gray, ultra thick, >=stealth] (4,0) -- (4,2.2);
						\draw[->, color=gray, ultra thick, >=stealth] (4,2) -- (4,4.2);
						\draw[->, color=gray, ultra thick, >=stealth] (4,4) -- (4,6.2);
						\draw[-, color=gray, ultra thick] (4,6) -- (4,8);
						\draw[->, color=gray, ultra thick, >=stealth] (6,0) -- (6,2.2);
						\draw[->, color=gray, ultra thick, >=stealth] (6,2) -- (6,4.2);
						\draw[->, color=gray, ultra thick, >=stealth] (6,4) -- (6,6.2);
						\draw[-, color=gray, ultra thick] (6,6) -- (6,8);
						\draw[->, color=gray, ultra thick, >=stealth] (8,0) -- (8,2.2);
						\draw[->, color=gray, ultra thick, >=stealth] (8,2) -- (8,4.2);
						\draw[->, color=gray, ultra thick, >=stealth] (8,4) -- (8,6.2);
						\draw[-, color=gray, ultra thick] (8,6) -- (8,8);
						\draw[->, color=gray, ultra thick, >=stealth] (10,0) -- (10,2.2);
						\draw[->, color=gray, ultra thick, >=stealth] (10,2) -- (10,4.2);
						\draw[->, color=gray, ultra thick, >=stealth] (10,4) -- (10,6.2);
						\draw[-, color=gray, ultra thick] (10,6) -- (10,8);
						\draw[->, color=gray, ultra thick, >=stealth] (-1,1) -- (1.2,1);
						\draw[->, color=gray, ultra thick, >=stealth] (1,1) -- (3.2,1);
						\draw[->, color=gray, ultra thick, >=stealth] (3,1) -- (5.2,1);
						\draw[->, color=gray, ultra thick, >=stealth] (5,1) -- (7.2,1);
						\draw[->, color=gray, ultra thick, >=stealth] (7,1) -- (9.2,1);
						\draw[-, color=gray, ultra thick] (9,1) -- (11,1);
						\draw[->, color=gray, ultra thick, >=stealth] (-1,3) -- (1.2,3);
						\draw[->, color=gray, ultra thick, >=stealth] (1,3) -- (3.2,3);
						\draw[->, color=gray, ultra thick, >=stealth] (3,3) -- (5.2,3);
						\draw[->, color=gray, ultra thick, >=stealth] (5,3) -- (7.2,3);
						\draw[->, color=gray, ultra thick, >=stealth] (7,3) -- (9.2,3);
						\draw[-, color=gray, ultra thick] (9,3) -- (11,3);
						\draw[->, color=gray, ultra thick, >=stealth] (-1,5) -- (1.2,5);
						\draw[->, color=gray, ultra thick, >=stealth] (1,5) -- (3.2,5);
						\draw[->, color=gray, ultra thick, >=stealth] (3,5) -- (5.2,5);
						\draw[->, color=gray, ultra thick, >=stealth] (5,5) -- (7.2,5);
						\draw[->, color=gray, ultra thick, >=stealth] (7,5) -- (9.2,5);
						\draw[-, color=gray, ultra thick] (9,5) -- (11,5);
						\draw[->, color=gray, ultra thick, >=stealth] (-1,7) -- (1.2,7);
						\draw[->, color=gray, ultra thick, >=stealth] (1,7) -- (3.2,7);
						\draw[->, color=gray, ultra thick, >=stealth] (3,7) -- (5.2,7);
						\draw[->, color=gray, ultra thick, >=stealth] (5,7) -- (7.2,7);
						\draw[->, color=gray, ultra thick, >=stealth] (7,7) -- (9.2,7);
						\draw[-, color=gray, ultra thick] (9,7) -- (11,7);
						\draw[->, ultra thick, >=stealth] (2,1) -- (2,2.2);
						\draw[->, ultra thick, >=stealth] (2,2.0) -- (2,4.2);
						\draw[-, ultra thick] (2,4.0) -- (2,5.0);
						\draw[->, ultra thick, >=stealth] (0.0,3) -- (1.2,3);
						\draw[->, ultra thick, >=stealth] (1.0,3) -- (3.2,3);
						\draw[-, ultra thick] (3.0,3) -- (4.0,3);
						\draw[->, ultra thick, >=stealth] (6,5.0) -- (6,6.2);
						\draw[-, ultra thick] (6,6.0) -- (6,7.0);
						\draw[->, ultra thick, >=stealth] (8,5.0) -- (8,6.2);
						\draw[-, ultra thick] (8,6.0) -- (8,7.0);
						\draw[->, ultra thick, >=stealth] (6.0,5) -- (7.2,5);
						\draw[-, ultra thick] (7.0,5) -- (8.0,5);
						\draw[->, ultra thick, >=stealth] (6.0,7) -- (7.2,7);
						\draw[-, ultra thick] (7.0,7) -- (8.0,7);
						\draw[->, ultra thick, >=stealth] (6.0,3) -- (7.2,3);
						\draw[-, ultra thick] (7.0,3) -- (8.0,3);
						\draw[color=black,fill=white] (2,3) circle (0.3);
						\node [] (2,3) at (2,3) {$ \gamma $};
						\node [right] (2.1,4) at (2.1,4) {$ a $};
						\node [below] (3,2.9) at (3,2.9) {$ b $};
						\node [left] (1.9,2) at (1.9,2) {$ c $};
						\node [above] (1,3.1) at (1,3.1) {$ d $};
						\node [above] (7,7.1) at (7,7.1) {$ r $};
						\node [right] (8.1,6) at (8.1,6) {$ s $};
						\node [below] (7,4.9) at (7,4.9) {$ t $};
						\node [left] (5.9,6) at (5.9,6) {$ u $};
						\draw[color=black,fill=white] (6,3) circle (0.3);
						\node [] (6,3) at (6,3) {$ \alpha $};
						\draw[color=black,fill=white] (8,3) circle (0.3);
						\node [] (8,3) at (8,3) {$ \beta $};
						\node [below] (7.0,2.9) at (7.0,2.9) {$ \ell $};
					\end{tikzpicture}
				\end{center}
				\caption{\label{QMDv-rede} Piece of $ \mathcal{L} _{2} $ that supports the $ D_{M} \left( G \right) $ models where we see: (i) the rose-coloured sector ($ S_{v} $) centred in the $ v $-th vertex; (ii) the baby blue coloured sector ($ S_{f} $) highlighting the $ f $-th face; and (iii) the light orange coloured sector ($ S_{\ell } $) that is centred in the $ \ell $-th edge. Here, the highlighted edges (in black colour) correspond to Hilbert subspaces $ \mathfrak{H} _{\left\vert G \right\vert } $ in which the vertex (rose-coloured sector), face (baby blue coloured sector) and link (light orange coloured sector) operators mentioned in (\ref{ac-qdmv-operators}) and (\ref{b-qdmv-operator}) act effectively. Note that, as $ S_{v} $ and $ S_{\ell } $ contain vertices in their interior, an analogous comment applies to the vertices highlighted with Greek letters: i.e., according to what can be seen in Figure \ref{QMDv-operators-components}, these vertices correspond to Hilbert subspaces $ \mathfrak{H} _{M} $ in which, for instance, only the vertex and link operators act effectively.}
			\end{figure}
			together with those of the face operator
			\begin{equation}
				B_{f} = B^{\left( e \right) } _{f} \ . \label{b-qdmv-operator}
			\end{equation}
			Here, $ e $ should be interpreted as the neutral element of $ G $.
		
			Note that, by virtue of all the operators in (\ref{ac-qdmv-operators}) and (\ref{b-qdmv-operator}) being expressed by using Kronecker deltas, it is not difficult to conclude that, no matter what they do, $ 0 $ and $ 1 $ are the only values they return by acting on each vertex, face or edge of $ \mathcal{L} _{2} $. And since these same expressions (\ref{ac-qdmv-operators}) and (\ref{b-qdmv-operator}) are such that
			\begin{equation*}
				\left[ A_{v} , B_{f} \right] = \left[ A_{v} , C_{\ell } \right] = \left[ B_{f} , C_{\ell } \right] = 0 \ ,
			\end{equation*}
			and
			\begin{eqnarray*}
				A_{v^{\prime }} \circ A_{v^{\prime \prime }} & = & A_{v^{\prime }} \cdot \delta \left( v^{\prime } , v^{\prime \prime } \right) \ , \\
				B_{f^{\prime }} \circ B_{f^{\prime \prime }} & = & B_{f^{\prime }} \cdot \delta \left( f^{\prime } , f^{\prime \prime } \right) \quad \textnormal{and} \\
				C_{\ell ^{\prime }} \circ C_{\ell ^{\prime \prime }} & = & C_{\ell ^{\prime }} \cdot \delta \left( \ell ^{\prime } , \ell ^{\prime \prime } \right) \ ,
			\end{eqnarray*}
			all this allows us to assert that $ A_{v} $, $ B_{f} $ and $ C_{\ell } $ are three projectors: i.e., $ A_{v} $, $ B_{f} $ and $ C_{\ell } $ are three operators that allow us to assert that, when		
			\begin{equation}
				A_{v} \left\vert \xi _{0} \right\rangle = \left\vert \xi _{0} \right\rangle \ , \ \ B_{f} \left\vert \xi _{0} \right\rangle = \left\vert \xi _{0} \right\rangle \ \ \textnormal{and} \ \ C_{\ell } \left\vert \xi _{0} \right\rangle = \left\vert \xi _{0} \right\rangle \label{QDMv-ground-state}
			\end{equation}
			hold for all the $ N_{v} $ vertices, $ N_{f} $ faces and $ N_{\ell } $ edges of $ \mathcal{L} _{2} $, the state $ \left\vert \xi _{0} \right\rangle $ of this lattice system belongs to the Hilbert subspace
			\begin{equation*}
				\mathfrak{H} ^{\left( 0 \right) } _{D_{M} \left( G \right) } \subset \mathfrak{H} _{D_{M} \left( G \right) } = \underbrace{\mathfrak{H} _{\left\vert G \right\vert } \otimes \ldots \otimes \mathfrak{H} _{\left\vert G \right\vert }} _{N_{\ell } \ \textnormal{\tiny{times}}} \ \otimes \ \underbrace{\mathfrak{H} _{M} \otimes \ldots \otimes \mathfrak{H} _{M}} _{N_{v} \ \textnormal{\tiny{times}}}
			\end{equation*}
			whose states have the lowest possible energy. From the point of view of (\ref{H-qdmv}), this is equivalent to saying that $ \left\vert \xi _{0} \right\rangle $ is an eigenstate of $ H_{D_{M} \left( G \right) } $ that has an eigenvalue equal to
			\begin{equation}
				E_{0} = - \left( \mathcal{J} _{A} N_{v} + \mathcal{J} _{B} N_{f} + \mathcal{J} _{C} N_{\ell } \right) \label{e0-energy}
			\end{equation}
			because the $ D_{M} \left( G \right) $ models are examples of frustation-free models [\citen{miguel-1}].
			
			\subsubsection{A few words about the existence of additional vertex, face and edge projectors}
			
				Since we just talked about the projectivity of $ A_{v} $, $ B_{f} $ and $ C_{\ell } $, it is important to open a small parenthesis here to tell you, the reader, that they are not the only projectors of these $ D_{M} \left( G \right) $ models. After all, by noting that $ G $ is a finite group that has [\citen{lang}] 
				\begin{itemize}
					\item $ R $ distinct (non-equivalent) conjugacy classes $ \mathsf{C} _{L} = \left\{ h g_{L} h^{-1} : h \in G \right\} $, and (therefore)
					\item $ R $ distinct (non-isomorphic) irreducible representations $ \rho _{1+J} : G \rightarrow GL _{\left\vert G \right\vert } \left( \mathds{C} \right) $ over the complex numbers,
				\end{itemize}
				it is not difficult to see that these three operators are nothing more than mere special cases of others
				\begin{equation}
					A_{v,J} = \frac{1}{\vert G \vert } \sum _{g \in G} \chi _{1+J} \left( g^{-1} \right) \cdot A^{\left( g \right) } _{v} \ , \ \ B_{f,L} \equiv \sum _{g \in \mathsf{C} _{L}} B^{\left( g \right) } _{f} \quad \textnormal{and} \quad C_{\ell , \Lambda } \equiv C^{\left( \Lambda \right) } _{\ell } \ , \label{all-operators}
				\end{equation}
				where $ \chi _{1+J} \left( g \right) = \mathrm{Tr} \left[ \rho _{1+J}  \left( g \right) \right] $, $ J , L = 0 , 1 , \ldots , R-1 $ and $ \Lambda = 0 , 1 , \ldots , M-1 $. And since all these operators
				\begin{itemize}
					\item[\textbf{(a)}] have eigenvalues equal to $ 0 $ and $ 1 $,
					
					\item[\textbf{(b)}] satisfy the relations
					\begin{equation}
						\left[ A_{v^{\prime } , J^{\prime }} , B_{f^{\prime } , L^{\prime }} \right] = \left[ A_{v^{\prime } , J^{\prime }} , C_{\ell ^{\prime } , \Lambda ^{\prime }} \right] = \left[ B_{f^{\prime } , L^{\prime }} , C_{\ell ^{\prime } , \Lambda ^{\prime }} \right] = 0 \ , \label{dm-solubility}
					\end{equation}
					\begin{eqnarray*}
						A_{v^{\prime } , J^{\prime }} \circ A_{v^{\prime \prime } , J^{\prime \prime }} & = & A_{v^{\prime } , J^{\prime }} \cdot \delta \left( v^{\prime } , v^{\prime \prime } \right) \cdot \delta \left( J^{\prime } , J^{\prime \prime } \right) \ , \\
						B_{f^{\prime } , L^{\prime }} \circ B_{f^{\prime \prime } , L^{\prime \prime }} & = & B_{f^{\prime } , L^{\prime }} \cdot \delta \left( f^{\prime } , f^{\prime \prime } \right) \cdot \delta \left( L^{\prime } , L^{\prime \prime } \right) \quad \textnormal{and} \\
						C_{\ell ^{\prime } , \Lambda ^{\prime }} \circ C_{\ell ^{\prime \prime } , \Lambda ^{\prime \prime }} & = & C_{\ell ^{\prime } , \Lambda ^{\prime }} \cdot \delta \left( \ell ^{\prime } , \ell ^{\prime \prime } \right) \cdot \delta \left( \Lambda ^{\prime } , \Lambda ^{\prime \prime } \right)
					\end{eqnarray*}
					not only for all the values of $ J^{\prime \left( \prime \right) } , L^{\prime \left( \prime \right) } = 0 , 1 , \ldots , R-1 $ and $ \Lambda ^{\prime \left( \prime \right) } = 0 , 1 , \ldots , M-1 $, but also for all the vertices, faces and edges of $ \mathcal{L} _{2} $, and
					
					\item[\textbf{(c)}] are such that
					\begin{equation*}
						\sum ^{R-1} _{J=0} A_{v,J} = \mathds{1} _{v} \ , \ \ \sum ^{R-1} _{L=0} B_{f,L} = \mathds{1} _{f} \quad \textnormal{and} \quad \sum ^{M-1} _{\Lambda =0} C_{\ell , \Lambda } = \mathds{1} _{\ell } \ ,
					\end{equation*}
					where $ \mathds{1} _{v} $, $ \mathds{1} _{f} $ and $ \mathds{1} _{\ell } $ are identity operators that act effectively on the $ v $-th vertex, $ f $-th face and $ \ell $-th edge of $ \mathcal{L} _{2} $ respectively\footnote{This notation \textquotedblleft $ \mathds{1} _{v} $\textquotedblright , \textquotedblleft $ \mathds{1} _{f} $\textquotedblright \hspace*{0.01cm} and \textquotedblleft $ \mathds{1} _{\ell } $\textquotedblright \hspace*{0.01cm} is being used here only to be consistent with some expressions that will be presented later. Nevertheless, as all these operators act effectively on the $ v $-th vertex, $ f $-th face and $ \ell $-th edge of $ \mathcal{L} _{2} $ respectively, all of them can be interpreted, in fact, as an identity operator $ \mathds{1} _{\mathcal{L} _{2}} $ that acts on all lattice edges simultaneously.},
				\end{itemize}
				it is correct to assert that 
				\begin{eqnarray*}
					\mathfrak{A} & = & \left\{ A_{v,0} \ , \ A_{v,1} \ , \ \ldots \ , \ A_{v, R-1} \right\} \ , \\
					\mathfrak{B} & = & \left\{ B_{f,0} \ , \ B_{f,1} \ , \ \ldots \ , \ B_{f, R-1} \right\} \quad \textnormal{and} \\
					\mathfrak{C} & = & \left\{ C_{\ell ,0} \ , \ C_{\ell ,1} \ , \ \ldots \ , \ C_{\ell ,M-1} \right\}
				\end{eqnarray*}
				are three complete sets of orthogonal projectors onto $ \mathfrak{H} _{D_{M} \left( G \right) } $. And why is it important to open this parenthesis here? Because, in addition to (\ref{dm-solubility}) ensures that the $ D_{M} \left( G \right) $ models are exactly solvable [\citen{miguel-1}], the fact that these models have been defined by using all these projectors is in full agreement with the requirements of Quantum Mechanics [\citen{gottfried}]. After all, by noting that 
				\begin{equation*}
					A_{v} = A_{v,0} \ , \ \ B_{f} = B_{f,0} \quad \textnormal{and} \quad C_{\ell } = C_{\ell , 0}
				\end{equation*}
				because
				\begin{itemize}
					\item $ \chi _{1} \left( g \right) = 1 $ holds for all the elements of $ G $, and
					\item $ \mathsf{C} _{0} $ is the conjugacy class of $ e $,
				\end{itemize}
				this is precisely what allows us to assert that a state $ \left\vert \xi _{0} \right\rangle $, which satisfies (\ref{QDMv-ground-state}) for all the $ N_{v} $ vertices, $ N_{f} $ faces and $ N_{\ell } $ edges of $ \mathcal{L} _{2} $, belongs to $ \mathfrak{H} ^{\left( 0 \right) } $. In plain English, it is the existence of these projectors (\ref{all-operators}) that allow us to decompose the $ D_{M} \left( G \right) $ Hilbert space into the direct sum
				\begin{equation}
					\mathfrak{H} _{D_{M} \left( G \right) } = \mathfrak{H} ^{\left( 0 \right) } _{D_{M} \left( G \right) } \oplus \mathfrak{H} ^{\perp } _{D_{M} \left( G \right) } \ , \label{soma-direta}
				\end{equation}
				where $ \mathfrak{H} ^{\left( 0 \right) } _{D_{M} \left( G \right) } $ and $ \mathfrak{H} ^{\perp } _{D_{M} \left( G \right) } $ are the orthogonal subspaces that contain all the $ D_{M} \left( G \right) $ vacuum and non-vacuum states respectively [\citen{mf-errata}].
				
			\subsubsection{Understanding what the vertex and face operators are}
			
				Of course, given that we ended the last paragraph by citing these $ D_{M} \left( G \right) $ vacuum and non-vacuum states, it is very important that we explain how they can be produced. But, before we do that, it seems more interesting to pay attention to what these operators $ A_{v,J} $, $ B_{f,L} $ and $ C_{\ell , \Lambda } $ do in addition to being projectors. And since we have already said that one of the ideas behind these $ D_{M} \left( G \right) $ models is, for instance, to mimic some lattice gauge theories where matter fields are present, it is worth to say that this mimicry is mostly done by the operator $ A_{v,0} $ that defines the Hamiltonian (\ref{H-qdmv}).
				
				As a matter of fact, in order to understand how $ A_{v,0} $ does this, one of the things that we need to do is understand how its components $ A^{\left( g \right) } _{v} $ act on $ \mathcal{L} _{2} $. And in accordance with what Figure \ref{QMDv-operators-components} shows us, these components are operators that, by acting on $ S_{v} $, change\footnote{Not only here, but elsewhere in this paper, we will use the indices $ v $ and $ \ell $ whenever necessary to emphasize that $ \left\vert \alpha \right\rangle $ and $ \left\vert g \right\rangle $ are associated with the $ v $-th vertex and the $ \ell $-th edge of $ \mathcal{L} _{2} $ respectively.\label{comment-v-index}}
				\begin{itemize}
					\item the matter fields $ \left\vert \alpha \right\rangle _{v} $ for other $ \left\vert \gamma \right\rangle _{v} = \left\vert \mu \left( g , \alpha \right) \right\rangle _{v} $, and
					\item the gauge fields $ \left\vert g^{\prime } \right\rangle _{\ell } $ for other
					\begin{equation*}
						\left\vert g \cdot g^{\prime } \right\rangle _{\ell } \quad \textnormal{or} \quad \left\vert g^{\prime } \cdot g^{-1} \right\rangle _{\ell }
					\end{equation*}
					if the orientation of the $ \ell $-th edge points into or out of the $ v $-th vertex respectively.
				\end{itemize}
				That is, regardless of whether or not these components change the matter fields that appear on the lattice vertices, they change all the gauge fields of $ S_{v} $ when $ g \neq e $. In this way, as the result of the action of $ A_{v,0} $ comes from the sum of all the transformations that $ A^{\left( g \right) } _{v} $ is capable of doing, it is not wrong to conclude, for instance, that $ A_{v,0} $ averages out the possible transformations that $ A^{\left( g \right) } _{v} $ is able to do by using all elements of $ G $ [\citen{mf-lattice-gauge}]. And since the only difference between $ A_{v,0} $ and the others $ A_{v,J} $ is due to the characters $ \chi _{1+J} \left( g \right) $, which can take on values other than $ 1 $ when $ J \neq 0 $ and $ g \neq e $, it is also not wrong to extend this conclusion to all these operators $ A_{v,J} $. That is, even though $ \chi _{1+J} \left( g \right) $ can take on complex values other than $ 1 $ when $ J \neq 0 $ and $ g \neq e $, each of these operators $ A_{v,J} $ computes some kind of \textquotedblleft exotic\textquotedblright \hspace*{0.01cm} weighted average, when $ J \neq 0 $, for all the transformations that $ A^{\left( g \right) } _{v} $ is capable of doing.
				
				Regardless of the \textquotedblleft exoticity\textquotedblright \hspace*{0.01cm} of what has just been said, the fact is that, since we already said that the qudits assigned to the lattice edges must be interpreted as the $ D_{M} \left( G \right) $ lattice gauge fields, this allows us to conclude that all the transformations that $ A^{\left( g \right) } _{v} $ performs are naturally lattice gauge transformations. But while this conclusion is correct, a relevant question is: how can we justify this conclusion in a more fundamental way? And this is a relevant question that, for instance, can be answered by explaining what is the actual role that the operators $ B_{f,L} $ have in these models. After all, in order to explain this actual role of $ B_{f,L} $, it is imperative to note that the product $ a^{-1} b^{-1} cd $, which appears explicitly in the definition of $ B^{\left( g \right) } _{f} $, is one of the \emph{holonomies} that can be calculated by using all the gauge qudits around the $ f $-th lattice face [\citen{mf-lattice-gauge}]. And since the sum that defines each $ B_{f,L} $ is constrained to the fact that this operator always returns [\citen{mf-lattice-gauge}]
				\begin{itemize}
					\item $ 1 $, if the holonomy of the lattice face on which it acts effectively belongs to $ \mathsf{C} _{L} $, and
					\item $ 0 $, otherwise,
				\end{itemize}
				it is natural to conclude that each $ B_{f,L} $ can be interpreted as a kind of \textquotedblleft holonomy meter\textquotedblright \hspace*{0.01cm}.
				
				Note that, as a consequence of this natural conclusion, $ B_{f,0} $ can be recognized as an operator that measures only flat connections: i.e., it measures only trivial holonomies that are characterized by $ e $ along the faces. Thus, by remembering that
				\begin{itemize}
					\item it is precisely this $ B_{f,0} $ that make up the Hamiltonian (\ref{H-qdmv}), and
					\item the smallest energy eigenvalue (\ref{e0-energy}) is obtained only when, for instance, all the face holonomies of $ \mathcal{L} _{2} $ belong to $ \mathsf{C} _{0} $,
				\end{itemize}
				we can assert that, when these $ D_{M} \left( G \right) $ models are in their ground states, $ \mathcal{L} _{2} $ is locally flat. Of course, the converse of this assertion is not true since, for instance, the $ D_{M} \left( G \right) $ models have non-vacuum states $ \left\vert \xi \right\rangle $ where
				\begin{equation*}
					B_{f} \left\vert \xi \right\rangle = \left\vert \xi \right\rangle
				\end{equation*}
				hold for all the $ N_{f} $ faces of $ \mathcal{L} _{2} $. But the act of interpreting these face operators as \textquotedblleft holonomy meters\textquotedblright \hspace*{0.01cm} is quite revealing: after all, as (\ref{dm-solubility}) shows us that $ A_{v,J} $ and $ B_{f,L} $ commute among them for all values of $ v $, $ f $, $ J $ and $ L $, this means that all the holonomies measured by $ B_{f,L} $ continue to belong to the same $ \mathsf{C} _{L} $ after the action of $ A_{v,J} $ on $ \mathcal{L} _{2} $. That is, since these face holonomies can be associated with local estimates of how curved is the $ 2 $-dimensional manifold $ \mathcal{M} _{2} $ that $ \mathcal{L} _{2} $ discretizes, all these local (non-)deformations are preserved under the transformations that all the operators $ A_{v,J} $ are capable of doing [\citen{mf-lattice-gauge}]. In this fashion, as this geometric point of view is analogous to the one that underlies all the continuous gauge theories [\citen{mf-gauge,gitman}], it is precisely this that allows us to assert that the action of these operators $ A^{\left( g \right) } _{v} $ and $ A_{v,J} $ can be interpreted as lattice gauge transformations. Note that this is one of the things that explains, for instance, the fact that one of the $ D_{M} \left( G \right) $ vacuum states is
				\begin{equation}
					\bigl\vert \xi ^{\left( 0 \right) } _{0} \bigr\rangle = \prod _{v \in \mathcal{L}} A_{v} \underbrace{\ \left\vert e \right\rangle \otimes \ldots \otimes \left\vert e \right\rangle \ } _{N_{\ell } \ \textnormal{times}} \otimes \underbrace{\ \left\vert 0 \right\rangle \otimes \ldots \otimes \left\vert 0 \right\rangle \ } _{N_{v} \ \textnormal{times}} \label{vacuo-1} 
				\end{equation}
				because, in addition to all the face holonomies of \textquotedblleft seed\textquotedblright
				\begin{equation*}
					\underbrace{\ \left\vert e \right\rangle \otimes \ldots \otimes \left\vert e \right\rangle \ } _{N_{\ell } \ \textnormal{times}} \otimes \underbrace{\ \left\vert 0 \right\rangle \otimes \ldots \otimes \left\vert 0 \right\rangle \ } _{N_{v} \ \textnormal{times}}
				\end{equation*}
				belonging to $ \mathsf{C} _{0} $, the action of the operator $ \prod _{v \in \mathcal{L}} A_{v} $ is unable to change them. 
				
			\subsubsection{\label{correspondence-principle} What does the correspondence principle has to tell us about these $ D_{M} \left( G \right) $ models?}
			
				Given what has just been said about all these vertex and face operators, it is quite clear that there is a symbiosis between them with regard, for instance, to the interpretation of the $ D_{M} \left( G \right) $ models as examples of the lattice gauge theories. Nevertheless, it is worth noting that, despite what has been said to be correct, we still have not talked about the role that the group action $ \mu $ have in these models. And although it seems strange to conclude that the operators $ A_{v,J} $ perform lattice gauge transformations by disregarding, for instance, what they do on the matter fields, the truth is that this conclusion was inherited from the $ D \left( G \right) $ models: i.e., this conclusion was inherited from lattice gauge models where there are no matter fields [\citen{mf-lattice-gauge}].
				
				In order to understand this last comment, it is important to note, at least, two things and the first one is that the $ D \left( G \right) $ models, whose Hamiltonian is
				\begin{equation}
					H_{D \left( G \right) } = - \mathcal{J} _{A} \sum _{v  \in \mathcal{L} _{2}} A_{v,0} - \mathcal{J} _{B} \sum _{f \in \mathcal{L} _{2}} \ B_{f,0} \ , \label{hg-hamiltonian}
				\end{equation}
				are defined by using vertex ($ A_{v,J} $) and face ($ B_{f,L} $) projectors that, despite being similar to those in (\ref{all-operators}), have a small difference: the components of these $ D \left( G \right) $ projectors are those in Figure \ref{kuperberg-figure}.
				\begin{figure}[t!]
					\begin{center}
		    	   			\begin{tikzpicture}[
							scale=0.3,
							equation/.style={thin},
							trans/.style={thin,shorten >=0.5pt,shorten <=0.5pt,>=stealth},
							flecha/.style={thin,->,shorten >=0.5pt,shorten <=0.5pt,>=stealth}
							]
							\draw[equation] (-9.8,0.15) -- (-9.8,0.15) node[midway,right] {$ A^{\left( g \right) } _{v} $};
							\draw[trans] (-6.8,2.3) -- (-6.8,-2.3) node[above=2pt,right=-1pt] {};
							\draw[trans] (-0.1,2.3) -- (1.1,-0.06) node[above=2pt,right=-1pt] {};
							\draw[trans] (-0.1,-2.3) -- (1.1,0.06) node[above=2pt,right=-1pt] {};
							\draw[flecha] (-5.1,0.0) -- (-3.9,0.0) node[above=2pt,right=-1pt] {};
							\draw[flecha] (-4.1,0.0) -- (-1.6,0.0) node[above=2pt,right=-1pt] {};
							\draw[trans] (-1.8,0.0) -- (-0.9,0.0) node[above=2pt,right=-1pt] {};
							\draw[flecha] (-3.0,-1.4) -- (-3.0,-0.5) node[above=2pt,right=-1pt] {};
							\draw[flecha] (-3.0,-0.7) -- (-3.0,1.0) node[above=2pt,right=-1pt] {};
							\draw[trans] (-3.0,0.8) -- (-3.0,1.4) node[above=2pt,right=-1pt] {};
							\draw[equation] (-3.0,2.1) -- (-3.0,2.1) node[midway] {$ a $};
							\draw[equation] (0.6,0.0) -- (0.6,0.0) node[midway,left] {$ b $};
							\draw[equation] (-3.0,-2.1) -- (-3.0,-2.1) node[midway] {$ c $};
							\draw[equation] (-6.7,0.0) -- (-6.7,0.0) node[midway,right] {$ d $};
							\draw[equation] (2.5,-0.07) -- (2.5,-0.07) node[midway] {$ = $};
							\draw[trans] (4.3,2.3) -- (4.3,-2.3) node[above=2pt,right=-1pt] {};
							\draw[trans] (13.6,2.3) -- (14.8,-0.06) node[above=2pt,right=-1pt] {};
							\draw[trans] (13.6,-2.3) -- (14.8,0.06) node[above=2pt,right=-1pt] {};
							\draw[flecha] (7.9,0.0) -- (9.1,0.0) node[above=2pt,right=-1pt] {};
							\draw[flecha] (8.9,0.0) -- (11.4,0.0) node[above=2pt,right=-1pt] {};
							\draw[trans] (11.2,0.0) -- (12.1,0.0) node[above=2pt,right=-1pt] {};
							\draw[flecha] (10.0,-1.4) -- (10.0,-0.5) node[above=2pt,right=-1pt] {};
							\draw[flecha] (10.0,-0.7) -- (10.0,1.0) node[above=2pt,right=-1pt] {};
							\draw[trans] (10.0,0.8) -- (10.0,1.4) node[above=2pt,right=-1pt] {};
							\draw[equation] (10.1,2.0) -- (10.1,2.0) node[midway] {$ ga $};
							\draw[equation] (14.2,-0.2) -- (14.2,-0.2) node[midway,left] {$ gb $};
							\draw[equation] (10.1,-2.0) -- (10.1,-2.0) node[midway] {$ cg^{-1}  $};
							\draw[equation] (4.4,0.0) -- (4.4,0.0) node[midway,right] {$ dg^{-1}  $};
						\end{tikzpicture} \\
						\begin{tikzpicture}[
							scale=0.3,
							equation/.style={thin},
							trans/.style={thin,shorten >=0.5pt,shorten <=0.5pt,>=stealth},
							flecha/.style={thin,->,shorten >=0.5pt,shorten <=0.5pt,>=stealth}
							]
							\draw[equation] (-8.8,0.0) -- (-8.8,0.0) node[midway,right] {$ B^{\left( h \right) } _{f} $};
							\draw[trans] (-5.6,2.3) -- (-5.6,-2.3) node[above=2pt,right=-1pt] {};
							\draw[trans] (-0.1,2.3) -- (1.1,-0.06) node[above=2pt,right=-1pt] {};
							\draw[trans] (-0.1,-2.3) -- (1.1,0.06) node[above=2pt,right=-1pt] {};
							\draw[flecha] (-4.7,1.2) -- (-2.1,1.2) node[above=2pt,right=-1pt] {};
							\draw[trans] (-2.3,1.2) -- (-0.1,1.2) node[above=2pt,right=-1pt] {};
							\draw[flecha] (-4.7,-1.2) -- (-2.1,-1.2) node[above=2pt,right=-1pt] {};
							\draw[trans] (-2.3,-1.2) -- (-0.1,-1.2) node[above=2pt,right=-1pt] {};
							\draw[flecha] (-3.8,-2.0) -- (-3.8,0.3) node[above=2pt,right=-1pt] {};
							\draw[trans] (-3.8,0.0) -- (-3.8,2.0) node[above=2pt,right=-1pt] {};
							\draw[flecha] (-1.0,-2.0) -- (-1.0,0.3) node[above=2pt,right=-1pt] {};
							\draw[trans] (-1.0,0.0) -- (-1.0,2.0) node[above=2pt,right=-1pt] {};
							\draw[equation] (-2.3,2.0) -- (-2.3,2.0) node[midway] {$ a $};				
							\draw[equation] (0.7,0.0) -- (0.7,0.0) node[midway,left] {$ d $};
							\draw[equation] (-2.3,-2.0) -- (-2.3,-2.0) node[midway] {$ c $};
							\draw[equation] (-5.5,0.0) -- (-5.5,0.0) node[midway,right] {$ b $};
							\draw[equation] (6.7,-0.07) -- (6.7,-0.07) node[midway] {$ =  \ \delta \left( h , a^{-1} b^{-1} cd \right) $};
							\draw[trans] (12.2,2.3) -- (12.2,-2.3) node[above=2pt,right=-1pt] {};
							\draw[trans] (17.7,2.3) -- (18.8,-0.06) node[above=2pt,right=-1pt] {};
							\draw[trans] (17.7,-2.3) -- (18.8,0.06) node[above=2pt,right=-1pt] {};
							\draw[flecha] (13.0,1.2) -- (15.6,1.2) node[above=2pt,right=-1pt] {};
							\draw[trans] (15.4,1.2) -- (17.6,1.2) node[above=2pt,right=-1pt] {};
							\draw[flecha] (13.0,-1.2) -- (15.6,-1.2) node[above=2pt,right=-1pt] {};
							\draw[trans] (15.4,-1.2) -- (17.6,-1.2) node[above=2pt,right=-1pt] {};
							\draw[flecha] (13.9,-2.0) -- (13.9,0.3) node[above=2pt,right=-1pt] {};
							\draw[trans] (13.9,0.0) -- (13.9,2.0) node[above=2pt,right=-1pt] {};
							\draw[flecha] (16.7,-2.0) -- (16.7,0.3) node[above=2pt,right=-1pt] {};
							\draw[trans] (16.7,0.0) -- (16.7,2.0) node[above=2pt,right=-1pt] {};
							\draw[equation] (15.4,2.0) -- (15.4,2.0) node[midway] {$ a $};
							\draw[equation] (18.4,0.0) -- (18.4,0.0) node[midway,left] {$ d $};
							\draw[equation] (15.4,-2.0) -- (15.4,-2.0) node[midway] {$ c $};
							\draw[equation] (12.2,0.0) -- (12.2,0.0) node[midway,right] {$ b $};
						\end{tikzpicture}
					\end{center}
					\caption{\label{kuperberg-figure} Definition of the components $ A^{\left( g \right) } _{v} $ and $ B^{\left( h \right) } _{f} $, which define the $ D \left( G \right) $ vertex and face operators, in terms of their effective action on $ \mathcal{L} _{2} $. Note that, since these $ D \left( G \right) $ vertex and face operators are also expressed as $ A_{v,J} = \frac{1}{\vert G \vert } \sum _{g \in G} \chi _{1+J} \left( g^{-1} \right) \cdot A^{\left( g \right) } _{v} $ and $ B_{f,L} \equiv \sum _{g \in \mathsf{C} _{L}} B^{\left( g \right) } _{f} $ respectively, the definition of these components makes it clear, for instance, that the $ D \left( G \right) $ and $ D_{M} \left( G \right) $ face operators are exactly the same.}
				\end{figure}
				And why is it important to note this first thing? Because the only difference between the $ D \left( G \right) $ and $ D_{M} \left( G \right) $ vertex operators concerns precisely the action on the matter fields. After all, even if there is (were) some matter field assigned with the vertices of $ \mathcal{L} _{2} $, Figure \ref{kuperberg-figure} makes it quite clear that the $ D \left( G \right) $ vertex operators are (would be) unable to act on these matter fields. In this fashion, as the $ D \left( G \right) $ and $ D_{M} \left( G \right) $ vertex operators act on the gauge fields in exactly the same way, this inability of the $ D \left( G \right) $ vertex operators allows us to conclude that they can actually be interpreted as $ D_{M} \left( G \right) $ vertex operators that are \textquotedblleft blind\textquotedblright \hspace*{0.01cm} to the matter fields, regardless of whether these matter fields are on the lattice vertices or not.
				
				Note that, as strange as this interpretation may sound at first glance, there is no way not to recognize that it makes sense because, by bearing in mind that
				\begin{itemize}
					\item quantum-computational models try/need to model some reality that can be physically implemented, and
					\item the $ D_{M} \left( G \right) $ models are intentionally defined to be seen as generalizations of the $ D \left( G \right) $ ones,
				\end{itemize}
				there must be a (mathematical) correspondence between these two classes of lattice models in such a way that the $ D \left( G \right) $ models can be recovered as special cases of the $ D_{M} \left( G \right) $ ones. In plain English, these two classes of lattice models need to respect the same kind of correspondence principle that the most diverse physical theories respect when they are formulated [\citen{costa}]. Therefore, as the $ D \left( G \right) $ vertex operators are \textquotedblleft blind\textquotedblright \hspace*{0.01cm} to the presence of matter fields, two good ways to make the $ D_{M} \left( G \right) $ vertex operators become \textquotedblleft blind\textquotedblright \hspace*{0.01cm} to these matter fields seems to be imposing that
				\begin{itemize}
					\item the group action be such that $ \mu \left( g , \alpha \right) = \alpha $ for all $ g \in G $ because, in this case, there will be no change on the matter fields, or/and
					\item $ M $ is equal to $ 1 $ since, in this other case, the $ D_{M} \left( G \right) $ vertex operators will be unable to make any changes on the matter fields due to lack of options.
				\end{itemize}
				
			\subsubsection{The presence of the Potts model}
			
				But, before we say whether (or which of) these two \textquotedblleft good ways\textquotedblright \hspace*{0.01cm} really work to recover the $ D \left( G \right) $ models as special cases of the $ D_{M} \left( G \right) $ ones, we need to remember that the $ D_{M} \left( G \right) $ Hamiltonian (\ref{H-qdmv}) is not defined only by vertex and face operators: it is also defined by a link operator $ C_{\ell , 0} $ whose meaning, in addition to not having been discussed so far, is starred by $ \mu $. And with regard to the components $ C^{\left( \alpha \right) } _{\ell } $ that appear in Figure \ref{QMDv-operators-components}, it is possible to assert that they are defined in this way only to make it possible to interpret these $ D_{M} \left( G \right) $ models in terms of a coupling of the $ D \left( G \right) $ models with the \emph{Potts models} [\citen{potts,wu}]. As a matter of fact, by analysing a hypothetical situation where, for instance, $ \mathcal{J} _{A} $ and $ \mathcal{J} _{B} $ could be taken as null, it is not difficult to see that the Hamiltonian (\ref{H-qdmv}) would reduce to
				\begin{equation}
					H_{D_{M} \left( G \right) } = - \mathcal{J} _{C} \sum _{\ell \in \mathcal{L} _{2}} C_{\ell } \ , \label{potts-reduction}
				\end{equation}
				which is \textquotedblleft exactly\textquotedblright \hspace*{0.01cm} the same expression as the interaction Hamiltonian of a Potts model.
				
				It is obvious that you, the reader, may argue that, due to the presence of $ \mu $ in the Kronecker deltas of $ C^{\left( \alpha \right) } _{\ell } $, there is a deep difference between (\ref{potts-reduction}) and the interaction Hamiltonian operator of the original Potts model. And although this argument is perfectly correct, it is worth stressing, once again, what we have said in the last paragraph: the $ D_{M} \left( G \right) $ models should be interpreted in terms of a coupling of the $ D \left( G \right) $ models with the Potts ones. That is, the $ D \left( G \right) $ and Potts models cannot be disconnected from each other in order to define these $ D_{M} \left( G \right) $ models: these two classes of models must be coupled to each other and the main responsible for this coupling is $ \mu $ because, due to its presence in the vertex and link operators, it allows to check whether the matter fields have been manipulated or not.
				
				Observe that, since all the elements of $ \mathcal{B} _{m} $ are orthonormal vectors and, therefore,
				\begin{equation}
					\delta \left( \mu \left( a , \alpha \right) , \beta \right) = \left\langle \hspace*{0.04cm} \mu \left( a , \alpha \right) \vert \hspace*{0.04cm} \beta \hspace*{0.04cm} \right\rangle \ , \label{inner-prod}
				\end{equation}
				it is not difficult to see, for example, that the smallest eigenvalues of (\ref{potts-reduction}) are associated with the eigenstates where the matter fields are aligned from the $ \mu $ point of view. That is, (\ref{inner-prod}) is a result that allows us to interpret all the link operators $ C_{\ell , \Lambda } $ as \emph{comparators}, since each component $ C^{\left( \alpha \right) } _{\ell } $ measures the alignment of two neighbouring matter fields (i.e., of two matter fields that are assigned with the two boundary vertices of the $ \ell $-th edge) from this $ \mu $ point of view. And as these matter fields also need to be aligned in the ground state of the $ D_{M} \left( G \right) $ models where $ \mathcal{J} _{A} $, $ \mathcal{J} _{B} $ and $ \mathcal{J} _{C} $ are positive parameters, it is not difficult to see that the only way to make the $ D_{M} \left( G \right) $ vertex operators \textquotedblleft blind\textquotedblright \hspace*{0.01cm} to the matter fields (and, therefore, recover the $ D \left( G \right) $ models as a special case of the $ D_{M} \left( G \right) $ ones) is by taking $ M = 1 $. After all, since (\ref{inner-prod}) will always be equal to $ 1 $ when $ M = 1 $ because $ \mu \left( a , 0 \right) = 0 $, the $ D_{M} \left( G \right) $ Hamiltonian will reduce to
				\begin{equation}
					\left. H_{D_{M} \left( G \right) } \right\vert _{M=1} = - \mathcal{J} _{A} \sum _{v  \in \mathcal{L} _{2}} A_{v} - \mathcal{J} _{B} \sum _{f \in \mathcal{L} _{2}} \ B_{f} - \left( \mathcal{J} _{C} N_{\ell } \right) \cdot \mathds{1} _{\mathcal{L} _{2}} \label{H1-qdmv}
				\end{equation}
				in this case, a result that is identical to (\ref{hg-hamiltonian}) except for one constant. But since, whenever we measure any energy, we are always actually measuring energy differences, the presence of this constant does not prevent us from recognizing that (\ref{H1-qdmv}) is, indeed, one of the possible Hamiltonian of the $ D \left( G \right) $ models [\citen{kitaev-1,pachos,naaij}]. 
		
		\subsection{\label{QDMv-properties} General properties of the $ D_{M} \left( G \right) $ models}
		
			For the sake of completeness, it is also important to point out that, in the case of the $ D_{M} \left( G \right) $ non-vacuum states, all the energy excitations that characterize them are produced through the action of operators $ W^{\left( J , L , \Lambda \right) } _{\ell } $ and $ W^{\left( J , \Lambda \right) } _{v} $ that, due to the projectivity of $ A_{v,J} $, $ B_{f,L} $ and $ C_{\ell , \Lambda } $, are respectively such that
			\begin{subequations} \label{DG-quasiparticles-creation}
				\begin{align}
					W^{\left( J , L , \Lambda \right) } _{\ell } \circ A_{v,0} & = A_{v,J} \circ W^{\left( J , L , \Lambda \right) } _{\ell } \ , \label{DG-quasiparticles-creation-a} \\
					W^{\left( J , L , \Lambda \right) } _{\ell } \circ B_{f,0} & = B_{f,L} \circ W^{\left( J , L , \Lambda \right) } _{\ell } \ , \label{DG-quasiparticles-creation-b} \\
					W^{\left( J , L , \Lambda \right) } _{\ell } \circ C_{\ell ,0} & = C_{\ell , \Lambda } \circ W^{\left( J , L , \Lambda \right) } _{\ell } \ , \quad \textnormal{and} \label{DG-quasiparticles-creation-c}
				\end{align}
			\end{subequations}
			\begin{subequations} \label{matter-quasiparticles-creation}
				\begin{align}
					W^{\left( J , \Lambda \right) } _{v} \circ A_{v,0} & = A_{v, J} \circ W^{\left( J , \Lambda \right) } _{v} \quad \textnormal{and} \label{matter-quasiparticles-creation-a} \\
					W^{\left( J , \Lambda \right) } _{v} \circ C_{\ell ,0} & = C_{\ell , \Lambda } \circ W^{\left( J , \Lambda \right) } _{v} \label{matter-quasiparticles-creation-c} \ .
				\end{align}
			\end{subequations}
			And by remembering, once again, that
			\begin{itemize}
				\item quantum-computational models try/need to model some reality that can be physically implemented, and
				\item these $ D_{M} \left( G \right) $ models were defined as the computational analogues of some lattice gauge theories,
			\end{itemize}
			it becomes quite clear that, at least, all the fusion rules
			\begin{eqnarray*}
				q^{\left( J^{\prime } , L^{\prime } , \Lambda ^{\prime } \right) } \times q^{\left( J^{\prime \prime } , L^{\prime \prime } , \Lambda ^{\prime \prime } \right) } & = & q^{\left( J^{\prime \prime } , L^{\prime \prime } , \Lambda ^{\prime \prime } \right) } \times q^{\left( J^{\prime } , L^{\prime } , \Lambda ^{\prime } \right) } \ , \\
					q^{\left( J^{\prime } , L^{\prime } , \Lambda ^{\prime } \right) } \times Q^{\left( J^{\prime \prime } , \Lambda ^{\prime } \right) } & = & \ Q^{\left( J^{\prime \prime } , \Lambda ^{\prime } \right) } \times q^{\left( J^{\prime } , L^{\prime } , \Lambda ^{\prime } \right) } \quad \textnormal{and} \\
					Q^{\left( J^{\prime } , \Lambda ^{\prime } \right) } \times Q^{\left( J^{\prime \prime } , \Lambda ^{\prime \prime } \right) } & = & Q^{\left( J^{\prime \prime } , \Lambda ^{\prime \prime } \right) } \times Q^{\left( J^{\prime } , \Lambda ^{\prime } \right) }
			\end{eqnarray*}
			need to be satisfied so that all the $ D_{M} \left( G \right) $ energy excitations $ q^{\left( J , L , \Lambda \right) } $ and $ Q^{\left( J , \Lambda \right) } $, which are locally produced by the action of the operators $ W^{\left( J , L , \Lambda \right) } _{\ell } $ and $ W^{\left( J , \Lambda \right) } _{v} $ respectively, can be interpreted as quasiparticles.
			
			By the way, since we are talking about quasiparticles, it is important to point out that, due to the correspondence principle $ \left. D_{M} \left( G \right) \right\vert _{M=1} = D \left( G \right) $, it is not difficult to conclude that all the $ D \left( G \right) $ quasiparticles are also included, in some way, in the $ D_{M} \left( G \right) $ models. However, as these two classes of lattice models are not equal when $ M \neq 1 $, something different must happen to these quasiparticles that were inherited from the $ D \left( G \right) $ models when $ \mu $ is, for example, a non-trivial group action.
			
			\subsubsection{\label{confined-tcm}The Toric Code coupled to matter fields as an example}	
			
				In order to begin to understanding what is different about the quasiparticles that were inherited from the $ D \left( G \right) $ models, it is interesting to analyse the general properties of the cyclic Abelian $ D_{2} \left( \mathds{Z} _{2} \right) $ model: i.e., it is interesting to analyse the general properties of a lattice model that, when defined by using an $ \mathcal{L} _{2} $ that discretizes a two-dimensional torus, can be recognized as a Toric Code coupled to the Ising model. And according to what was discussed above, it is not difficult to see that the matrix representations of the $ D_{2} \left( \mathds{Z} _{2} \right) $ vertex, face and link operators are given by [\citen{james}]
				\begin{subequations} \label{abc-tc}
					\begin{align}
						A_{v,J} = \frac{1}{2} \sum _{g \in \mathds{Z} _{2}} \left( -1 \right) ^{Jg} \cdot M _{v} \left( g \right) \prod _{\ell ^{\prime } \in S_{v}} \left( \sigma ^{x} _{\ell ^{\prime }} \right) ^{g} \ , \label{a-tc} \\
						B_{f,L} = \frac{1}{2} \sum _{g \in \mathds{Z} _{2}} \left( -1 \right) ^{Lg} \prod _{\ell ^{\prime } \in S_{f}} \left( \sigma ^{z} _{\ell ^{\prime }} \right) ^{g} \quad \textnormal{and} \label{b-tc} \\
						C_{\ell ,\Lambda } = \frac{1}{2} \sum _{g \in \mathds{Z} _{2}} \left( -1 \right) ^{\Lambda g} \cdot M _{\ell } \left( g \right) \prod _{v \in S_{\ell }} \left( \sigma ^{z} _{v} \right) ^{g} \label{c-tc}
					\end{align}
				\end{subequations}
				respectively, where [\citen{miguel-1,pramod-wrong}]
				\begin{subequations} \label{matter-action-22}
					\begin{align}
						M _{v} \left( g \right) & = \left( \sigma ^{x} _{v} \right) ^{g} \quad \textnormal{and} \label{matter-action-22a} \\
						M _{\ell } \left( g \right) & = \left( \sigma ^{z} _{\ell } \right) ^{g} \ . \label{matter-action-22b}
					\end{align}
				\end{subequations}
				After all, since the set $ \left\{ M \left( g \right) : g \in G \right\} $ is composed of matrices that represent the gauge group $ G $ and, consequently, the group action $ \mu $ [\citen{pramod-wrong}]\label{M-set-representation}, something that is no longer difficult to observe is, for instance, that all these operators (\ref{abc-tc}) satisfy all the conditions \textbf{(a)}, \textbf{(b)} and \textbf{(c)}.
				
				By the way, another thing that is no longer difficult to observe is that, due to the fact that (\ref{a-tc}) and (\ref{b-tc}) are represented with the help of the Pauli matrices $ \sigma ^{x} $ and $ \sigma ^{z} $ [\citen{arfken}], the operators $ W^{\left( J , L , \Lambda \right) } _{\ell } $ that produce quasiparticles in this $ D_{2} \left( \mathds{Z} _{2} \right) $ model can be represented by\footnote{\label{footnote-qft} Here, we think it is better to write this representation as (\ref{D2-gauge-quasiparticles}), rather than the one
				\begin{equation*}
					W^{\left( 1 , 0 , 0 \right) } _{\ell } = \sigma ^{z} _{\ell } \ , \ \ W^{\left( 0 , 1 , 1 \right) } _{\ell } = \sigma ^{x} _{\ell } \quad \textnormal{and} \quad W^{\left( 1 , 1 , 1 \right) } _{\ell } = \textnormal{\textquotedblleft } \ \sigma ^{y} _{\ell } \ \textnormal{\textquotedblright } = \sigma ^{x} _{\ell } \circ \sigma ^{z} _{\ell } = \sigma ^{z} _{\ell } \circ \sigma ^{x} _{\ell } 
				\end{equation*}
				we wrote in Ref. [\citen{mf-pedagogical}], because it places more emphasis on the fact that these operators need to be expressed in terms of those that compose the $ D_{2} \left( \mathds{Z} _{2} \right) $ Hamiltonian. After all, it is always good to remember that, as well as in QFT (where Hamiltonians can be expressed in the \emph{Fock representation} by using the creation $ a^{\dagger } $ and annihilation $ a $ operators [\citen{itzykson}]), the entire $ D \left( \mathds{Z} _{2} \right) $ energy spectrum can also be well understood from [\citen{mf-pedagogical}]
				\begin{itemize}
					\item the knowledge of the ground state of these models, and
					\item the excitations produced by the action of the operators that compose its Hamiltonian on this ground state.
				\end{itemize}
				Note that write (\ref{D2-gauge-quasiparticles}) is also very welcome because, when $ J = L = \Lambda = 0 $, the operator $ W^{\left( J , L , \Lambda \right) } _{\ell } $ can be identified as those that produces (a pair of) vacuum quasiparticles.}				
				\begin{equation}
					W^{\left( J , L , \Lambda \right) } _{\ell } = \left( \sigma ^{z} _{\ell } \right) ^{J} \circ \left( \sigma ^{x} _{\ell } \right) ^{L} \quad \textnormal{or} \quad W^{\left( J , L , \Lambda \right) } _{\ell } = \left( \sigma ^{x} _{\ell } \right) ^{L} \circ \left( \sigma ^{z} _{\ell } \right) ^{J} \ . \label{D2-gauge-quasiparticles}
				\end{equation}
				That is, these operators $ W^{\left( J , L , \Lambda \right) } _{\ell } $ have the same expression as those that produce quasiparticles in the $ D \left( \mathds{Z} _{2} \right) $ model and, as in this same model, these operators always produce pairs of quasiparticles that are detectable by the face and link operators: scilicet,
				\begin{itemize}
					\item $ A_{v} $ detects one quasiparticle $ q^{\left( 1 , 0 , 0 \right) } $, on each vertex that delimits the $ \ell $-th edge of $ \mathcal{L} _{2} $, after
					\begin{equation*}
						W^{\left( 1 , 0 , 0 \right) } _{\ell } = \sigma ^{z} _{\ell }
					\end{equation*}
					acts on this edge and changes the lattice gauge field from $ \left\vert 0 \right\rangle $ to $ \left\vert 1 \right\rangle $, and
					\item $ B_{f} $ indicates that
					\begin{equation*}
						W^{\left( 0 , 1 , 1 \right) } _{\ell } = \sigma ^{x} _{\ell } \ ,
					\end{equation*}
					by acting on this $ \ell $-th edge, produces one quasiparticle $ q^{\left( 0 , 1 , 1 \right) } $ on each face that shares this edge.
				\end{itemize}
				And, if we really recognize that the operators (\ref{D2-gauge-quasiparticles}) are the same ones that produce the $ D \left( \mathds{Z} _{2} \right) $ quasiparticles, it is not difficult to conclude, for instance, that $ W^{\left( 1 , 1 , 1 \right) } _{\ell } $ also produces another pair of quasiparticles because the fusion
				\begin{equation*}
					q^{\left( 1 , 0 , 0 \right) } \times q^{\left( 0 , 1 , 1 \right) }
				\end{equation*}
				results in a non-elementary quasiparticle, which is the dyon $ q^{\left( 1 , 1 , 1 \right) } $.
				
			\subsubsection{On the confinement of the $ D_{2} \left( \mathds{Z} _{2} \right) $ magnetic quasiparticles}	
				
				Given what we said in the previous paragraph, perhaps you, the reader, are wondering about the fact that, while $ W^{\left( 1 , 0 , 0 \right) } _{\ell } $ is indexed by $ \Lambda = 0 $, the operator $ W^{\left( 0 , 1 , 1 \right) } _{\ell } $ is indexed by $ \Lambda = 1 $. After all, if these two operators actually produce the same quasiparticles of the $ D \left( \mathds{Z} _{2} \right) $ model, should not the quasiparticles $ q^{\left( 1 , 0 , 0 \right) } $ and $ q^{\left( 0 , 1 , 1 \right) } $ be duals of each other and, therefore, should not these two operators/quasiparticles be indexed by the same value of $ \Lambda $? And if you are asking this, know that this is an excellent question, since it gives us the opportunity to answer what is different about these quasiparticles that were inherited from the $ D \left( \mathds{Z} _{2} \right) $ model.
				
				The simplest way to answer this is by noting that, in accordance with (\ref{DG-quasiparticles-creation}), every pair of quasiparticles, which is produced by an operator $ W^{\left( J , L , \Lambda \right) } _{\ell } $ where $ \Lambda $ is non-zero, is also detectable by the operator $ C_{\ell } $. That is to say, while a pair of quasiparticles produced by $ W^{\left( 1 , 0 , 0 \right) } _{\ell } $ is not detected by $ C_{\ell } $, a pair of quasiparticles produced by $ W^{\left( 0 , 1 , 1 \right) } _{\ell } $ is. And from the point of view of the $ D_{2} \left( \mathds{Z} _{2} \right) $ Hamiltonian, this means that, if a pair of quasiparticles is produced by $ W^{\left( 1 , 0 , 0 \right) } _{\ell } $ over what was once a vacuum state, it will raise the energy of this lattice system to
				\begin{equation*}
					E_{e} = E_{0} + 2 \mathcal{J} _{A} \ ,
				\end{equation*}
				whereas the production made by $ W^{\left( 0 , 1 , 1 \right) } _{\ell } $ over the same vacuum state will raise the energy to
				\begin{equation*}
					E_{m} = E_{0} + 2 \mathcal{J} _{A} + \mathcal{J} _{C} \ . 
				\end{equation*}
				
				By the way, since we touched on the subject of energy, it is important to continue exploring this subject to explain what else is different about these pairs of quasiparticles. And as strange as this may seem at first, the main difference is directly associated with the possibility of transporting these quasiparticles over $ \mathcal{L} _{2} $.
				\begin{figure}[!t]
					\begin{center}
						\tikzstyle myBG=[line width=3pt,opacity=1.0]
						\newcommand{\drawLatticeLine}[2]
						{
							\draw[gray,very thick] (#1) -- (#2);
						}
						\newcommand{\drawLatticeLineFlex}[2]
						{
							\draw[->,gray,very thick,>=stealth] (#1) -- (#2);
						}
						\newcommand{\drawDashedLine}[2]
						{
							\draw[gray!30,dashed,very thick] (#1) -- (#2);
						}
						\newcommand{\drawExcitedLine}[2]{
							\draw[black,ultra thick] (#1) -- (#2);
						}
						\newcommand{\graphLinesHorizontal}
						{
							\drawLatticeLineFlex{2,1}{2,3.1};
							\drawLatticeLine{2,3}{2,5};
							\drawLatticeLineFlex{4,1}{4,3.1};
							\drawLatticeLine{4,3}{4,5};
							\drawLatticeLineFlex{6,1}{6,3.1};
							\drawLatticeLine{6,3}{6,5};
							\drawLatticeLineFlex{8,1}{8,3.1};
							\drawLatticeLine{8,3}{8,5};
							\drawLatticeLineFlex{10,1}{10,3.1};
							\drawLatticeLine{10,3}{10,5};
							\drawLatticeLineFlex{12,1}{12,3.1};
							\drawLatticeLine{12,3}{12,5};
							\drawLatticeLineFlex{1,2}{3.1,2};
							\drawLatticeLineFlex{3,2}{5.1,2};
							\drawLatticeLineFlex{5,2}{7.1,2};
							\drawLatticeLineFlex{7,2}{9.1,2};
							\drawLatticeLineFlex{9,2}{11.1,2};
							\drawLatticeLine{11,2}{13,2};
							\drawLatticeLineFlex{1,4}{3.1,4};
							\drawLatticeLineFlex{3,4}{5.1,4};
							\drawLatticeLineFlex{5,4}{7.1,4};
							\drawLatticeLineFlex{7,4}{9.1,4};
							\drawLatticeLineFlex{9,4}{11.1,4};
							\drawLatticeLine{11,4}{13,4};
							\drawExcitedLine{2,4}{4,4};
						}
						\begin{tikzpicture}
							\graphLinesHorizontal;
							\draw[color=red!70,fill=yellow!70] (2,4) circle (1.4ex);
							\draw[color=red!70,fill=red!70] (2,4) circle (1.2ex);
							\node [] (2,4) at (2,4) {$ {\textcolor{white}{\boldsymbol{-}}} $};
							\draw[color=red!70,fill=yellow!70] (4,4) circle (1.4ex);
							\draw[color=red!70,fill=red!70] (4,4) circle (1.2ex);
							\node [] (4,4) at (4,4) {$ {\textcolor{white}{\boldsymbol{+}}} $};
						\end{tikzpicture}
					\end{center} \bigskip
					\begin{center}
						\tikzstyle myBG=[line width=3pt,opacity=1.0]
						\newcommand{\drawLatticeLine}[2]
						{
							\draw[gray,very thick] (#1) -- (#2);
						}
						\newcommand{\drawLatticeLineFlex}[2]
						{
							\draw[->,gray,very thick,>=stealth] (#1) -- (#2);
						}
						\newcommand{\drawDashedLine}[2]
						{
							\draw[gray!30,dashed,very thick] (#1) -- (#2);
						}
						\newcommand{\drawExcitedLine}[2]{
							\draw[black,ultra thick] (#1) -- (#2);
						}
						\newcommand{\graphLinesHorizontal}
						{
							\drawLatticeLineFlex{2,1}{2,3.1};
							\drawLatticeLine{2,3}{2,5};
							\drawLatticeLineFlex{4,1}{4,3.1};
							\drawLatticeLine{4,3}{4,5};
							\drawLatticeLineFlex{6,1}{6,3.1};
							\drawLatticeLine{6,3}{6,5};
							\drawLatticeLineFlex{8,1}{8,3.1};
							\drawLatticeLine{8,3}{8,5};
							\drawLatticeLineFlex{10,1}{10,3.1};
							\drawLatticeLine{10,3}{10,5};
							\drawLatticeLineFlex{12,1}{12,3.1};
							\drawLatticeLine{12,3}{12,5};
							\drawLatticeLineFlex{1,2}{3.1,2};
							\drawLatticeLineFlex{3,2}{5.1,2};
							\drawLatticeLineFlex{5,2}{7.1,2};
							\drawLatticeLineFlex{7,2}{9.1,2};
							\drawLatticeLineFlex{9,2}{11.1,2};
							\drawLatticeLine{11,2}{13,2};
							\drawLatticeLineFlex{1,4}{3.1,4};
							\drawLatticeLineFlex{3,4}{5.1,4};
							\drawLatticeLineFlex{5,4}{7.1,4};
							\drawLatticeLineFlex{7,4}{9.1,4};
							\drawLatticeLineFlex{9,4}{11.1,4};
							\drawLatticeLine{11,4}{13,4};
							\drawExcitedLine{4,4}{4,2};
						}
						\begin{tikzpicture}
							\graphLinesHorizontal;
							\draw[color=red!70,fill=yellow!70] (2,4) circle (1.4ex);
							\draw[color=red!70,fill=red!70] (2,4) circle (1.2ex);
							\node [] (2,4) at (2,4) {$ {\textcolor{white}{\boldsymbol{-}}} $};
							\draw[color=red!70,fill=white] (4,4) circle (1.4ex);
							\draw[color=red!70,fill=yellow!70] (4,2) circle (1.4ex);
							\draw[color=red!70,fill=red!70] (4,2) circle (1.2ex);
							\node [] (4,2) at (4,2) {$ {\textcolor{white}{\boldsymbol{+}}} $};
						\end{tikzpicture}
					\end{center}
					\caption{\label{e-transporte} Piece of the same lattice region of $ \mathcal{L} _{2} $ at two different times, where we see a pair of quasiparticles $ q^{\left( 1 , 0, 0 \right) } $ and $ q^{\left( 1 , 0 , 0 \right) } _{-} $ (red outlined and purposely indexed with the \textquotedblleft $ + $\textquotedblright \hspace*{0.01cm} and \textquotedblleft $ - $\textquotedblright \hspace*{0.01cm} symbols respectively) of the $ D_{2} \left( \mathds{Z} _{2} \right) $ model. In the first instant $ t_{1} $ (above) we see these quasiparticles at the positions where they were produced due to the action of $ W^{\left( 1 , 0 , 0 \right) } _{\ell } $: i.e., on the two vertices that delimit the $ \ell $-th lattice edge (highlighted in black colour) on which $ W^{\left( 1 , 0 , 0 \right) } _{\ell } $ acts. In the second instant $ t_{2} > t_{1} $ (below) we have these same quasiparticles, but after one of them has been transported to one of the vertices that delimit the $ \ell ^{\prime } $-th lattice edge (which is also highlighted in black colour). After all, as these quasiparticles are their own anti-quasiparticles, the action of a new operator $ W^{\left( 1 , 0 , 0 \right) } _{\ell ^{\prime }} $ produces two new ones and, therefore, leads to a fusion $ q^{\left( 1 , 0, 0 \right) } _{+} \times q^{\left( 1 , 0 , 0 \right) } _{-} $ at the vertex (highlighted in white colour) that delimits these two lattice edges.}
				\end{figure}
				After all, since $ q^{\left( 1 , 0 , 0 \right) } $ can be interpreted as its own anti-quasiparticle (because $ \sigma ^{x,y,z} = \left( \sigma ^{x,y,z} \right) ^{-1} $), it is not difficult to see that an operator
				\begin{equation*}
					O \left( \boldsymbol{\gamma } \right) = \prod _{\ell ^{\prime } \in \boldsymbol{\gamma }} W^{\left( 1 , 0 , 0 \right) } _{\ell ^{\prime }}
				\end{equation*} 
				can transport it to another lattice vertex. For this, it is enough that $ \boldsymbol{\gamma } $ is a set of lattice edges that describes a continuous path that, as shown in Figure \ref{e-general-transporte},
				\begin{figure}[!t]
					\begin{center}
						\tikzstyle myBG=[line width=3pt,opacity=1.0]
						\newcommand{\drawLatticeLine}[2]
						{
							\draw[gray,very thick] (#1) -- (#2);
						}
						\newcommand{\drawLatticeLineFlex}[2]
						{
							\draw[->,gray,very thick,>=stealth] (#1) -- (#2);
						}
						\newcommand{\drawDashedLine}[2]
						{
							\draw[gray!30,dashed,very thick] (#1) -- (#2);
						}
						\newcommand{\drawExcitedLine}[2]{
							\draw[black,ultra thick] (#1) -- (#2);
						}
						\newcommand{\graphLinesHorizontal}
						{
							\drawLatticeLineFlex{2,1}{2,3.1};
							\drawLatticeLine{2,3}{2,5};
							\drawLatticeLineFlex{4,1}{4,3.1};
							\drawLatticeLine{4,3}{4,5};
							\drawLatticeLineFlex{6,1}{6,3.1};
							\drawLatticeLine{6,3}{6,5};
							\drawLatticeLineFlex{8,1}{8,3.1};
							\drawLatticeLine{8,3}{8,5};
							\drawLatticeLineFlex{10,1}{10,3.1};
							\drawLatticeLine{10,3}{10,5};
							\drawLatticeLineFlex{12,1}{12,3.1};
							\drawLatticeLine{12,3}{12,5};
							\drawLatticeLineFlex{1,2}{3.1,2};
							\drawLatticeLineFlex{3,2}{5.1,2};
							\drawLatticeLineFlex{5,2}{7.1,2};
							\drawLatticeLineFlex{7,2}{9.1,2};
							\drawLatticeLineFlex{9,2}{11.1,2};
							\drawLatticeLine{11,2}{13,2};
							\drawLatticeLineFlex{1,4}{3.1,4};
							\drawLatticeLineFlex{3,4}{5.1,4};
							\drawLatticeLineFlex{5,4}{7.1,4};
							\drawLatticeLineFlex{7,4}{9.1,4};
							\drawLatticeLineFlex{9,4}{11.1,4};
							\drawLatticeLine{11,4}{13,4};
							\drawExcitedLine{4,4}{4,2};
							\drawExcitedLine{4,2}{12,2};
							\drawExcitedLine{12,2}{12,4};
							\drawExcitedLine{12,4}{8,4};
						}
						\begin{tikzpicture}
							\graphLinesHorizontal;
							\draw[color=red!70,fill=yellow!70] (2,4) circle (1.4ex);
							\draw[color=red!70,fill=red!70] (2,4) circle (1.2ex);
							\node [] (2,4) at (2,4) {$ {\textcolor{white}{\boldsymbol{-}}} $};
							\draw[color=red!70,fill=white] (4,4) circle (1.4ex);
							\draw[color=red!70,fill=yellow!70] (8,4) circle (1.4ex);
							\draw[color=red!70,fill=red!70] (8,4) circle (1.2ex);
							\node [] (8,4) at (8,4) {$ {\textcolor{white}{\boldsymbol{+}}} $};
						\end{tikzpicture}
					\end{center}
					\caption{\label{e-general-transporte} Here, we see a transport of quasiparticles that is a little more general than the one shown in Figure \ref{e-general-transporte}, but that illustrates how the action of an operator $ O \left( \boldsymbol{\gamma } \right) $ works. After all, since an operator $ W^{\left( 1 , 0 , 0 \right) } _{\ell ^{\prime }} $ can be used to transport a quasiparticle to a neighbouring vertex, the operator $ O \left( \boldsymbol{\gamma } \right) $ makes use of this possibility to transport this same quasiparticle to more distant vertices. And a good way to understand this is to realize that the path $ \boldsymbol{\gamma } $ (which is highlighted in black colour) is composed of edges that are neighbours two by two because, as $ W^{\left( 1 , 0 , 0 \right) } _{\ell ^{\prime }} $ produces two quasiparticles by acting on an edge, $ \prod _{\ell ^{\prime } \in \boldsymbol{\gamma }} W^{\left( 1 , 0 , 0 \right) } _{\ell ^{\prime }} $ leads to a fusion $ q^{\left( 1 , 0, 0 \right) } _{+} \times q^{\left( 1 , 0 , 0 \right) } _{-} $ at all vertices that are shared by two edges in $ \boldsymbol{\gamma } \cup \left\{ \ell \right\} $. Thus, if we consider the same initial situation as in Figure \ref{e-general-transporte}, where two quasiparticles were produced by $ W^{\left( 1 , 0 , 0 \right) } _{\ell } $, all these fusions cause one of these quasiparticles to be transported to the vertex that delimits only $ \boldsymbol{\gamma } $ and not the $ \ell $-th lattice edge.} 
				\end{figure}
				is delimited by the same vertex where this $ q^{\left( 1 , 0 , 0 \right) } $ is. However, although it is also possible to transport a quasiparticle $ q^{\left( 0 , 1 , 1 \right) } $ to another lattice face by using an analogous operator 
				\begin{equation*}
					O \left( \boldsymbol{\gamma } ^{\ast } \right) = \prod _{\ell ^{\prime } \in \boldsymbol{\gamma } ^{\ast }} W^{\left( 0 , 1 , 1 \right) } _{\ell ^{\prime }} \ ,
				\end{equation*}
				where $ \boldsymbol{\gamma } ^{\ast } $ is a continuous dual path like the one shown in Figure \ref{confinamento-fluxo},
				\begin{figure}[!t]
					\begin{center}
						\tikzstyle myBG=[line width=3pt,opacity=1.0]
						\newcommand{\drawLatticeLine}[2]
						{
							\draw[gray,very thick] (#1) -- (#2);
						}
						\newcommand{\drawLatticeLineFlex}[2]
						{
							\draw[->,gray,very thick,>=stealth] (#1) -- (#2);
						}
						\newcommand{\drawDashedExcitedLine}[2]{
							\draw[black,dashed,ultra thick] (#1) -- (#2);
						}
						\newcommand{\graphLinesHorizontal}
						{
							\drawLatticeLineFlex{2,1}{2,3.1};
							\drawLatticeLine{2,3}{2,5};
							\drawLatticeLineFlex{4,1}{4,3.1};
							\drawLatticeLine{4,3}{4,5};
							\drawLatticeLineFlex{6,1}{6,3.1};
							\drawLatticeLine{6,3}{6,5};
							\drawLatticeLineFlex{8,1}{8,3.1};
							\drawLatticeLine{8,3}{8,5};
							\drawLatticeLineFlex{10,1}{10,3.1};
							\drawLatticeLine{10,3}{10,5};
							\drawLatticeLineFlex{12,1}{12,3.1};
							\drawLatticeLine{12,3}{12,5};
							\drawLatticeLineFlex{1,2}{3.1,2};
							\drawLatticeLineFlex{3,2}{5.1,2};
							\drawLatticeLineFlex{5,2}{7.1,2};
							\drawLatticeLineFlex{7,2}{9.1,2};
							\drawLatticeLineFlex{9,2}{11.1,2};
							\drawLatticeLine{11,2}{13,2};
							\drawLatticeLineFlex{1,4}{3.1,4};
							\drawLatticeLineFlex{3,4}{5.1,4};
							\drawLatticeLineFlex{5,4}{7.1,4};
							\drawLatticeLineFlex{7,4}{9.1,4};
							\drawLatticeLineFlex{9,4}{11.1,4};
							\drawLatticeLine{11,4}{13,4};
							\drawDashedExcitedLine{3,3}{5,3};
						}
						\begin{tikzpicture}
							\graphLinesHorizontal;
							\draw[color=blue!70,fill=yellow!70] (3,3) circle (1.4ex);
							\draw[color=blue!70,fill=blue!70] (3,3) circle (1.2ex);
							\node [] (3,3) at (3,3) {$ {\textcolor{white}{\boldsymbol{-}}} $};
							\draw[color=blue!70,fill=yellow!70] (5,3) circle (1.4ex);
							\draw[color=blue!70,fill=blue!70] (5,3) circle (1.2ex);
							\node [] (5,3) at (5,3) {$ {\textcolor{white}{\boldsymbol{+}}} $};
							\draw[color=black,fill=orange!70] (4,3) circle (0.7ex);
						\end{tikzpicture}
					\end{center} \bigskip
					\begin{center}
						\tikzstyle myBG=[line width=3pt,opacity=1.0]
						\newcommand{\drawLatticeLine}[2]
						{
							\draw[gray,very thick] (#1) -- (#2);
						}
						\newcommand{\drawLatticeLineFlex}[2]
						{
							\draw[->,gray,very thick,>=stealth] (#1) -- (#2);
						}
						\newcommand{\drawDashedExcitedLine}[2]{
							\draw[black,dashed,ultra thick] (#1) -- (#2);
						}
						\newcommand{\graphLinesHorizontal}
						{
							\drawLatticeLineFlex{2,1}{2,3.1};
							\drawLatticeLine{2,3}{2,5};
							\drawLatticeLineFlex{4,1}{4,3.1};
							\drawLatticeLine{4,3}{4,5};
							\drawLatticeLineFlex{6,1}{6,3.1};
							\drawLatticeLine{6,3}{6,5};
							\drawLatticeLineFlex{8,1}{8,3.1};
							\drawLatticeLine{8,3}{8,5};
							\drawLatticeLineFlex{10,1}{10,3.1};
							\drawLatticeLine{10,3}{10,5};
							\drawLatticeLineFlex{12,1}{12,3.1};
							\drawLatticeLine{12,3}{12,5};
							\drawLatticeLineFlex{1,2}{3.1,2};
							\drawLatticeLineFlex{3,2}{5.1,2};
							\drawLatticeLineFlex{5,2}{7.1,2};
							\drawLatticeLineFlex{7,2}{9.1,2};
							\drawLatticeLineFlex{9,2}{11.1,2};
							\drawLatticeLine{11,2}{13,2};
							\drawLatticeLineFlex{1,4}{3.1,4};
							\drawLatticeLineFlex{3,4}{5.1,4};
							\drawLatticeLineFlex{5,4}{7.1,4};
							\drawLatticeLineFlex{7,4}{9.1,4};
							\drawLatticeLineFlex{9,4}{11.1,4};
							\drawLatticeLine{11,4}{13,4};
							\drawDashedExcitedLine{5,3}{11,3};
						}
						\begin{tikzpicture}
							\graphLinesHorizontal;
							\draw[color=blue!70,fill=yellow!70] (3,3) circle (1.4ex);
							\draw[color=blue!70,fill=blue!70] (3,3) circle (1.2ex);
							\node [] (3,3) at (3,3) {$ {\textcolor{white}{\boldsymbol{-}}} $};
							\draw[color=blue!70,fill=white] (5,3) circle (1.2ex);
							\draw[color=blue!70,fill=yellow!70] (11,3) circle (1.4ex);
							\draw[color=blue!70,fill=blue!70] (11,3) circle (1.2ex);
							\node [] (11,3) at (11,3) {$ {\textcolor{white}{\boldsymbol{+}}} $};
							\draw[color=black,fill=orange!70] (4,3) circle (0.7ex);
							\draw[color=black,fill=orange!70] (6,3) circle (0.7ex);
							\draw[color=black,fill=orange!70] (8,3) circle (0.7ex);
							\draw[color=black,fill=orange!70] (10,3) circle (0.7ex);
						\end{tikzpicture}
					\end{center}
					\caption{\label{confinamento-fluxo} Piece of the same lattice region of $ \mathcal{L} _{2} $ at two different times in the $ D_{2} \left( \mathds{Z} _{2} \right) $ model. In the first instant $ t_{1} $ (above) we have a pair of quasiparticles $ q^{\left( 0 , 1 , 1 \right) } _{+} $ and $ q^{\left( 0 , 1 , 1 \right) } _{-} $ (blue outlined and purposely indexed with the \textquotedblleft $ + $\textquotedblright \hspace*{0.01cm} and \textquotedblleft $ - $\textquotedblright \hspace*{0.01cm} symbols respectively), which were produced by the action of a single operator $ W^{\left( 0 , 1 , 1 \right) } _{\ell } $. Here, the single orange dot corresponds to the unique vacuum violation that is detected by $ C_{\ell , 0} $. In the second instant $ t_{2} > t_{1} $ (below) we have these same quasiparticles, but after one of them has been transported away from the other due to the action of $ W^{\left( 0 , 1 , 1 \right) } _{\ell ^{\prime }} $. This operator acts on all the edges that intersect the dual path $ \boldsymbol{\gamma } ^{\ast } _{1} $ (highlighted in dashed black). Observe that, in this latter case, we have new (three) orange dots: one for each edge involved in this transport, making clear the linearity that is related to the growth of the system energy in this transport.}
				\end{figure}
				something different happens: in the same way that the link operator (that appear in the $ D_{2} \left( \mathds{Z} _{2} \right) $ Hamiltonian) detects the action of $ W^{\left( 0 , 1 , 1 \right) } _{\ell } $ when pairs of quasiparticles $ q^{\left( 0 , 1 , 1 \right) } $ are produced, this link operator also detects the action of each operator $ W^{\left( 0 , 1 , 1 \right) } _{\ell ^{\prime }} $ that participates in this transport process and, therefore, increases the energy of this lattice system to
				\begin{equation}
					E_{\mathrm{inc}} = E_{m} + \left( \mathsf{n} _{\ell ^{\prime }} \cdot \mathcal{J} _{C} \right) \ . \label{aumento}
				\end{equation}
				Here, $ \mathsf{n} _{\ell ^{\prime }} $ is the number of edges or, equivalently, of operators $ W^{\left( 0 , 1 , 1 \right) } _{\ell ^{\prime }} $ involved with this transport. In this way, and by remembering, once again, that quantum-computational models try/need to model some reality that can be physically implemented, we need to ignore the fact that this transport is mathematically possible and consider that all these quasiparticles $ q^{\left( J , 1 , 1 \right) } $ are \emph{confined} (i.e., that all they \textquotedblleft cannot\textquotedblright \hspace*{0.01cm} be separately transported).
			
			\subsubsection{A hadronic analogy}
			
				In order to understand why we need to consider that all these quasiparticles $ q^{\left( J , 1 , 1 \right) } $ are confined, it is crucial to note that, if the motion of any (quasi)particle increases/decreases the energy of a physical system to which it belongs, this prevents us from finding out the statistics of this (quasi)particle by changing its spatial position [\citen{salinas}]. Note that this is not the case for the quasiparticle $ q^{\left( 1 , 0 , 0 \right) } $ that, as in the $ D \left( \mathds{Z} _{2} \right) $ model, can be recognized as a boson [\citen{mf-errata,pachos}]. But although this increase (\ref{aumento}) prevents us from finding out the statistics of $ q^{\left( 0 , 1 , 1 \right) } $ by changing its spatial position, we need to make an important observation here: after all, and as suggested in Figure \ref{neutron-analogy},
				\begin{figure}[!t]
					\begin{center}
						\tikzstyle myBG=[line width=3pt,opacity=1.0]
						\newcommand{\drawLatticeLine}[2]
						{
							\draw[gray,very thick] (#1) -- (#2);
						}
						\newcommand{\drawLatticeLineFlex}[2]
						{
							\draw[->,gray,very thick,>=stealth] (#1) -- (#2);
						}
						\newcommand{\graphLinesHorizontal}
						{
							\drawLatticeLineFlex{2,1}{2,3.1};
							\drawLatticeLine{2,3}{2,5};
							\drawLatticeLineFlex{4,1}{4,3.1};
							\drawLatticeLine{4,3}{4,5};
							\drawLatticeLineFlex{6,1}{6,3.1};
							\drawLatticeLine{6,3}{6,5};
							\drawLatticeLineFlex{8,1}{8,3.1};
							\drawLatticeLine{8,3}{8,5};
							\drawLatticeLineFlex{10,1}{10,3.1};
							\drawLatticeLine{10,3}{10,5};
							\drawLatticeLineFlex{12,1}{12,3.1};
							\drawLatticeLine{12,3}{12,5};
							\drawLatticeLineFlex{1,2}{3.1,2};
							\drawLatticeLineFlex{3,2}{5.1,2};
							\drawLatticeLineFlex{5,2}{7.1,2};
							\drawLatticeLineFlex{7,2}{9.1,2};
							\drawLatticeLineFlex{9,2}{11.1,2};
							\drawLatticeLine{11,2}{13,2};
							\drawLatticeLineFlex{1,4}{3.1,4};
							\drawLatticeLineFlex{3,4}{5.1,4};
							\drawLatticeLineFlex{5,4}{7.1,4};
							\drawLatticeLineFlex{7,4}{9.1,4};
							\drawLatticeLineFlex{9,4}{11.1,4};
							\drawLatticeLine{11,4}{13,4};
						}
						\begin{tikzpicture}
							\graphLinesHorizontal;
							\draw[color=blue!70,fill=yellow!70] (3,3) circle (1.4ex);
							\draw[color=blue!70,fill=blue!70] (3,3) circle (1.2ex);
							\node [] (3,3) at (3,3) {$ {\textcolor{white}{\boldsymbol{-}}} $};
							\draw[color=blue!70,fill=yellow!70] (11,3) circle (1.4ex);
							\draw[color=blue!70,fill=blue!70] (11,3) circle (1.2ex);
							\node [] (11,3) at (11,3) {$ {\textcolor{white}{\boldsymbol{+}}} $};
							\draw[color=black,fill=orange!70] (4,3) circle (0.7ex);
							\draw[color=black,fill=orange!70] (6,3) circle (0.7ex);
							\draw[color=black,fill=orange!70] (8,3) circle (0.7ex);
							\draw[color=black,fill=orange!70] (10,3) circle (0.7ex);
						\end{tikzpicture}
					\end{center} \bigskip
					\begin{center}
						\tikzstyle myBG=[line width=3pt,opacity=1.0]
						\newcommand{\drawLatticeLine}[2]
						{
							\draw[gray,very thick] (#1) -- (#2);
						}
						\newcommand{\drawLatticeLineFlex}[2]
						{
							\draw[->,gray,very thick,>=stealth] (#1) -- (#2);
						}
						\newcommand{\drawDashedExcitedLine}[2]{
							\draw[black,dashed,ultra thick] (#1) -- (#2);
						}
						\newcommand{\graphLinesHorizontal}
						{
							\drawLatticeLineFlex{2,1}{2,3.1};
							\drawLatticeLine{2,3}{2,5};
							\drawLatticeLineFlex{4,1}{4,3.1};
							\drawLatticeLine{4,3}{4,5};
							\drawLatticeLineFlex{6,1}{6,3.1};
							\drawLatticeLine{6,3}{6,5};
							\drawLatticeLineFlex{8,1}{8,3.1};
							\drawLatticeLine{8,3}{8,5};
							\drawLatticeLineFlex{10,1}{10,3.1};
							\drawLatticeLine{10,3}{10,5};
							\drawLatticeLineFlex{12,1}{12,3.1};
							\drawLatticeLine{12,3}{12,5};
							\drawLatticeLineFlex{1,2}{3.1,2};
							\drawLatticeLineFlex{3,2}{5.1,2};
							\drawLatticeLineFlex{5,2}{7.1,2};
							\drawLatticeLineFlex{7,2}{9.1,2};
							\drawLatticeLineFlex{9,2}{11.1,2};
							\drawLatticeLine{11,2}{13,2};
							\drawLatticeLineFlex{1,4}{3.1,4};
							\drawLatticeLineFlex{3,4}{5.1,4};
							\drawLatticeLineFlex{5,4}{7.1,4};
							\drawLatticeLineFlex{7,4}{9.1,4};
							\drawLatticeLineFlex{9,4}{11.1,4};
							\drawLatticeLine{11,4}{13,4};
							\drawDashedExcitedLine{3,3}{9,3};
						}
						\begin{tikzpicture}
							\graphLinesHorizontal;
							\draw[color=blue!70,fill=white] (3,3) circle (1.2ex);
							\draw[color=blue!70,fill=yellow!70] (9,3) circle (1.4ex);
							\draw[color=blue!70,fill=blue!70] (9,3) circle (1.2ex);
							\node [] (9,3) at (9,3) {$ {\textcolor{white}{\boldsymbol{-}}} $};
							\draw[color=blue!70,fill=yellow!70] (11,3) circle (1.4ex);
							\draw[color=blue!70,fill=blue!70] (11,3) circle (1.2ex);
							\node [] (11,3) at (11,3) {$ {\textcolor{white}{\boldsymbol{+}}} $};
							\draw[color=black,fill=orange!70] (10,3) circle (0.7ex);
						\end{tikzpicture}
					\end{center}
					\caption{\label{neutron-analogy} In this figure, we present again (above) the same situation presented in Figure \ref{confinamento-fluxo}, which depicts the linear increase in the energy of this lattice system when we transport one of the quasiparticles $ q^{\left( 0 , 1 , 1 \right) } $ away from another. And we present this same situation again because, as $ W^{\left( 0 , 1 , 1 \right) } _{\ell } \circ W^{\left( 0 , 1 , 1 \right) } _{\ell } = \mathds{1} _{\ell } $, it is not difficult to see that, if an operator $ \prod _{\ell ^{\prime } \in \boldsymbol{\gamma } ^{\ast } _{2}} W^{\left( 0 , 1 , 1 \right) } _{\ell ^{\prime }} $ acts on the path $ \boldsymbol{\gamma } ^{\ast } _{2} $ (which is highlighted in dashed black), it will erase the energy track produced by the transport of Figure \ref{confinamento-fluxo}. Therefore, as the final result of this action leads us to the same initial configuration as in Figure \ref{confinamento-fluxo}, it is not wrong to say that, when we transport two of these quasiparticles together, this energy of this lattice system is preserved.}
				\end{figure}
				if, instead of transporting just one of these quasiparticles, we transport the two quasiparticles of a pair produced by $ W^{\left( 0 , 1 , 1 \right) } _{\ell } $ together, this new transport will be done without any increase in the energy of this lattice system. And this is an important observation because it legitimizes the interpretation that these quasiparticles $ q^{\left( 0 , 1 , 1 \right) } $ are confined in this $ D_{2} \left( \mathds{Z} _{2} \right) $ model.	
				
				In view of the legitimacy of this confinement interpretation, it is interesting to point out that this confinement of quasiparticles is, in some way, analogous to the phenomenon of quark confinement [\citen{wilson-quarks}]. After all, although it is not impossible to move one quark away from another/others inside a hadron, it is well known that, as this quark moves away from another/others, the potential energy of this hadronic system increases until the moment that Nature produces an additional meson (i.e., a hadron usually composed of a quark-antiquark pair) in order to conserve the energy of the system\footnote{In plain English, Nature prefers to convert this energy increase into mass-energy of a new quark-antiquark pair and this is exactly what, for example, justifies the appearance of the jets (i.e., spray of new hadrons) in the various experiments involving the collision of high-energy hadrons [\citen{sterman}].} [\citen{quigg}]. Thus, by noting that
				\begin{itemize}
					\item the potential energy between a quark and an antiquark in a meson increases linearly with the distance between them [\citen{cern-dzierba}], and
					\item the action of $ W^{\left( J , L , \Lambda \right) } _{\ell } $ produces one pair, which is composed of one quasiparticle $ q^{\left( J , L , \Lambda \right) } _{+} $ and its anti-quasiparticle $ q^{\left( J , L , \Lambda \right) } _{-} $, where before it was a vacuum,
				\end{itemize}
				this analogy seems to endorse the confinement of quasiparticles mentioned above because the energy (\ref{aumento}), which is associated with a pair of quasiparticles $ q^{\left( 0 , 1 , 1 \right) } _{+} $ and $ q^{\left( 0 , 1 , 1 \right) } _{-} $, also increases linearly with the distance between them.
				
				Given all these facts, it seems to make a lot of sense to say that, from the perspective of elementary particle physics, the action of $ W^{\left( J , 1 , 1\right) } _{\ell } $ produces a kind of prototype of a meson. And this perspective becomes even more interesting when we notice, for instance, that the most general $ D \left( G \right) $ quasiparticles are interpreted as \emph{anyons}: i.e., they are not necessarily interpreted as bosons or fermions. After all, as the concept of anyon arose from the advent of the \emph{Aharonov-Bohm Effect} [\citen{aha-bohm}] (because it is possible to realize such anyonic statistics for systems where one electric particle rotates around one punctual magnetic field on a $ 2 $-dimensional surface [\citen{lerda-anyons}]), the $ D \left( G \right) $ quasiparticles which are detectable in the vertices were baptized as type $ e $ (\underline{e}lectric), while those that are detectable in the faces ended up being denoted as type $ m $ (\underline{m}agnetic) [\citen{mf-pedagogical}]. In other words, we are facing a scenario that, due to the same correspondence principle mentioned in Subsubsection \ref{correspondence-principle}, allows us to recognize $ q^{\left( J , 0 , 0 \right) } $ and $ q^{\left( 0 , L , \Lambda \right) } $ as electric and magnetic quasiparticles respectively. And, in plain English, this is exactly what seems to reinforce that $ W^{\left( J , 1 , 1 \right) } _{\ell } $ actually produces a kind of prototype of a meson in the $ D_{2} \left( \mathds{Z} _{2} \right) $ model since, for instance, there are several works that explore the hypothesis that quarks are magnetically confined [\citen{abrik-1,abrik-2,nielsen-olesen,nambu-1,eguchi,wyld,nambu-2,hooft}]. 
				
			\subsubsection{\label{32-model}Is there any $ D_{M} \left( \mathds{Z} _{N} \right) $ model whose ground state degeneracy depends on $ \pi _{1} \left( \mathcal{M} _{2} \right) $?}
			
				Although we have just discussed this confinement of quasiparticles only in the context of the $ D_{2} \left( \mathds{Z} _{2} \right) $ model, it is important to understand how this confinement appears in more general $ D_{M} \left( \mathds{Z} _{N} \right) $ models (i.e., in the $ D_{M} \left( \mathds{Z} _{N} \right) $ models where $ M $ or/and $ N $ is/are greater than $ 2 $). And the best way to understand this is by noting that, unlike what happens with the $ D \left( \mathds{Z} _{N} \right) $ models, the $ D_{M} \left( \mathds{Z} _{N} \right) $ ground state degeneracy does not necessarily depend on the order of $ \pi _{1} \left( \mathcal{M} _{2} \right) $\label{not-necessarily}.
				
				Note that the $ D_{2} \left( \mathds{Z} _{2} \right) $ model is a good example of this fact because, when
				\begin{equation}
					O_{1} \bigl( \overline{\boldsymbol{\gamma }} ^{\ast } \bigr) = \prod _{\ell ^{\prime } \in \overline{\boldsymbol{\gamma }} ^{\ast }} W^{\left( 0 , 1 , 1 \right) } _{\ell ^{\prime }} \label{cinturao}
				\end{equation}
				acts on a set $ \overline{\boldsymbol{\gamma }} ^{\ast } $ of edges that intersect any closed dual path, it always leads to a non-vacuum state and, therefore, the $ D_{2} \left( \mathds{Z} _{2} \right) $ ground state is independent of the order of $ \pi _{1} \left( \mathcal{M} _{2} \right) $. However, when we analyse the Toric Code coupled to matter fields with $ M > 2 $, we may find a situation that is somewhat different. This is what happens, for instance, with the $ D_{3} \left( \mathds{Z} _{2} \right) $ model, whose vertex operators can always be represented as
				\begin{equation*}
					A_{v,J} = \frac{1}{2} \sum _{g \in \mathds{Z} _{2}} \left( -1 \right) ^{Jg} \cdot M _{v} \left( g \right) \prod _{\ell ^{\prime } \in S_{v}} \left( \sigma ^{x} _{\ell ^{\prime }} \right) ^{g} \ , 
				\end{equation*} 
				where [\citen{miguel-1,pramod-wrong}]
				\begin{equation}
					M_{v} \left( 0 \right) = \begin{pmatrix}
						\mathds{1} & \mathbf{0} \\
						\mathbf{0} ^{\mathrm{T}} & 1
					\end{pmatrix} \ \ \textnormal{and} \ \
					M_{v} \left( 1 \right) = \begin{pmatrix}
						\sigma ^{x} & \mathbf{0} \\
						\mathbf{0} ^{\mathrm{T}} & 1
					\end{pmatrix} \ , \label{action-32}
				\end{equation}
				and where $ \mathds{1} $ is an identity matrix and $ \mathbf{0} $ is a zero column matrix. After all, by noting that the set
				\begin{equation}
					\mathfrak{Fix} _{\mu }  = \left\{ \bigl\vert \alpha \bigr\rangle _{v} \in \mathfrak{H} _{M} : M_{v} \left( g \right) \bigl\vert \alpha \bigr\rangle _{v} = \bigl\vert \alpha \bigr\rangle _{v} \right. \textnormal{for all} \left. g \in \mathds{Z} _{3} \right\} \label{fixed-set}
				\end{equation}
				of points of $ \mathfrak{H} _{M} $ that are fixed by the group action $ \mu $ is non-empty, this allows to conclude that the $ D_{3} \left( \mathds{Z} _{2} \right) $ ground state degeneracy is dependent of the order of $ \pi _{1} \left( \mathcal{M} _{2} \right) $. And in order to understand this conclusion, the first thing we need to do here is to note that, as
				\begin{itemize}
					\item all the $ D_{M} \left( \mathds{Z} _{2} \right) $ face operators are represented by (\ref{b-tc}), and
					\item all the $ D_{M} \left( \mathds{Z} _{2} \right) $ link operators cannot perform any permutation between the gauge or matter fields,
				\end{itemize}
				the matrices in (\ref{action-32}) allow us to recognize that this $ D_{3} \left( \mathds{Z} _{2} \right) $ model has, at least, two vacuum states\footnote{Note that (\ref{32-ground-state-a}) is just a more streamlined way of writing the same vacuum state (\ref{vacuo-1}).}
				\begin{subequations} \label{32-ground-state}
					\begin{align}
						\bigl\vert \xi ^{\left( 0 \right) } _{0} \bigr\rangle & = \prod _{v^{\prime }} A_{v^{\prime }} \left( \bigotimes _{\ell \in \mathcal{L} _{2}} \left\vert e \right\rangle \right) \otimes \left( \bigotimes _{v \in \mathcal{L} _{2}} \left\vert 0 \right\rangle \right) \ \ \textnormal{and} \label{32-ground-state-a} \\
						\bigl\vert \xi ^{\left( 2 \right) } _{0} \bigr\rangle & = \prod _{v^{\prime }} A_{v^{\prime }} \left( \bigotimes _{\ell \in \mathcal{L} _{2}} \left\vert e \right\rangle \right) \otimes \left( \bigotimes _{v \in \mathcal{L} _{2}} \left\vert 2 \right\rangle \right) \label{32-ground-state-b}
					\end{align}
				\end{subequations}
				because there is no transformation, which can be expressed as a product of the operators $ A_{v} $, $ B_{f} $ and $ C_{\ell } $, that can connect these two vacuum states. And once all the $ D_{M} \left( G \right) $ models reduce to the $ D \left( G \right) $ models when $ M = 1 $, the second thing we need to do here is to note that the same operators (\ref{D2-gauge-quasiparticles}) produce (pairs of) quasiparticles in this $ D_{3} \left( \mathds{Z} _{2} \right) $ model. By the way, since we touched on the subject of the quasiparticle production, it is also important to note that the operators
				\begin{equation}
					W^{\left( g , 0 \right) } _{v} = M_{v} \left( g \right) \label{matter-operators-32}
				\end{equation}
				also need to be listed among those are able to produce (matter) excitations in this model because, among other things, they define the $ D_{3} \left( \mathds{Z} _{2} \right) $ Hamiltonian\footnote{This comment is in full agreement with the one we already made in the footnote on page \pageref{footnote-qft}.}. However, despite the action of these operators (\ref{matter-operators-32}) on the vacuum state (\ref{32-ground-state-a}) being identical to that of the operators (\ref{matter-action-22a}) on the unique $ D_{2} \left( \mathds{Z} _{2} \right) $ vacuum state (\ref{vacuo-1}), there is a \textquotedblleft problem\textquotedblright \hspace*{0.01cm} here\label{problem}: after all, as (\ref{action-32}) permutes $ \left\vert 0 \right\rangle _{v} \leftrightarrow \left\vert 1 \right\rangle _{v} $ but fixes $ \left\vert 2 \right\rangle _{v} $, these operators (\ref{matter-operators-32}) are completely unable to produce any (matter) excitation on the vacuum state (\ref{32-ground-state-b}). That is, the only operators that compose the $ D_{3} \left( \mathds{Z} _{2} \right) $ Hamiltonian and, therefore, can excite this vacuum state (\ref{32-ground-state-b}) are the operators (\ref{D2-gauge-quasiparticles}).
				
				Nevertheless, it is precisely this \textquotedblleft problem\textquotedblright \hspace*{0.01cm} that, for instance, allows us to make some important observations, and the first one concerns the relationship between the $ D_{3} \left( \mathds{Z} _{2} \right) $ ground state degeneracy and the cardinality of (\ref{fixed-set}). After all, as
				\begin{itemize}
					\item $ \left\vert \mathfrak{Fix} _{\mu } \right\vert = 1 $ because (\ref{action-32}) fixes only $ \left\vert 2 \right\rangle _{v} $, and
					\item the fixation $ M_{v} \left( g \right) \bigl\vert 2 \bigr\rangle _{v} = \bigl\vert 2 \bigr\rangle _{v} $ makes the $ D_{3} \left( \mathds{Z} _{2} \right) $ vertex operator $ C_{\ell ,0} $ unable to detect any energy excitation produced by $ W^{\left( J , 1 , 1 \right) } _{\ell } $ on the vacuum state (\ref{32-ground-state-b}),
				\end{itemize}
				it becomes clear that all the quasiparticles produced by $ W^{\left( 0 , 1 , 1 \right) } _{\ell } $ on this vacuum state (\ref{32-ground-state-b}) are not confined. Thus, by
				\begin{itemize}
					\item noting that the action of (\ref{cinturao}) on the vacuum state (\ref{32-ground-state-b}) does not lead to an excited state, and
					\item considering that
					\begin{equation*}
						\left\{ \ \mathcal{C} ^{\ast } _{1} \ , \ \mathcal{C} ^{\ast } _{2} \ , \ \ldots , \ \mathcal{C} ^{\ast } _{s-1} \ , \ \mathcal{C} ^{\ast } _{s} \ \right\} 
					\end{equation*} 
					is a non-empty set that contains all the non-contractile curves that generate $ \pi _{1} \left( \mathcal{M} _{2} \right) $,
				\end{itemize}
				we conclude that all the vacuum states
				\begin{equation*}
					\bigl\vert \xi ^{\left( 2 , \vec{\lambda } \right) } _{0} \bigr\rangle  = \prod ^{s} _{p=1} \left[ O_{1} \bigl( \overline{\boldsymbol{\gamma }} ^{\ast } _{p} \bigr) \right] ^{\lambda _{p}} \bigl\vert \xi ^{\left( 2 \right) } _{0} \bigr\rangle \ , 
				\end{equation*} 
				where $ \vec{\lambda } = \left( \lambda _{1} , \lambda _{2} , \ldots , \lambda _{s-1} , \lambda _{s} \right) \neq \left( 0 , 0 , \ldots , 0 , 0 \right) $, are topologically independent of each other and, by definition, with respect to the vacuum states (\ref{32-ground-state}) \footnote{This condition $ \vec{\lambda } \neq \vec{0} $ is of paramount importance because, when $ \vec{\lambda } = \vec{0} $, the vacuum state $ \bigl\vert \xi ^{\left( 2 , \vec{\lambda } \right) } _{0} \bigr\rangle $ is reduced to (\ref{32-ground-state-b}).} [\citen{mf-pedagogical}]. Here,
				\begin{itemize}
					\item $ \lambda _{p} = 0 , 1 $, and 
					\item $ \overline{\boldsymbol{\gamma }} ^{\ast } _{p} $ is a closed dual path (similar to what appears, in dashed black, in Figure \ref{confinamento-fluxo}) that should be interpreted as the discretization of $ \mathcal{C} ^{\ast } _{p} $.
				\end{itemize}
				
				But although we have used the $ D_{3} \left( \mathds{Z} _{2} \right) $ model to show that there is a $ D_{M} \left( \mathds{Z} _{N} \right) $ model whose ground state degeneracy depends on $ \pi _{1} \left( \mathcal{M} _{2} \right) $, it is not difficult to show that this also happens with many other $ D_{M} \left( \mathds{Z} _{N} \right) $ models. And in order to show this, it is enough to analyse all the $ D_{M} \left( \mathds{Z} _{N} \right) $ models where
				\begin{equation}
					\mathfrak{Fix} _{\mu }  = \left\{ \bigl\vert \alpha \bigr\rangle _{v} \in \mathfrak{H} _{M} : M_{v} \left( g \right) \bigl\vert \alpha \bigr\rangle _{v} = \bigl\vert \alpha \bigr\rangle _{v} \right. \textnormal{for all} \left. g \in \mathds{Z} _{N} \right\} \label{fix-general-set}
				\end{equation}
				is a non-empty set. After all, as these models have $ \left\vert \mathfrak{Fix} _{\mu } \right\vert $ vacuum states
				\begin{equation}
					\bigl\vert \xi ^{\left( \alpha \right) } _{0} \bigr\rangle = \prod _{v^{\prime }} A_{v^{\prime }} \left( \bigotimes _{\ell \in \mathcal{L} _{2}} \left\vert e \right\rangle \right) \otimes \left( \bigotimes _{v \in \mathcal{L} _{2}} \left\vert \alpha \right\rangle \right) \label{vacuum-filling}
				\end{equation}
				whose matter fields cannot be manipulated by using
				\begin{equation*}
					W^{\left( g , 0 \right) } _{v} = M_{v} \left( g \right) \ ,
				\end{equation*}
				all the operators
				\begin{equation}
					O_{L} \left( \overline{\boldsymbol{\gamma }} ^{\ast } \right) = \prod _{\ell ^{\prime } \in \overline{\Gamma } ^{\ast } _{\circlearrowleft }} W^{\left( 0 , L , 0 \right) } _{\ell ^{\prime }} \prod _{\ell ^{\prime } \in \overline{\Gamma } ^{\ast } _{\circlearrowright }} \left( W^{\left( 0 , L , 0 \right) } _{\ell ^{\prime }} \right) ^{\dagger } \ , \label{cinturao-L}
				\end{equation}
				are completely unable to excite these $ \left\vert \mathfrak{Fix} _{\mu } \right\vert $ vacuum states\footnote{Note that, in the case of the $ D_{M} \left( \mathds{Z} _{2} \right) $ models, these operators (\ref{cinturao-L}) reduce to
				\begin{equation*}
					O_{1} \left( \overline{\boldsymbol{\gamma }} ^{\ast } \right) = \prod _{\ell ^{\prime } \in \overline{\boldsymbol{\gamma }} ^{\ast }} W^{\left( 0 , 1 , 0 \right) } _{\ell ^{\prime }}
				\end{equation*}
				because $ \bigl( W^{\left( 0 , 1 , 0 \right) } _{\ell ^{\prime }} \bigr) ^{\dagger } = W^{\left( 0 , 1 , 0 \right) } _{\ell ^{\prime }} = \sigma ^{x} _{\ell ^{\prime }} $. That is, if we ignore the fact that $ \overline{\boldsymbol{\gamma }} ^{\ast } $ is a non-contractile closed dual path, it is quite remarkable that this result \textquotedblleft coincides\textquotedblright \hspace*{0.01cm} with (\ref{cinturao}).} by acting on a non-contractile closed dual path $ \overline{\boldsymbol{\gamma }} ^{\ast } $ that crosses all the edges of a set $ \overline{\Gamma } ^{\ast } _{\circlearrowleft } \cup \overline{\Gamma } ^{\ast } _{\circlearrowright } $. Here, $ \overline{\Gamma } ^{\ast } _{\circlearrowleft } $ and $ \overline{\Gamma } ^{\ast } _{\circlearrowright } $ are two subsets, whose edges have some counterclockwise and clockwise orientations respectively, as shown in Figure \ref{dual-gamma-orientations}.
				\begin{figure}[!t]
					\begin{center}
						\tikzstyle myBG=[line width=3pt,opacity=1.0]
						\newcommand{\drawLatticeLine}[2]
						{
							\draw[gray,very thick] (#1) -- (#2);
						}
						\newcommand{\drawLatticeLineFlex}[2]
						{
							\draw[->,gray,very thick,>=stealth] (#1) -- (#2);
						}
						\newcommand{\drawOrientedBlueLine}[2]{
							\draw[->,blue!30,ultra thick,>=stealth] (#1) -- (#2);
						}
						\newcommand{\graphLinesHorizontal}
						{
							\drawOrientedBlueLine{3.7,4.5}{4.9,4.5};
							\drawLatticeLineFlex{2,1}{2,3.1};
							\drawLatticeLineFlex{2,3}{2,5.1};
							\drawLatticeLine{2,5}{2,7};
							\drawLatticeLineFlex{4,1}{4,3.1};
							\drawLatticeLineFlex{4,3}{4,5.1};
							\drawLatticeLine{4,5}{4,7};
							\drawLatticeLineFlex{6,1}{6,3.1};
							\drawLatticeLineFlex{6,3}{6,5.1};
							\drawLatticeLine{6,5}{6,7};
							\drawLatticeLineFlex{8,1}{8,3.1};
							\drawLatticeLineFlex{8,3}{8,5.1};
							\drawLatticeLine{8,5}{8,7};
							\drawLatticeLineFlex{10,1}{10,3.1};
							\drawLatticeLineFlex{10,3}{10,5.1};
							\drawLatticeLine{10,5}{10,7};
							\drawLatticeLineFlex{12,1}{12,3.1};
							\drawLatticeLineFlex{12,3}{12,5.1};
							\drawLatticeLine{12,5}{12,7};
							\drawLatticeLineFlex{1,2}{3.1,2};
							\drawLatticeLineFlex{3,2}{5.1,2};
							\drawLatticeLineFlex{5,2}{7.1,2};
							\drawLatticeLineFlex{7,2}{9.1,2};
							\drawLatticeLineFlex{9,2}{11.1,2};
							\drawLatticeLine{11,2}{13,2};
							\drawLatticeLineFlex{1,4}{3.1,4};
							\drawLatticeLineFlex{3,4}{5.1,4};
							\drawLatticeLineFlex{5,4}{7.1,4};
							\drawLatticeLineFlex{7,4}{9.1,4};
							\drawLatticeLineFlex{9,4}{11.1,4};
							\drawLatticeLine{11,4}{13,4};
							\drawLatticeLineFlex{1,6}{3.1,6};
							\drawLatticeLineFlex{3,6}{5.1,6};
							\drawLatticeLineFlex{5,6}{7.1,6};
							\drawLatticeLineFlex{7,6}{9.1,6};
							\drawLatticeLineFlex{9,6}{11.1,6};
							\drawLatticeLine{11,6}{13,6};
						}
						\begin{tikzpicture}
							\graphLinesHorizontal;
							\draw[color=blue!70,fill=yellow!70] (3,5) circle (1.4ex);
							\draw[color=blue!70,fill=blue!70] (3,5) circle (1.2ex);
							\node [] (3,5) at (3,5) {$ {\textcolor{white}{\boldsymbol{-}}} $};
							\draw[color=blue!70,fill=yellow!70] (5,5) circle (1.4ex);
							\draw[color=blue!70,fill=blue!70] (5,5) circle (1.2ex);
							\node [] (5,5) at (5,5) {$ {\textcolor{white}{\boldsymbol{+}}} $};
							\draw[color=black,fill=orange!70] (4,5) circle (0.7ex);
						\end{tikzpicture}
					\end{center} \bigskip
					\begin{center}
						\tikzstyle myBG=[line width=3pt,opacity=1.0]
						\newcommand{\drawLatticeLine}[2]
						{
							\draw[gray,very thick] (#1) -- (#2);
						}
						\newcommand{\drawLatticeLineFlex}[2]
						{
							\draw[->,gray,very thick,>=stealth] (#1) -- (#2);
						}
						\newcommand{\drawDashedLine}[2]
						{
							\draw[gray!30,dashed,very thick] (#1) -- (#2);
						}
						\newcommand{\drawOrientedBlueLine}[2]{
							\draw[->,blue!30,ultra thick,>=stealth] (#1) -- (#2);
						}
						\newcommand{\drawOrientedRedLine}[2]{
							\draw[->,red!30,ultra thick,>=stealth] (#1) -- (#2);
						}
						\newcommand{\drawDashedExcitedLine}[2]{
							\draw[black!30,dashed,ultra thick] (#1) -- (#2);
						}
						\newcommand{\graphLinesHorizontal}
						{
							\drawDashedExcitedLine{5,5}{11,5};
							\drawDashedExcitedLine{11,5}{11,3};
							\drawDashedExcitedLine{11,3}{7,3};
							\drawOrientedBlueLine{5.7,4.5}{6.9,4.5};
							\drawOrientedBlueLine{7.7,4.5}{8.9,4.5};
							\drawOrientedBlueLine{9.7,4.5}{10.9,4.5};
							\drawOrientedBlueLine{10.5,4.3}{10.5,3.1};
							\drawOrientedRedLine{10.3,2.5}{9.1,2.5};
							\drawOrientedRedLine{8.3,2.5}{7.1,2.5};
							\drawLatticeLineFlex{2,1}{2,3.1};
							\drawLatticeLineFlex{2,3}{2,5.1};
							\drawLatticeLine{2,5}{2,7};
							\drawLatticeLineFlex{4,1}{4,3.1};
							\drawLatticeLineFlex{4,3}{4,5.1};
							\drawLatticeLine{4,5}{4,7};
							\drawLatticeLineFlex{6,1}{6,3.1};
							\drawLatticeLineFlex{6,3}{6,5.1};
							\drawLatticeLine{6,5}{6,7};
							\drawLatticeLineFlex{8,1}{8,3.1};
							\drawLatticeLineFlex{8,3}{8,5.1};
							\drawLatticeLine{8,5}{8,7};
							\drawLatticeLineFlex{10,1}{10,3.1};
							\drawLatticeLineFlex{10,3}{10,5.1};
							\drawLatticeLine{10,5}{10,7};
							\drawLatticeLineFlex{12,1}{12,3.1};
							\drawLatticeLineFlex{12,3}{12,5.1};
							\drawLatticeLine{12,5}{12,7};
							\drawLatticeLineFlex{1,2}{3.1,2};
							\drawLatticeLineFlex{3,2}{5.1,2};
							\drawLatticeLineFlex{5,2}{7.1,2};
							\drawLatticeLineFlex{7,2}{9.1,2};
							\drawLatticeLineFlex{9,2}{11.1,2};
							\drawLatticeLine{11,2}{13,2};
							\drawLatticeLineFlex{1,4}{3.1,4};
							\drawLatticeLineFlex{3,4}{5.1,4};
							\drawLatticeLineFlex{5,4}{7.1,4};
							\drawLatticeLineFlex{7,4}{9.1,4};
							\drawLatticeLineFlex{9,4}{11.1,4};
							\drawLatticeLine{11,4}{13,4};
							\drawLatticeLineFlex{1,6}{3.1,6};
							\drawLatticeLineFlex{3,6}{5.1,6};
							\drawLatticeLineFlex{5,6}{7.1,6};
							\drawLatticeLineFlex{7,6}{9.1,6};
							\drawLatticeLineFlex{9,6}{11.1,6};
							\drawLatticeLine{11,4}{13,4};
							\drawLatticeLine{11,6}{13,6};
						}
						\begin{tikzpicture}
							\graphLinesHorizontal;
							\draw[color=blue!70,fill=yellow!70] (7,3) circle (1.4ex);
							\draw[color=blue!70,fill=blue!70] (7,3) circle (1.2ex);
							\node [] (7,3) at (7,3) {$ {\textcolor{white}{\boldsymbol{+}}} $};
							\draw[color=blue!70,fill=yellow!70] (3,5) circle (1.4ex);
							\draw[color=blue!70,fill=blue!70] (3,5) circle (1.2ex);
							\node [] (3,5) at (3,5) {$ {\textcolor{white}{\boldsymbol{-}}} $};
							\draw[color=blue!70,fill=white] (5,5) circle (1.2ex);
							\draw[color=black,fill=orange!70] (4,5) circle (0.7ex);
							\draw[color=black,fill=orange!70] (6,5) circle (0.7ex);
							\draw[color=black,fill=orange!70] (8,5) circle (0.7ex);
							\draw[color=black,fill=orange!70] (10,5) circle (0.7ex);
							\draw[color=black,fill=orange!70] (11,4) circle (0.7ex);
							\draw[color=black,fill=orange!70] (10,3) circle (0.7ex);
							\draw[color=black,fill=orange!70] (8,3) circle (0.7ex);
						\end{tikzpicture}
					\end{center} 
					\caption{\label{dual-gamma-orientations} Here, we see a piece of the same lattice region, at two different times, to explain the meaning of the edge subsets $ \Gamma ^{\ast } _{\circlearrowleft } $ and $ \Gamma ^{\ast } _{\circlearrowright } $. In the first instant $ t_{1} $ (above) we see a pair of magnetic quasiparticles, which was produced around the $ \ell $-th oriented edge. Observe the presence of a light blue arrow, which points from the quasiparticle \textquotedblleft $ - $\textquotedblright \hspace*{0.01cm} to \textquotedblleft $ + $\textquotedblright : this arrow and the $ \ell $-th edge define an ordered basis, whose orientation is $ \circlearrowleft $. In the second instant $ t_{2} > t_{1} $ (below) we see these same quasiparticles, but after one of them has been transported away from the other along a dual path $ \boldsymbol{\gamma } ^{\ast } $ (highlighted in dashed light black). Observe again the presence of arrows in this figure: in the case of the light blue arrows, they refer to the same basis \textquotedblleft $ \circlearrowleft $\textquotedblright ; now, in the case of the light red arrows, each of them defines, with the lattice edges that they intersect, another basis that has an inverse orientation $ \circlearrowright $. Alongside these two observations, it is also important to note that, as $ W^{\left( 0 , L , \Lambda \right) } _{\ell ^{\prime }} $ always produce pairs of \textquotedblleft quasiparticle\textquotedblright \hspace*{0.01cm} and \textquotedblleft anti-quasiparticle\textquotedblright \hspace*{0.01cm} in the $ D_{M} \left( \mathds{Z} _{N} \right) $ models, they are also mathematically capable of transporting all these quasiparticles as long as they are used in a clever way. This clever way is by using $ O_{L} \left( \boldsymbol{\gamma } ^{\ast } \right) = \prod _{\ell ^{\prime } \in \Gamma ^{\ast } _{\circlearrowleft }} W^{\left( 0 , L , 0 \right) } _{\ell ^{\prime }} \prod _{\ell ^{\prime } \in \Gamma ^{\ast } _{\circlearrowright }} \left( W^{\left( 0 , L , 0 \right) } _{\ell ^{\prime }} \right) ^{\dagger } $ as long as the edge subsets $ \Gamma ^{\ast } _{\circlearrowleft } $ and $ \Gamma ^{\ast } _{\circlearrowright } $ contain only the edges that define \textquotedblleft $ \circlearrowleft $\textquotedblright \hspace*{0.01cm} and \textquotedblleft $ \circlearrowright $\textquotedblright \hspace*{0.01cm} respectively.}
				\end{figure}				
				In other words, all the $ D_{M} \left( \mathds{Z} _{N} \right) $ models where $ \left\vert \mathfrak{Fix} _{\mu } \right\vert \neq 0 $ have a set of vacuum states
				\begin{equation*}
					\bigl\vert \xi ^{\left( \alpha , \vec{\lambda } , \vec{L} \right) } _{0} \bigr\rangle = \prod ^{s} _{p=1} \left[ O_{L_{p}} \left( \overline{\boldsymbol{\gamma }} ^{\ast } _{p} \right) \right] ^{\lambda _{p}} \bigl\vert \xi ^{\left( \alpha \right) } _{0} \bigr\rangle
				\end{equation*} 
				that is degenerate as a function of $ \pi _{1} \left( \mathcal{M} _{2} \right) $, where
				\begin{eqnarray*}
					\vec{L} & = & \left( L_{1} , L_{2} , \ldots , L_{s-1} , L_{s} \right) \quad \textnormal{and} \\
					\vec{\lambda } & = & \left( \lambda _{1} , \lambda _{2} , \ldots , \lambda _{s-1} , \lambda _{s} \right) \neq \left( 0 , 0 , \ldots , 0 , 0 \right) \ ,
				\end{eqnarray*}
				with $ \lambda _{p} = 0,1 $.
			
			\subsubsection{\label{dirac-comment} Another interesting analogy} 
			
				By continuing to take advantage of this scenario, where each of the vacuum states (\ref{vacuum-filling}) is defined by filling all the lattice vertices with the same matter field $ \left\vert \alpha \right\rangle $, it is interesting to analyse the differences between the $ D_{M} \left( \mathds{Z} _{N} \right) $ models that have a trivial group action $ \mu $ from those that do not. And something that is not difficult to see is that, when $ \mu $
				\begin{itemize}
					\item is a trivial group action, all the $ D_{M} \left( \mathds{Z} _{N} \right) $ vacuum states 
					\begin{equation}
						\bigl\vert \xi ^{\left( \alpha \right) } _{0} \bigr\rangle = \prod _{v^{\prime }} A_{v^{\prime }} \left( \bigotimes _{\ell \in \mathcal{L} _{2}} \left\vert e \right\rangle \right) \otimes \left( \bigotimes _{v \in \mathcal{L} _{2}} \left\vert \alpha \right\rangle \right) \ , \label{vacuum-states-patological}
					\end{equation}
					are independent of each other because, from the algebraic point of view, $ \mu $ defines $ M $ orbits containing only one element (i.e., $ 1 $-cycles) [\citen{james}], and
					\item is a non-trivial group action, some of these vacuum states (\ref{vacuum-states-patological}) (or perhaps all of them) can be connected by using some transformation, which can be expressed as a product of the operators $ A_{v} $, $ B_{f} $ and $ C_{\ell } $ because $ \mu $ defines some orbit(s) containing more than one element (i.e., $ k $-cycles where $ k > 1 $).
				\end{itemize}
				As a consequence of this, we can assert that, when $ \mu $
				\begin{itemize}
					\item is a trivial group action, the $ D_{M} \left( \mathds{Z} _{N} \right) $ ground state is $ \mathfrak{n} $-fold degenerated, where
					\begin{equation}
						\mathfrak{n} = \left\vert \mathfrak{Fix} _{\mu } \right\vert \cdot \mathfrak{d} _{D \left( \mathds{Z} _{N} \right) } \label{degeneracy-trivial-dmn}
					\end{equation}
					is the product between the cardinality of (\ref{fix-general-set}) (which, in this case, is equal to $ M $) and the number $ \mathfrak{d} _{D \left( \mathds{Z} _{N} \right) } $ of vacuum states in the $ D \left( \mathds{Z} _{N} \right) $ ground state\label{ddg-notation}, and
					\item is a non-trivial group action, the $ \mathfrak{n} $-fold degeneracy of the $ D_{M} \left( \mathds{Z} _{N} \right) $ ground state is characterized by
					\begin{equation}
						\mathfrak{n} = \mathfrak{n} _{\mathrm{orb}} + \left\vert \mathfrak{Fix} _{\mu } \right\vert \cdot \mathfrak{d} _{D \left( \mathds{Z} _{N} \right) } \label{degeneracy-non-trivial-dmn}
					\end{equation}
					because $ \mu $ can also define $ \mathfrak{n} _{\mathrm{orb}} $ orbits containing more than one element.
				\end{itemize}
				
				Of course, and for the sake of completeness, it is worth remarking that the degeneracy degrees of the more general $ D_{M} \left( G \right) $ ground states, where $ G $ is not necessarily an Abelian group, can be calculated as
				\begin{equation}
					\mathfrak{n} = \mathrm{Tr} \left( \prod _{v \in \mathcal{L} _{2}} A_{v} \prod _{f \in \mathcal{L} _{2}} B_{f} \prod _{\ell \in \mathcal{L} _{2}} C_{\ell } \right) \ . \label{dmg-general-ground-states}
				\end{equation}
				That is, all those degeneracy degrees, which were mentioned in the previous paragraph, can also be obtained through (\ref{dmg-general-ground-states}). By the way, and also for the sake of completeness, it is also important to point out that, although we have only paid attention to these different $ D_{M} \left( \mathds{Z} _{N} \right) $ vacuum states, which can be defined by filling all the vertices of $ \mathcal{L} _{2} $ with the same matter field, it is not difficult to conclude that this may also happen with other $ D_{M} \left( G \right) $ models where $ G $ is not necessarily an Abelian group. And in accordance with what was discussed in the previous Subsubsection, the key condition for this to occur is that
				\begin{equation*}
					\mathfrak{Fix} _{\mu }  = \left\{ \bigl\vert \alpha \bigr\rangle _{v} \in \mathfrak{H} _{M} : M_{v} \left( g \right) \bigl\vert \alpha \bigr\rangle _{v} = \bigl\vert \alpha \bigr\rangle _{v} \right. \textnormal{for all} \left. g \in G \right\}
				\end{equation*}
				is a non-empty set.
				
				But while this allows us to infer that the ground state degeneracy of these $ D_{M} \left( G \right) $ models also depends, in some way, on the second group of homology $ \mathcal{H} _{2} \left( \mathcal{M} _{2} \right) $ \citen{kinsey}] \footnote{There is a connection (between the ground state degeneracy of the $ D_{M} \left( G \right) $ models and the homology group $ \mathcal{H} _{2} \left( \mathcal{M} _{2} \right) $) that can be perfectly exploited by using, for instance, a mathematical induction on what was presented in Ref. \citen{mf-pedagogical}]. Nevertheless, as the pedagogical discussion of this connection deserves a paper dedicated only to this (even because writing it here would make the present manuscript even longer than it already is), we will postpone this for now.}, this also allows us to see something that appears to be a little more relevant. After all, and regardless of whether $ G $ is an Abelian group or not, all these $ D_{M} \left( G \right) $ vacuum states, which are defined by filling all the vertices of $ \mathcal{L} _{2} $ with the same matter field, are quite similar to the one that was proposed by P. A. M. Dirac in 1929 [\citen{dirac-sea}], who claimed that the vacuum could be interpreted as an infinite \textquotedblleft sea\textquotedblright \hspace*{0.01cm} of particles. And this seems to be more relevant because this similarity is also observed, for instance, in the Abelian $ H_{M} / \mathds{C} \left( \mathds{Z} _{N} \right) $ models that were discussed in Ref. [\citen{mf-errata}], which can be interpreted as special cases of the $ D_{M} \left( G \right) $ models where $ \mathcal{J} _{B} = 0 $.
				
				In order to understand this similarity, it is crucial to keep in mind that we can only go from one vacuum state $ \bigl\vert \xi ^{\left( \alpha ^{\prime } \right) } _{0} \bigr\rangle $ to another $ \bigl\vert \xi ^{\left( \alpha ^{\prime \prime } \right) } _{0} \bigr\rangle $ if we are able to make exchanges $ \left\vert \alpha ^{\prime } \right\rangle _{v} \rightarrow \left\vert \alpha ^{\prime \prime } \right\rangle _{v} $ over all the lattice vertices. And this is a task that needs to be done with the help of some operator $ W^{\left( J , \Lambda \right) } _{v} $ that, by performing this exchange on the $ v $-th lattice vertex, produces a quasiparticle $ Q^{\left( J , \Lambda \right) } $ there. Note that this is precisely the situation of the operator
				\begin{equation*}
					W^{\left( 1 , 0 \right) } _{v} = \begin{pmatrix}
						\sigma ^{x} & \mathbf{0} \\
						\mathbf{0} ^{\mathrm{T}} & 1
					\end{pmatrix}
				\end{equation*}
				that, being one of the operators that define the $ D_{3} \left( \mathds{Z} _{2} \right) $ Hamiltonian, manages to excite, for instance, the vacuum state (\ref{32-ground-state-a}) of the $ D_{3} \left( \mathds{Z} _{2} \right) $ model. In this fashion, by remembering that
				\begin{itemize}
					\item $ W^{\left( 1 , 0 \right) } _{v} $ is the same operator $ M_{v} \left( 1 \right) $ that defines the component $ A^{\left( 1 \right) } _{v} $ of the $ D_{3} \left( \mathds{Z} _{2} \right) $ vertex operators, and
					\item this $ A^{\left( 1 \right) } _{v} $ performs lattice gauge transformations that are incapable of changing any state of the $ D_{3} \left( \mathds{Z} _{2} \right) $ model, 
				\end{itemize}
				it is not difficult to conclude that, despite the action of $ W^{\left( 1 , 0 \right) } _{v} $ on a single lattice vertex produces a quasiparticle $ Q^{\left( 1 , 0 \right) } $, the action of this same operator on all the vertices at once keeps this lattice system in the same vacuum state (\ref{32-ground-state-a}) because, for instance,
				\begin{equation*}
					\prod _{v \in \mathcal{L} _{2}} W^{\left( 1 , 0 \right) } _{v} = \prod _{v \in \mathcal{L} _{2}} A^{\left( 1 \right) } _{v} \ .
				\end{equation*}
				In other words, if we analyse this $ D_{3} \left( \mathds{Z} _{2} \right) $ model by taking its vacuum state (\ref{32-ground-state-a}), we see that there is no difference between thinking this quasiparticle $ Q^{\left( 1 , 0 \right) } $ (which has a fusion rule that identifies it as its own anti-quasiparticle) as [\citen{mf-errata}]
				\begin{itemize}
					\item something real, in a situation where $ W^{\left( 1 , 0 \right) } _{v} $ acts on the $ v $-th vertex of a lattice that has all its vertices previously coated by vacuum quasiparticles $ Q^{\left( 0 , 0 \right) } $, or
					\item a hole, in a situation where this same $ W^{\left( 1 , 0 \right) } _{v} $ acts on the $ v $-th vertex of a lattice previously filled by quasiparticles $ Q^{\left( 1 , 0 \right) } $.
				\end{itemize}
			
			\subsubsection{Why can $ Q^{\left( J , \Lambda \right) } $ be interpreted as quasiparticles?}
			
				Of course, even though $ W^{\left( 1 , 0 \right) } _{v} $ is such that $ W^{\left( 1 , 0 \right) } _{v} \circ W^{\left( 1 , 0 \right) } _{v} = \mathds{1} _{v} $, the fact that it does not produce $ Q^{\left( 1 , 0 \right) } $ in pairs may be making you, the reader, question whether $ Q^{\left( 1 , 0 \right) } $ can actually be interpreted as a quasiparticle or not. After all, since these matter excitations are not produced in pairs of \textquotedblleft particle\textquotedblright \hspace*{0.01cm} and \textquotedblleft antiparticle\textquotedblright , it is not possible to transport them over $ \mathcal{L} _{2} $ analogous to what happens, for instance, with the $ D \left( G \right) $ quasiparticles (except by a \textquotedblleft teleport\textquotedblright \hspace*{0.01cm} operator
				\begin{equation*}
					W^{\left( 1 , 0 \right) } _{v^{\prime \prime }} \circ W^{\left( 1 , 0 \right) } _{v^{\prime }}
				\end{equation*}
				that transports it from one vertex $ v^{\prime } $ to another $ v^{\prime \prime } $ completely arbitrary) [\citen{mf-errata}].
				
				Yet, despite this impossibility of transporting $ Q^{\left( 1 , 0 \right) } $, it is interesting to note that, in addition to the fact that there is nothing preventing this $ D_{3} \left( \mathds{Z} _{2} \right) $ model from serving as a guide for the construction of other lattice model(s) that can support this transport, $ Q^{\left( 1 , 0 \right) } $ seems to have some electrical properties. And what allows us to have this perception about $ Q^{\left( 1 , 0 \right) } $ is that, in addition to it fuses (or, at least, overlaps) with all the electric quasiparticles inherited from the $ D \left( \mathds{Z}_{2} \right) $ model at the same lattice vertex, it also presents a kind of electrostatic interaction with another, which have the same flavour (i.e., which have the same $ \left( 1 , 0 \right) $ index), when $ \mu $ is a non-trivial group action. After all, when we have only two quasiparticles $ Q^{\left( 1 , 0 \right) } $ on two vertices $ v^{\prime } $ and $ v^{\prime \prime } $ of $ \mathcal{L} _{2} $, the energy of this system is equal to [\citen{mf-errata}]
				\begin{itemize}
					\item $ E_{0} + 6 \hspace*{0.04cm} \mathcal{J} _{C} $, when $ v^{\prime } $ and $ v^{\prime \prime } $ are neighbours, and
					\item $ E_{0} + 8 \hspace*{0.04cm} \mathcal{J} _{C} $, otherwise.
				\end{itemize}
				In view of all that we have just said, it is impossible not to recognize that, despite this impossibility of transporting $ Q^{\left( 1 , 0 \right) } $ actually prevents us from finding out its statistics by changing its spatial position, it is quite sensible to consider it as a quasiparticle. And since this electrostatic behaviour also shows up in the matter excitations that are produced in the $ D_{M} \left( G \right) $ models whose group actions are not trivial, it also becomes sensible to consider $ Q^{\left( J , \Lambda \right) } $ as quasiparticles. 
				
			\subsubsection{\label{general-non-abelian} On the presence of non-Abelian fusion rules in the Abelian $ D_{M} \left( \mathds{Z} _{N} \right) $ models}
			
				Note that this clarification, which we have just made about the interpretation of $ Q^{\left( J , \Lambda \right) } $ as quasiparticles, further reinforces the similarity between all the $ D_{M} \left( \mathds{Z} _{N} \right) $ vacuum states, which are defined by filling all the vertices with the same matter field (i.e., with the same matter excitation), and Dirac \textquotedblleft seas\textquotedblright . But, as we have also said that all the $ D_{M} \left( \mathds{Z} _{N} \right) $ quasiparticles are produced by the same operators that make up the $ D_{M} \left( \mathds{Z} _{N} \right) $ Hamiltonian, this requires us to answer the following question: given that all these vacuum states can be interpreted as different phases that coexist in the same energy regime, how is it possible to perform transitions between/among all these phases since, when $ \left\vert \mathfrak{Fix} _{\mu } \right\vert \neq 0 $, all the operators $ W^{\left( J , \Lambda \right) } _{v} $ that can be identified in the $ D_{M} \left( \mathds{Z} _{N} \right) $ Hamiltonian cannot excite $ \left\vert \mathfrak{Fix} _{\mu } \right\vert $ of these vacuum states?
				
				In order for us to understand the answer to this question, it is pedagogical to continue using the $ D_{3} \left( \mathds{Z} _{2} \right) $ model as an example, since it is such that $ \left\vert \mathfrak{Fix} _{\mu } \right\vert = 1 $. By the way, in order for us to really understand the answer to this question by using the $ D_{3} \left( \mathds{Z} _{2} \right) $ model, it is crucial to note that the relations (\ref{matter-quasiparticles-creation}) indicate that the matrix representations of $ W^{\left( J , K \right) } _{v} $ in this model need to be, at least, such that
				\begin{equation*}
					W^{\left( J , 0 \right) } _{v} = \begin{pmatrix}
						a_{J0} & b_{J0} & c_{J0} \\
						b_{J0} & a_{J0} & c_{J0} \\
						d_{J0} & d_{J0} & r_{J0}
					\end{pmatrix} \ \ \textnormal{and} \ \ W^{\left( J , 1 \right) } _{v} = \begin{pmatrix}
						a_{J1} & b_{J1} & c_{J1} \\
						- b_{J1} & - a_{J1} & - c_{J1} \\
						d_{J1} & - d_{J1} & 0
					\end{pmatrix} \ ,
				\end{equation*}
				whose entries must be interpreted as complex numbers. Nonetheless, according to what Ref. [\citen{mf-errata}] tells us about the $ H_{3} / \mathds{C} \left( \mathds{Z} _{2} \right) $ model, the only operator that manages to produce a quasiparticle that fosters transitions between the vacuum states (\ref{32-ground-state-a}) and (\ref{32-ground-state-b}) is
				\begin{equation}
					W^{\left( 2 , 0 \right) } _{v} = \begin{pmatrix}
						0 & 0 & 1 \\
						0 & 0 & 1 \\
						1 & 1 & \mathsf{a}
					\end{pmatrix} \ . \label{non-abelian-operator-h3z2}
				\end{equation}
				After all, by noting that $ \mathsf{a} $ is a complex number, the fact that\footnote{Here, we are considering the same single-qudit computational basis states of Ref. [\citen{mf-errata}], where the vector (ket) $ \left\vert n \right\rangle $, with $ n $ being a natural number, can be represented by a column matrix whose $ n $-th row contains the number $ 1 $ while the others are filled with the number $ 0 $.}
				\begin{equation*}
					W^{\left( 2 , 0 \right) } _{v} \bigl\vert 0 \bigr\rangle _{v} = W^{\left( 2 , 0 \right) } _{v} \bigl\vert 1 \bigr\rangle _{v} = \bigl\vert 2 \bigr\rangle _{v} \ \ \textnormal{and} \ \ W^{\left( 2 , 0 \right) } _{v} \bigl\vert 2 \bigr\rangle _{v} = \bigl\vert 0 \bigr\rangle _{v} + \bigl\vert 1 \bigr\rangle _{v} + \mathsf{a} \cdot \bigl\vert 2 \bigr\rangle _{v}
				\end{equation*} 
				makes it clear that, by considering that the vacuum states (\ref{32-ground-state-a}) and (\ref{32-ground-state-b}) correspond to two phases that can coexist in the same energy regime, it is possible to go from one phase to another, and vice versa, through
				\begin{itemize}
					\item an exchange $ W^{\left( 2 , 0 \right) } _{v} \bigl\vert 0 \bigr\rangle _{v} = \bigl\vert 2 \bigr\rangle _{v} $ on all the lattice vertices for a transition from (\ref{32-ground-state-a}) to (\ref{32-ground-state-b}), or
					\item exchanges, which can be carried out by using (several) combinations of the operators $ W^{\left( 1 , 0 \right) } _{v} $ and $ W^{\left( 2 , 0 \right) } _{v} $ that act on all the vertices of $ \mathcal{L} _{2} $, for a transition from (\ref{32-ground-state-b}) to (\ref{32-ground-state-a}).
				\end{itemize}
				In this fashion, by taking into account that the mission of $ W^{\left( 2 , 0 \right) } _{v} $ is to produce a quasiparticle $ Q^{\left( 2 , 0 \right) } $, this allows us to recognize, for instance, that (\ref{32-ground-state-b}) is also similar to a Dirac \textquotedblleft sea\textquotedblright . And what is special about this operator $ W^{\left( 2 , 0 \right) } _{v} $? What is special about it is that, as the composition
				\begin{equation*}
					W^{\left( 2 , 0 \right) } _{v} \circ W^{\left( 2 , 0 \right) } _{v} = \begin{pmatrix}
						1 & 1 & \mathsf{a} \\
						1 & 1 & \mathsf{a} \\
						\mathsf{a} & \mathsf{a} & 2 + \mathsf{a} ^{2}
					\end{pmatrix} = \underbrace{\begin{pmatrix}
						1 & 0 & 0 \\
						0 & 1 & 0 \\
						0 & 0 & 1
					\end{pmatrix}}_{W^{\left( 0 , 0 \right) } _{v}} + \underbrace{\begin{pmatrix}
						0 & 1 & 0 \\
						1 & 0 & 0 \\
						0 & 0 & 1
					\end{pmatrix}}_{W^{\left( 1 , 0 \right) } _{v}} + \ \mathsf{a} \underbrace{\begin{pmatrix}
						0 & 0 & 1 \\
						0 & 0 & 1 \\
						1 & 1 & \mathsf{a}
					\end{pmatrix}}_{W^{\left( 2 , 0 \right) } _{v}}
				\end{equation*}
				is associated with the fusion rule between two excitations $ Q^{\left( 2 , 0 \right) } $, it is clear that the $ D_{3} \left( \mathds{Z} _{2} \right) $ model can support \emph{non-Abelian} fusion rules [\citen{pachos,mf-errata}]. That is, the transition from (\ref{32-ground-state-b}) to (\ref{32-ground-state-a}) is performed by an operator
				\begin{equation*}
					\mathsf{F} = \prod _{v \in \mathcal{L} _{2}} W^{\left( 2 , 0 \right) } _{v}
				\end{equation*}
				that is composed of those that, by acting on each lattice vertex, produce a quasiparticle that presents a non-Abelian fusion rule
				\begin{equation*}
					Q^{\left( 2 , 0 \right) } \times Q^{\left( 2 , 0 \right) } = Q^{\left( 0 , 0 \right) } + Q^{\left( 1 , 0 \right) } + \mathsf{a} \cdot Q^{\left( 2 , 0 \right) }
				\end{equation*}
				with itself.
				
				Note that, although this operator (\ref{non-abelian-operator-h3z2}) is not included in the $ D_{3} \left( \mathds{Z} _{2} \right) $ Hamiltonian, its presence is legitimized, for instance, by Ref. [\citen{miguel-1}]. After all, although its authors have not discussed the need to make transitions among the five vacuum states
				\begin{eqnarray}
					\bigl\vert \xi ^{\left( 0 \right) } _{0} \bigr\rangle & = & \prod _{v^{\prime }} A_{v^{\prime }} \left( \bigotimes _{\ell \in \mathcal{L} _{2}} \left\vert e \right\rangle \right) \otimes \left( \bigotimes _{v \in \mathcal{L} _{2}} \left\vert 0 \right\rangle \right) \quad \textnormal{and} \nonumber \\
					\bigl\vert \xi ^{\left( 2 \right) } _{0} \bigr\rangle & = & \prod ^{s} _{p=1} \left[ O_{L_{p}} \left( \overline{\boldsymbol{\gamma }} ^{\ast } _{p} \right) \right] ^{\lambda _{p}} \prod _{v^{\prime }} A_{v^{\prime }} \left( \bigotimes _{\ell \in \mathcal{L} _{2}} \left\vert e \right\rangle \right) \otimes \left( \bigotimes _{v \in \mathcal{L} _{2}} \left\vert 2 \right\rangle \right) \label{DM-32-independent}
				\end{eqnarray}
				that define the $ D_{3} \left( \mathds{Z} _{2} \right) $ ground state (because there are four
				\begin{equation*}
					\vec{\lambda } = \left( 0 , 0 \right) \ , \ \  \vec{\lambda } = \left( 0 , 1 \right) \ , \ \ \vec{\lambda } = \left( 1 , 0 \right) \ \ \textnormal{and} \ \ \vec{\lambda } = \left( 1 , 1 \right)
				\end{equation*}
				that define four vacuum states (\ref{DM-32-independent}) that are topologically independent of each other), it observes that these five vacuum states can be rewritten by using another Hilbert basis, which allows us to recognize, for instance, that all the lattice vertices have the same matter field $ \left\vert 0 \right\rangle + \left\vert 1 \right\rangle + \mathsf{a} \cdot \left\vert 2 \right\rangle $ with $ \mathsf{a} = 1 $. And according to what was said on page \pageref{degeneracy-non-trivial-dmn}, this five-fold degeneracy of the $ D_{3} \left( \mathds{Z} _{2} \right) $ ground state is consistent with the result (\ref{degeneracy-non-trivial-dmn}), since
				\begin{equation*}
					\mathfrak{n} _{\mathrm{orb}} = 1 \ , \ \ \left\vert \mathfrak{Fix} _{\mu } \right\vert = 1 \ \ \textnormal{and} \ \ \mathfrak{d} _{D \left( \mathds{Z} _{N} \right) } = 4 \ .
				\end{equation*}
				
			\subsubsection{And what do phase transitions tell us about the $ q^{\left( J , L , \Lambda \right) } $ quasiparticles?} 
				
				For the sake of completeness, it is important to point out that, although we have used the $ D_{3} \left( \mathds{Z} _{2} \right) $ model as an example to show that the presence of these quasiparticles, which exhibit non-Abelian fusion rules, support phase transitions in the lowest energy state, it is not difficult to prove that this presence is also needed to support these phase transitions when $ \left\vert \mathfrak{Fix} _{\mu } \right\vert \geqslant 1 $. Incidentally, another thing that we can also prove is that, by exploring a different point of view, which involves recognizing that $ \mathcal{L} _{2} $ is an example of connected graph [\citen{pramod-suggestion,ricardo}], it is possible to interpret some $ D_{M} \left( \mathds{Z} _{N} \right) $ vacuum states as symmetry-protected topological (SPT) phases and, consequently, the transitions among them as some global symmetry breaking. But, since all these $ H_{M} / \mathds{C} \left( \mathds{Z} _{N} \right) $ models can be interpreted special cases of the $ D_{M} \left( G \right) $ ones, we will not go into the details of all these proofs here because they can be found, for instance, in Ref. [\citen{mf-errata}].
				
				Anyway, as we are talking about these phase transitions, it is important to take the opportunity to explain, for instance, what they tell us about the $ q^{\left( J , L , \Lambda \right) } $ quasiparticles. After all, despite these quasiparticles having been inherited from the $ D \left( G \right) $ models, it is quite clear that, in the non-trivial $ D_{M} \left( G \right) $ models (i.e., in the $ D_{M} \left( G \right) $ models that are defined by using a non-trivial group action), these quasiparticles acquire two new properties:
				\begin{itemize}
					\item they can fuse with the $ Q^{\left( J , \Lambda \right) } $ quasiparticles that are produced by manipulating the matter qudits, and
					\item at least part of the magnetic quasiparticles can be confined, similarly to what happens to the quarks that are confined in mesons.
				\end{itemize}
				Thus, given these two new properties, a question that naturally arises is: do these two new properties make these $ q^{\left( J , L , \Lambda \right) } $ quasiparticles very different from those in the $ D \left( G \right) $ models? And the answer to this question is no for a very simple reason: correspondence principle.
				
				In order to understand how the correspondence principle explain this answer, it is necessary to remember that, in the same way that we can define these \textquotedblleft trivial\textquotedblright \hspace*{0.01cm} $ D_{M} \left( G \right) $ models, we can also define the \textquotedblleft trivial\textquotedblright \hspace*{0.01cm} $ D_{M} \left( G \right) $ ones (i.e., we can also define the $ D_{M} \left( G \right) $ models by using a trivial group action). And despite these \textquotedblleft trivial\textquotedblright \hspace*{0.01cm} $ D_{M} \left( G \right) $ models not being very funny because the $ D \left( G \right) $ and Ising models that define them are decoupled, it is precisely this decoupling that, for instance, causes all their magnetic quasiparticles to become unconfined. Of course, one of the consequences of this lack of confinement is the fact that we can evaluate the spin-statistics of all these quasiparticles. But since the $ D \left( G \right) $ and Ising models that define the \textquotedblleft trivial\textquotedblright \hspace*{0.01cm} $ D_{M} \left( G \right) $ models are decoupled, fortunately we do not need to worry about doing this evaluation: after all, as this decoupling also implies that all the $ q^{\left( J , L , \Lambda \right) } $ quasiparticles of these \textquotedblleft trivial\textquotedblright \hspace*{0.01cm} $ D_{M} \left( G \right) $ models are unable to interact with all the $ Q^{\left( J , \Lambda \right) } $ quasiparticles, they just have the same properties inherited from the $ D \left( G \right) $ models. In other words, all the $ q^{\left( J , L , \Lambda \right) } $ quasiparticles of these \textquotedblleft trivial\textquotedblright \hspace*{0.01cm} $ D_{M} \left( G \right) $ models are exactly the same quasiparticles of the $ D \left( G \right) $ ones. In this fashion, by noting that
				\begin{itemize}
					\item the group action $ \mu $ defines how the gauge qudits change the values of the matter qudits, not the other way around, 
					\item all the $ q^{\left( J , L , \Lambda \right) } $ quasiparticles in the $ D_{M} \left( G \right) $ models are produced by the same operators that produce all the quasiparticles in the $ D \left( G \right) $ models, and
					\item the correspondence principle already requires that the $ D \left( G \right) $ vertex operators be interpreted as the $ D_{M} \left( G \right) $ vertex operators that are \textquotedblleft blind\textquotedblright \hspace*{0.01cm} to the matter qudits,
				\end{itemize}
				it is correct to extend what we said at the end of the last paragraph to the $ q^{\left( J , L , \Lambda \right) } $ quasiparticles of all the $ D_{M} \left( G \right) $ models, whether they are \textquotedblleft trivial\textquotedblright \hspace*{0.01cm} or not. After all, since all the vacuum states
				\begin{equation*}
					\bigl\vert \xi ^{\left( \alpha , \vec{\lambda } \right) } _{0} \bigr\rangle  = \prod ^{s} _{p=1} \left[ O_{1} \bigl( \overline{\boldsymbol{\gamma }} ^{\ast } _{p} \bigr) \right] ^{\lambda _{p}} \prod _{v^{\prime }} A_{v^{\prime }} \left( \bigotimes _{\ell \in \mathcal{L} _{2}} \left\vert e \right\rangle \right) \otimes \left( \bigotimes _{v \in \mathcal{L} _{2}} \left\vert \alpha \right\rangle \right) \ , 
				\end{equation*}
				where $ M_{v} \left( g \right) \bigl\vert \alpha \bigr\rangle _{v} = \bigl\vert \alpha \bigr\rangle _{v} $, are mere replicas of the $ D \left( G \right) $ vacuum states, this shows that, just as the $ D \left( G \right) $ models need to be recovered as a special case of the $ D_{M} \left( G \right) $ ones by taking $ M = 1 $, the correspondence principle also requires that this happens when $ M > 2 $. 
	
		\section{\label{QDMf-construction}A dualization procedure on the $ D_{M} \left( G \right) $ models}
		
			Given that the general properties of the $ D_{M} \left( G \right) $ models are now quite clear, it is really time for us to turn over a new leaf to (finally!) pay attention to the construction of a new class of lattice gauge models ($ D^{K} \left( G \right) $) that are dual to the $ D_{M} \left( G \right) $ ones. After all, in addition to being valid to say that these new $ D^{K} \left( G \right) $ models can give us some clue as to how it might be possible to get a lattice model that is self-dual, it is also valid to say that the $ D_{M} \left( G \right) $ models previously discussed give us a tremendous advantage in this dual context. And by considering, for instance, the content of Figure \ref{dual-figure},
			\begin{figure}[!t]
				\begin{center}					
					\begin{tikzpicture}
						\draw[color=blue!20,fill=blue!20] (6,5) rectangle (8,7);
						\draw[color=red!20,fill=red!20] (1,2) rectangle (3,4);
						\draw[color=green!20,fill=green!20] (6.5,1.5) rectangle (9.5,2.5);
						\draw[dotted, color=gray!30, ultra thick] (1,0) -- (1,8);
						\draw[dotted, color=gray!30, ultra thick] (3,0) -- (3,8);
						\draw[dotted, color=gray!30, ultra thick] (5,0) -- (5,8);
						\draw[dotted, color=gray!30, ultra thick] (7,0) -- (7,8);
						\draw[dotted, color=gray!30, ultra thick] (9,0) -- (9,8);
						\draw[dotted, color=gray!30, ultra thick] (-1,2) -- (11,2);
						\draw[dotted, color=gray!30, ultra thick] (-1,4) -- (11,4);
						\draw[dotted, color=gray!30, ultra thick] (-1,6) -- (11,6);
						\draw[->, color=gray, ultra thick, >=stealth] (0,0) -- (0,2.2);
						\draw[->, color=gray, ultra thick, >=stealth] (0,2) -- (0,4.2);
						\draw[->, color=gray, ultra thick, >=stealth] (0,4) -- (0,6.2);
						\draw[-, color=gray, ultra thick] (0,6) -- (0,8);
						\draw[->, color=gray, ultra thick, >=stealth] (2,0) -- (2,2.2);
						\draw[->, color=gray, ultra thick, >=stealth] (2,2) -- (2,4.2);
						\draw[->, color=gray, ultra thick, >=stealth] (2,4) -- (2,6.2);
						\draw[-, color=gray, ultra thick] (2,6) -- (2,8);
						\draw[->, color=gray, ultra thick, >=stealth] (4,0) -- (4,2.2);
						\draw[->, color=gray, ultra thick, >=stealth] (4,2) -- (4,4.2);
						\draw[->, color=gray, ultra thick, >=stealth] (4,4) -- (4,6.2);
						\draw[-, color=gray, ultra thick] (4,6) -- (4,8);
						\draw[->, color=gray, ultra thick, >=stealth] (6,0) -- (6,2.2);
						\draw[->, color=gray, ultra thick, >=stealth] (6,2) -- (6,4.2);
						\draw[->, color=gray, ultra thick, >=stealth] (6,4) -- (6,6.2);
						\draw[-, color=gray, ultra thick] (6,6) -- (6,8);
						\draw[->, color=gray, ultra thick, >=stealth] (8,0) -- (8,2.2);
						\draw[->, color=gray, ultra thick, >=stealth] (8,2) -- (8,4.2);
						\draw[->, color=gray, ultra thick, >=stealth] (8,4) -- (8,6.2);
						\draw[-, color=gray, ultra thick] (8,6) -- (8,8);
						\draw[->, color=gray, ultra thick, >=stealth] (10,0) -- (10,2.2);
						\draw[->, color=gray, ultra thick, >=stealth] (10,2) -- (10,4.2);
						\draw[->, color=gray, ultra thick, >=stealth] (10,4) -- (10,6.2);
						\draw[-, color=gray, ultra thick] (10,6) -- (10,8);
						\draw[->, color=gray, ultra thick, >=stealth] (-1,1) -- (1.2,1);
						\draw[->, color=gray, ultra thick, >=stealth] (1,1) -- (3.2,1);
						\draw[->, color=gray, ultra thick, >=stealth] (3,1) -- (5.2,1);
						\draw[->, color=gray, ultra thick, >=stealth] (5,1) -- (7.2,1);
						\draw[->, color=gray, ultra thick, >=stealth] (7,1) -- (9.2,1);
						\draw[-, color=gray, ultra thick] (9,1) -- (11,1);
						\draw[->, color=gray, ultra thick, >=stealth] (-1,3) -- (1.2,3);
						\draw[->, color=gray, ultra thick, >=stealth] (1,3) -- (3.2,3);
						\draw[->, color=gray, ultra thick, >=stealth] (3,3) -- (5.2,3);
						\draw[->, color=gray, ultra thick, >=stealth] (5,3) -- (7.2,3);
						\draw[->, color=gray, ultra thick, >=stealth] (7,3) -- (9.2,3);
						\draw[-, color=gray, ultra thick] (9,3) -- (11,3);
						\draw[->, color=gray, ultra thick, >=stealth] (-1,5) -- (1.2,5);
						\draw[->, color=gray, ultra thick, >=stealth] (1,5) -- (3.2,5);
						\draw[->, color=gray, ultra thick, >=stealth] (3,5) -- (5.2,5);
						\draw[->, color=gray, ultra thick, >=stealth] (5,5) -- (7.2,5);
						\draw[->, color=gray, ultra thick, >=stealth] (7,5) -- (9.2,5);
						\draw[-, color=gray, ultra thick] (9,5) -- (11,5);
						\draw[->, color=gray, ultra thick, >=stealth] (-1,7) -- (1.2,7);
						\draw[->, color=gray, ultra thick, >=stealth] (1,7) -- (3.2,7);
						\draw[->, color=gray, ultra thick, >=stealth] (3,7) -- (5.2,7);
						\draw[->, color=gray, ultra thick, >=stealth] (5,7) -- (7.2,7);
						\draw[->, color=gray, ultra thick, >=stealth] (7,7) -- (9.2,7);
						\draw[-, color=gray, ultra thick] (9,7) -- (11,7);
					\end{tikzpicture}
				\end{center}
				\caption{\label{dual-figure} Piece of the same oriented square lattice $ \mathcal{L} _{2} $ that supports the $ D_{M} \left( G \right) $ models, where the presence of its dual lattice $ \mathcal{L} ^{\ast } _{2} $ is now being highlighted by using dotted lines. Observe that, as in Figure \ref{QMDv-rede}, here we are also highlighting the same the rose ($ S_{v} $) and baby blue ($ S_{f} $) coloured sectors, which are centred in the $ v $-th vertex and $ f $-th face of $ \mathcal{L} _{2} $ respectively. Nevertheless, contrary to what happens in Figure \ref{QMDv-rede}, which shows an orange sector ($ S_{\ell } $) composed of an edge and the two vertices that limit it, here we see a new green sector ($ S^{\prime } _{\ell } $) that, despite being centred on an $ \ell $-th edge, does not contain the two vertices that limit this edge. But while these sectors $ S_{\ell } $ and $ S^{\prime } _{\ell } $ are different from the point of view of $ \mathcal{L} _{2} $, it is important to observe that, when we look at $ S^{\prime } _{\ell } $ from the perspective of $ \mathcal{L} ^{\ast } _{2} $, it is equivalent to what $ S_{\ell } $ is from the perspective of $ \mathcal{L} _{2} $: i.e., when we look at $ \mathcal{L} ^{\ast } _{2} $, it is quite clear that $ S^{\prime } _{\ell } $ is geometrically dual to $ S_{\ell } $, and vice versa, because $ S^{\prime } _{\ell } $ is composed of a dual edge and the two dual vertices that limit it. In this way, as $ \mathcal{L} ^{\ast } _{2} $ also shows us that $ S_{v} $ is geometrically dual to $ S_{f} $, and vice versa, it becomes clear that, for the $ D^{K} \left( G \right) $ models to be interpreted as duals to the $ D_{M} \left( G \right) $ ones, it is necessary that the $ D^{K} \left( G \right) $ Hamiltonian be analogous to (\ref{H-qdmv}), but with its vertex, face and link operators acting on $ S_{v} $, $ S_{f} $ and $ S^{\prime } _{\ell } $ respectively.}
			\end{figure}
			which illustrates the existence of a dual lattice $ \mathcal{L} ^{\ast } _{2} $ (i.e., a lattice whose vertices/faces are the faces/vertices of $ \mathcal{L} _{2} $), it becomes clear that one of the aspects of this tremendous advantage is in the fact that this geometric duality can be used as a first guide to define the $ D^{K} \left( G \right) $ models as the algebraic duals of the $ D_{M} \left( G \right) $ ones.
			
			In order to begin to show how this geometric duality can be used as this first guide, it is important to remember that the self-duality of the $ D \left( G \right) $ models is not only characterized by the fact that, for each quasiparticle detected by the vertex operator $ A_{v} $, there is always another one, with the same properties, that is detected by the face operator $ B_{f} $, and vice versa: this self-duality of the can also be characterized by the fact that these $ D \left( G \right) $ vertex and face operators can be interpreted as the duals of each other. In other words, although it is already clear that $ A_{v} $ and $ B_{f} $ effectively act on the edges that define the vertices and faces of $ \mathcal{L} _{2} $ respectively, it is not difficult to see, from this geometric dual point of view of Figure \ref{dual-figure}, that these same operators also effectively act on the edges that define the faces and vertices of $ \mathcal{L} ^{\ast } _{2} $ respectively. And this is an important reminder because, as with the $ D_{M} \left( G \right) $ models previously analysed, there must be a (mathematical) correspondence between the $ D \left( G \right) $ and $ D^{K} \left( G \right) $ models so that the first ones (i.e., the $ D \left( G \right) $ models) can be recovered as special cases of the second ones (i.e., the $ D^{K} \left( G \right) $ models). In this sense, and by bearing in mind all the dual issues that have already been observed with the aid of Figure \ref{dual-figure}, it is not difficult to conclude that, if these $ D^{K} \left( G \right) $ models really exist, their Hamiltonian operators must be expressed as
			\begin{equation}
				H_{D^{K} \left( G \right) } = - \mathcal{J} ^{\prime } _{A} \sum _{v \in \mathcal{L} _{2}} A^{\prime } _{v} - \mathcal{J} ^{\prime } _{B} \sum _{f \in \mathcal{L} _{2}} B^{\prime } _{f} - \mathcal{J} ^{\prime } _{C} \sum _{\ell \in \mathcal{L} _{2}} C^{\prime } _{\ell } \ . \label{H-qdmf}
			\end{equation}
			Here, $ \mathcal{J} ^{\prime } _{A} $, $ \mathcal{J} ^{\prime } _{B} $ and $ \mathcal{J} ^{\prime } _{C} $ are three positive parameters; and $ A^{\prime } _{v} $, $ B^{\prime } _{f} $ and $ C^{\prime } _{\ell } $ are the \textquotedblleft new\textquotedblright \hspace*{0.01cm} vertex, face and link operators respectively, whose definitions and properties will be discussed from now on.
			
			\subsection{Some considerations about the vertex, face and edge operators}
			
				Because these new $ D^{K} \left( G \right) $ models also need to be interpreted as generalizations of the $ D \left( G \right) $ models, an obvious fact that we have to keep in mind is that all these operators $ A^{\prime } _{v} $, $ B^{\prime } _{f} $ and $ C^{\prime } _{\ell } $ also need to act on the same Hilbert (sub)space
				\begin{equation*}
					\underbrace{\mathfrak{H} _{\left\vert G \right\vert } \otimes \ldots \otimes \mathfrak{H} _{\left\vert G \right\vert }} _{N_{\ell } \ \textnormal{\tiny{times}}} 
				\end{equation*} 
				that was already associated with $ \mathcal{L} _{2} $ in the $ D \left( G \right) $ and $ D_{M} \left( G \right) $ models. And given that the obviousness of this fact lies in our desire to recognize the $ D_{M} \left( G \right) $ and $ D^{K} \left( G \right) $ models as duals of each other, it is worth noting that, because the sectors $ S_{v} $ and $ S_{f} $ are duals of each other, the $ D^{K} \left( G \right) $ vertex operators need to be exactly the same as the $ D \left( G \right) $ models. That is, since
				\begin{itemize}
					\item the face operators of the $ D \left( G \right) $ and $ D_{M} \left( G \right) $ models are the same, and
					\item the $ D^{K} \left( G \right) $ vertex operators need to be dual to the $ D_{M} \left( G \right) $ face operators,
				\end{itemize}
				these $ D^{K} \left( G \right) $ vertex operators need to be defined as
				\begin{equation}
					A^{\prime } _{v,J} = \frac{1}{\vert G \vert } \sum _{g \in G} \chi _{1+J} \left( g^{-1} \right) \cdot A^{\left( g \right) } _{v} \ , \label{a-prime}
				\end{equation}
				whose components $ A^{\left( g \right) } _{v} $ are exactly the same as those defined in Figure \ref{kuperberg-figure}.
				
				As a matter of fact, a good panorama of how the $ D^{K} \left( G \right) $ vertex, face and link operators act on $ \mathcal{L} _{2} $ can be understood with the help of Figure \ref{QMDf-rede},
				\begin{figure}[!t]
					\begin{center}					
						\begin{tikzpicture}
							\draw[color=blue!20,fill=blue!20] (6,5) rectangle (8,7);
							\draw[color=red!20,fill=red!20] (1,2) rectangle (3,4);
							\draw[color=green!20,fill=green!20] (6.5,1.5) rectangle (9.5,2.5);
							\draw[->, color=gray, ultra thick, >=stealth] (0,0) -- (0,2.2);
							\draw[->, color=gray, ultra thick, >=stealth] (0,2) -- (0,4.2);
							\draw[->, color=gray, ultra thick, >=stealth] (0,4) -- (0,6.2);
							\draw[-, color=gray, ultra thick] (0,6) -- (0,8);
							\draw[->, color=gray, ultra thick, >=stealth] (2,0) -- (2,2.2);
							\draw[->, color=gray, ultra thick, >=stealth] (2,2) -- (2,4.2);
							\draw[->, color=gray, ultra thick, >=stealth] (2,4) -- (2,6.2);
							\draw[-, color=gray, ultra thick] (2,6) -- (2,8);
							\draw[->, color=gray, ultra thick, >=stealth] (4,0) -- (4,2.2);
							\draw[->, color=gray, ultra thick, >=stealth] (4,2) -- (4,4.2);
							\draw[->, color=gray, ultra thick, >=stealth] (4,4) -- (4,6.2);
							\draw[-, color=gray, ultra thick] (4,6) -- (4,8);
							\draw[->, color=gray, ultra thick, >=stealth] (6,0) -- (6,2.2);
							\draw[->, color=gray, ultra thick, >=stealth] (6,2) -- (6,4.2);
							\draw[->, color=gray, ultra thick, >=stealth] (6,4) -- (6,6.2);
							\draw[-, color=gray, ultra thick] (6,6) -- (6,8);
							\draw[->, color=gray, ultra thick, >=stealth] (8,0) -- (8,2.2);
							\draw[->, color=gray, ultra thick, >=stealth] (8,2) -- (8,4.2);
							\draw[->, color=gray, ultra thick, >=stealth] (8,4) -- (8,6.2);
							\draw[-, color=gray, ultra thick] (8,6) -- (8,8);
							\draw[->, color=gray, ultra thick, >=stealth] (10,0) -- (10,2.2);
							\draw[->, color=gray, ultra thick, >=stealth] (10,2) -- (10,4.2);
							\draw[->, color=gray, ultra thick, >=stealth] (10,4) -- (10,6.2);
							\draw[-, color=gray, ultra thick] (10,6) -- (10,8);
							\draw[->, color=gray, ultra thick, >=stealth] (-1,1) -- (1.2,1);
							\draw[->, color=gray, ultra thick, >=stealth] (1,1) -- (3.2,1);
							\draw[->, color=gray, ultra thick, >=stealth] (3,1) -- (5.2,1);
							\draw[->, color=gray, ultra thick, >=stealth] (5,1) -- (7.2,1);
							\draw[->, color=gray, ultra thick, >=stealth] (7,1) -- (9.2,1);
							\draw[-, color=gray, ultra thick] (9,1) -- (11,1);
							\draw[->, color=gray, ultra thick, >=stealth] (-1,3) -- (1.2,3);
							\draw[->, color=gray, ultra thick, >=stealth] (1,3) -- (3.2,3);
							\draw[->, color=gray, ultra thick, >=stealth] (3,3) -- (5.2,3);
							\draw[->, color=gray, ultra thick, >=stealth] (5,3) -- (7.2,3);
							\draw[->, color=gray, ultra thick, >=stealth] (7,3) -- (9.2,3);
							\draw[-, color=gray, ultra thick] (9,3) -- (11,3);
							\draw[->, color=gray, ultra thick, >=stealth] (-1,5) -- (1.2,5);
							\draw[->, color=gray, ultra thick, >=stealth] (1,5) -- (3.2,5);
							\draw[->, color=gray, ultra thick, >=stealth] (3,5) -- (5.2,5);
							\draw[->, color=gray, ultra thick, >=stealth] (5,5) -- (7.2,5);
							\draw[->, color=gray, ultra thick, >=stealth] (7,5) -- (9.2,5);
							\draw[-, color=gray, ultra thick] (9,5) -- (11,5);
							\draw[->, color=gray, ultra thick, >=stealth] (-1,7) -- (1.2,7);
							\draw[->, color=gray, ultra thick, >=stealth] (1,7) -- (3.2,7);
							\draw[->, color=gray, ultra thick, >=stealth] (3,7) -- (5.2,7);
							\draw[->, color=gray, ultra thick, >=stealth] (5,7) -- (7.2,7);
							\draw[->, color=gray, ultra thick, >=stealth] (7,7) -- (9.2,7);
							\draw[-, color=gray, ultra thick] (9,7) -- (11,7);
							\draw[->, ultra thick, >=stealth] (2,1) -- (2,2.2);
							\draw[->, ultra thick, >=stealth] (2,2.0) -- (2,4.2);
							\draw[-, ultra thick] (2,4.0) -- (2,5.0);
							\draw[->, ultra thick, >=stealth] (0.0,3) -- (1.2,3);
							\draw[->, ultra thick, >=stealth] (1.0,3) -- (3.2,3);
							\draw[-, ultra thick] (3.0,3) -- (4.0,3);
							\draw[->, ultra thick, >=stealth] (6,5.0) -- (6,6.2);
							\draw[-, ultra thick] (6,6.0) -- (6,7.0);
							\draw[->, ultra thick, >=stealth] (8,5.0) -- (8,6.2);
							\draw[-, ultra thick] (8,6.0) -- (8,7.0);
							\draw[->, ultra thick, >=stealth] (6.0,5) -- (7.2,5);
							\draw[-, ultra thick] (7.0,5) -- (8.0,5);
							\draw[->, ultra thick, >=stealth] (6.0,7) -- (7.2,7);
							\draw[-, ultra thick] (7.0,7) -- (8.0,7);
							\draw[->, ultra thick, >=stealth] (8,1.0) -- (8,2.2);
							\draw[-, ultra thick] (8,2.0) -- (8,3.0);
							\node [right] (2.1,4.2) at (2.1,4.2) {$ a $};
							\node [below] (3.2,2.9) at (3.2,2.9) {$ b $};
							\node [left] (1.9,1.8) at (1.9,1.8) {$ c $};
							\node [above] (0.8,3.1) at (0.8,3.1) {$ d $};
							\draw[color=black,fill=white] (7,6) circle (0.3);
							\node [] (7,6) at (7,6) {$ \tilde{\gamma } $};
							\node [right] (7,7.3) at (7,7.3) {$ r $};
							\node [below] (8.3,6) at (8.3,6) {$ s $};
							\node [left] (7,4.7) at (7,4.7) {$ t $};
							\node [above] (5.7,6) at (5.7,6) {$ u $};
							\draw[color=black,fill=white] (7,2) circle (0.3);
							\node [] (7,2) at (7,2) {$ \tilde{\alpha } $};
							\draw[color=black,fill=white] (9,2) circle (0.3);
							\node [] (9,2) at (9,2) {$ \tilde{\beta } $};
							\node [right] (8.0,1.7) at (8.0,1.7) {$ \ell $};
						\end{tikzpicture}
					\end{center}
					\caption{\label{QMDf-rede} Replica of the previous figure to illustrate how $ \mathcal{L} _{2} $ can be used to support the $ D^{K} \left( G \right) $ models. Although, here, there are no longer the dotted lines of $ \mathcal{L} ^{\ast } _{2} $, we see the same the rose ($ S_{v} $) and baby blue ($ S_{f} $) coloured sectors that were highlighted in Figure \ref{QMDv-rede}, which are centred in the $ v $-th vertex and $ f $-th face of $ \mathcal{L} _{2} $ respectively, and a new sector ($ S^{\prime } _{\ell } $), which is centred in the $ \ell $-th edge of $ \mathcal{L} _{2} $, is highlighted in light green. Note that, when we compare the present piece with the one shown in Figure \ref{QMDv-rede}, it turns out to be clear that this light green coloured sector can be interpreted as the geometric dual of the light orange coloured sector of the Figure \ref{QMDv-rede}. After all, in the same way that the light orange coloured sector is defined by one edge and its end vertices, on which there are two matter fields $ \left\vert \alpha \right\rangle $ and $ \left\vert \beta \right\rangle $, the light green coloured sector is also defined by one (dual) edge and its end (dual) vertices, which also support two new matter fields $ \left\vert \tilde{\alpha } \right\rangle $ and $ \bigl\vert \tilde{\beta } \bigr\rangle $. These new matter fields will be denoted in this way (i.e., by using a tilde symbol) only for the convenience of distinguishing them from the previous matter fields.}
				\end{figure}	
				which can be interpreted as a new version of Figure \ref{dual-figure} where, despite $ \mathcal{L} ^{\ast } _{2} $ having been \textquotedblleft strangely\textquotedblright \hspace*{0.01cm} hidden, the presence of the $ D^{K} \left( G \right) $ gauge and matter fields is now being highlighted. Of course, by looking at this Figure \ref{QMDf-rede}, perhaps you, the reader, are wondering why we have \textquotedblleft strangely\textquotedblright \hspace*{0.01cm} hidden $ \mathcal{L} ^{\ast } _{2} $. And if you are really asking this question, the best answer that we can give you is that $ \mathcal{L} ^{\ast } _{2} $ was used only as a mere guide so that we could start to see/explore this duality: after all, remember that the lattice, which is the main protagonist of all these $ D \left( G \right) $, $ D_{M} \left( G \right) $ and $ D^{K} \left( G \right) $ models, is $ \mathcal{L} _{2} $. In this way, since the vertices of $ \mathcal{L} ^{\ast } _{2} $ correspond to the faces of $ \mathcal{L} _{2} $, this is precisely what explains why $ D^{K} \left( G \right) $ models need to be defined by assigning matter fields only to these faces.
				
				\subsubsection{How can we define the face operator $ B^{\prime } _{f} $?}
				
					Given our desire to recognize the $ D_{M} \left( G \right) $ and $ D^{K} \left( G \right) $ models as duals of each other, one thing we can already say about the $ D^{K} \left( G \right) $ face operators is that they obviously need to act on the matter fields. And in parallel to this, as we already know that the $ D \left( G \right) $ face operators measure the holonomies around the lattice faces, it is also correct to say that these $ D^{K} \left( G \right) $ face operators also need to do the \textquotedblleft same\textquotedblright \hspace*{0.01cm} thing. That is, these $ D^{K} \left( G \right) $ face operators need to measure how deformed are these lattice faces due to the presence of quasiparticles that, now, can be produced by manipulating gauge or/and matter qudits. Note that, since the correspondence principle also requires that the $ D \left( G \right) $ models be recovered as special cases of the $ D^{K} \left( G \right) $ ones in some limit that will become clear later on, these $ D^{K} \left( G \right) $ face operators need to measure the same holonomies as the $ D \left( G \right) $ ones when this limit is reached.
					
					Although we still do not know what the exact expressions of these $ D^{K} \left( G \right) $ face operators are, another thing that we can already say about them is that, in the same way that the operator $ B_{f} $ in (\ref{H-qdmv}) measures flat connections, the operator $ B_{f} $ in (\ref{H-qdmf}) also needs to measure the \textquotedblleft trivial holonomies\textquotedblright \hspace*{0.01cm} around the lattice faces. However, as Figure \ref{QMDf-rede} already makes it clear that there are matter fields on all these lattice faces, this \textquotedblleft trivial holonomy\textquotedblright \hspace*{0.01cm} may not be exactly the same trivial holonomy that $ B_{f} $ is able to measure, which explains the use of quotation marks. After all, as these matter fields must support the production of quasiparticles and, therefore, the presence of these quasiparticles will also be responsible for locally deforming the lattice in some way, these matter fields must be taken into account in the calculation of these \textquotedblleft trivial holonomies\textquotedblright . 
					
					Of course, even though we have just said a few words about these \textquotedblleft trivial holonomies\textquotedblright , everything we have said is still vague. And since we want to understand what the exact expressions of these $ D^{K} \left( G \right) $ face operators are, we need to stop being vague and present the exact definition of these \textquotedblleft trivial holonomies\textquotedblright . But before we present this definition, it is important to remember that all the different holonomies, which can be measured around the lattice faces by the $ D \left( G \right) $ and $ D_{M} \left( G \right) $ face operators, are defined as
					\begin{equation}
						h = a^{-1} b^{-1} cd \label{holonomy}
					\end{equation}
					by using the same binary operation that defines $ G $ as a group. And why is it important to remember this? Because this gives us a strong indication that it is perfectly possible to define something, which is very similar to a holonomy, as
					\begin{equation}
						h^{\prime } = \mathfrak{f} \left( \tilde{\alpha } \right) \cdot h = \mathfrak{f} \left( \tilde{\alpha } \right) \cdot a^{-1} b^{-1} cd \label{fake-holonomy}
					\end{equation}
					by using a function $ \mathfrak{f} : \tilde{S} \rightarrow G $ because this definition makes $ h^{\prime } $ an element of $ G $. Here, $ \tilde{S} $ is a set that indexes the elements of $ \mathcal{B} _{m} = \bigl\{ \left\vert \tilde{\alpha } \right\rangle : \tilde{\alpha } \in \tilde{S} \bigr\} $, which is the single-qudit computational basis of the Hilbert space $ \mathfrak{H} _{K} $ that supports the matter qudits that are assigned to the lattice faces.
					
					But despite the expression (\ref{fake-holonomy}) of this \textquotedblleft fake holonomy\textquotedblright \hspace*{0.01cm} makes some sense because, whatever the values of $ \mathfrak{f} \left( \tilde{\alpha } \right) $, it also allows us to define the same $ R $ distinct (non-equivalent) conjugacy classes
					\begin{equation*}
						\mathsf{C} _{L} = \bigl\{ h^{\prime } g_{L} \left( h^{\prime } \right) ^{-1} : h^{\prime } \in G \bigr\} \ ,
					\end{equation*}
					is there any mathematical result that guarantees that a function $ \mathfrak{f} $ can actually be used to make (\ref{fake-holonomy}) model all the possible face deformations in the presence of the matter fields? And the answer to this question is yes: not only does this mathematical result exist, but it also serves the purpose of interpreting the $ D_{M} \left( G \right) $ and $ D^{K} \left( G \right) $ models as duals of each other. After all, by remembering that the $ D_{M} \left( G \right) $ models were produced, by coupling the $ D \left( G \right) $ ones to matter fields allocated on the lattice vertices with the help of a group action $ \mu : G \times S \rightarrow S $, if we really want the $ D_{M} \left( G \right) $ vertex operators to be interpreted as the algebraic dual of the $ D^{K} \left( G \right) $ face operators and vice versa, the second ones (i.e., the $ D^{K} \left( G \right) $ face operators) need to be defined by using a \emph{co-action}
					\begin{equation}
						\tilde{\alpha } \mapsto \mathcal{F} \left( \tilde{\alpha } \right) = \mathfrak{f} \left( \tilde{\alpha } \right) \otimes \tilde{\alpha } \ , \label{co-action}
					\end{equation}
					where $ \tilde{\alpha } $ and $ \mathfrak{f} \left( \tilde{\alpha } \right) $ must be elements of $ \tilde{S} $ and $ G $ respectively. And since $ \mathfrak{f} $ exists, it is not absurd to use it to define (\ref{fake-holonomy}) and, therefore, the expressions of these $ D^{K} \left( G \right) $ face operators that need to be dual to those starred by a group action. In this fashion, by noting that
					\begin{itemize}
						\item this \textquotedblleft fake holonomy\textquotedblright \hspace*{0.01cm} (\ref{fake-holonomy}) can be reduced to the true holonomy (\ref{holonomy}) in some special cases, and
						\item the \textquotedblleft trivial holonomies\textquotedblright , which $ B^{\prime } _{f} $ needs to be able to measure, also need to be characterized by the neutral element of $ G $,
					\end{itemize}
					it seems convenient to define the $ D^{K} \left( G \right) $ face operators as
					\begin{equation}
						B^{\prime } _{f,h^{\prime }} \equiv B^{\prime \left( h^{\prime } \right) } _{f} \ , \label{b-prime}
					\end{equation}
					whose components $ B^{\prime \left( h^{\prime } \right) } _{f} $ are defined in Figure \ref{QMDf-operators-components}).
					\begin{figure}[!t]
						\begin{center}
							\begin{tikzpicture}[
								scale=0.3,
								equation/.style={thin},
								trans/.style={thin,shorten >=0.5pt,shorten <=0.5pt,>=stealth},
								flecha/.style={thin,->,shorten >=0.5pt,shorten <=0.5pt,>=stealth}
								]
								\draw[equation] (-10.8,0.0) -- (-10.8,0.0) node[midway,right] {$ B^{\prime \left( h^{\prime } \right) } _{f} $};	
								\draw[trans] (-7.1,2.3) -- (-7.1,-2.3) node[above=2pt,right=-1pt] {};
								\draw[trans] (-1.6,2.3) -- (-0.4,-0.06) node[above=2pt,right=-1pt] {};
								\draw[trans] (-1.6,-2.3) -- (-0.4,0.06) node[above=2pt,right=-1pt] {};
								\draw[flecha] (-6.2,1.2) -- (-3.6,1.2) node[above=2pt,right=-1pt] {};
								\draw[trans] (-3.8,1.2) -- (-1.6,1.2) node[above=2pt,right=-1pt] {};
								\draw[flecha] (-6.2,-1.2) -- (-3.6,-1.2) node[above=2pt,right=-1pt] {};
								\draw[trans] (-3.8,-1.2) -- (-1.6,-1.2) node[above=2pt,right=-1pt] {};
								\draw[flecha] (-5.3,-2.0) -- (-5.3,0.3) node[above=2pt,right=-1pt] {};
								\draw[trans] (-5.3,0.0) -- (-5.3,2.0) node[above=2pt,right=-1pt] {};
								\draw[flecha] (-2.5,-2.0) -- (-2.5,0.3) node[above=2pt,right=-1pt] {};
								\draw[trans] (-2.5,0.0) -- (-2.5,2.0) node[above=2pt,right=-1pt] {};
								\draw[trans,fill=white] (-3.85,0.0) circle (0.9);
								\draw[equation] (-3.8,2.0) -- (-3.8,2.0) node[midway] {$ a $};
								\draw[equation] (-0.8,0.0) -- (-0.8,0.0) node[midway,left] {$ d $};
								\draw[equation] (-3.8,-2.0) -- (-3.8,-2.0) node[midway] {$ c $};
								\draw[equation] (-7.0,0.0) -- (-7.0,0.0) node[midway,right] {$ b $};
								\draw[equation] (-3.85,0.0) -- (-3.85,0.0) node[midway] {$ \tilde{\alpha } $};
								\draw[equation] (7.2,-0.1) -- (7.2,-0.1) node[midway] {$ = \ \delta \bigl( h^{\prime } , \mathfrak{f} \left( \tilde{\alpha } \right) \cdot a^{-1} b^{-1} cd \bigr) $};
								\draw[trans] (14.5,2.3) -- (14.5,-2.3) node[above=2pt,right=-1pt] {};
								\draw[trans] (20.0,2.3) -- (21.1,-0.06) node[above=2pt,right=-1pt] {};
								\draw[trans] (20.0,-2.3) -- (21.1,0.06) node[above=2pt,right=-1pt] {};
								\draw[flecha] (15.3,1.2) -- (17.9,1.2) node[above=2pt,right=-1pt] {};
								\draw[trans] (17.7,1.2) -- (19.9,1.2) node[above=2pt,right=-1pt] {};
								\draw[flecha] (15.3,-1.2) -- (17.9,-1.2) node[above=2pt,right=-1pt] {};
								\draw[trans] (17.7,-1.2) -- (19.9,-1.2) node[above=2pt,right=-1pt] {};
								\draw[flecha] (16.2,-2.0) -- (16.2,0.3) node[above=2pt,right=-1pt] {};
								\draw[trans] (16.2,0.0) -- (16.2,2.0) node[above=2pt,right=-1pt] {};
								\draw[flecha] (19.0,-2.0) -- (19.0,0.3) node[above=2pt,right=-1pt] {};
								\draw[trans] (19.0,0.0) -- (19.0,2.0) node[above=2pt,right=-1pt] {};
								\draw[trans,fill=white] (17.65,0.0) circle (0.9);
								\draw[equation] (17.7,2.0) -- (17.7,2.0) node[midway] {$ a $};
								\draw[equation] (20.7,0.0) -- (20.7,0.0) node[midway,left] {$ d $};
								\draw[equation] (17.7,-2.0) -- (17.7,-2.0) node[midway] {$ c $};
								\draw[equation] (14.5,0.0) -- (14.5,0.0) node[midway,right] {$ b $};
								\draw[equation] (17.65,0.0) -- (17.65,0.0) node[midway] {$ \tilde{\alpha } $};	
						\end{tikzpicture}
					\end{center}
					\caption{\label{QMDf-operators-components} Definition of the components $ B^{\prime \left( h^{\prime } \right) } _{f} $ that define the $ D^{K} \left( G \right) $ face operators in terms of their effective action on $ \mathcal{L} _{2} $.}
				\end{figure}
				Here, as with all the operators in (\ref{H-qdmv}), the face operator that make up the $ D^{K} \left( G \right) $ Hamiltonian (\ref{H-qdmf}) is defined as $ B^{\prime } _{f} = B^{\prime \left( e \right) } _{f} $. 
			
			\subsubsection{How can we define the link operator $ C^{\prime } _{\ell } $?}
			
				In view of what we have just said, perhaps you, the reader, are wondering what guarantees the existence of $ \mathfrak{f} $. And this is an extremely relevant question whose answer can be well understood, for instance, by noticing that the $ D \left( G \right) $ models are Hamiltonian realizations of lattice gauge theories based (i) on an involutive \emph{Hopf algebra} $ \mathds{C} \left( G \right) $ [\citen{kuperberg}] and (ii) on \emph{finite quantum groupoids} (i.e., on a \emph{weak Hopf algebra}) [\citen{chang}]. More specifically, it is possible to affirm that the $ D \left( G \right) $ Hamiltonian realizes a representation of the Drinfeld's quantum double [\citen{drinfeld}] of these involutive Hopf algebras [\citen{buer}]. And why is it important to notice this? Because the underlying algebra with involution is a \emph{star-algebra} [\citen{polcino-1}] that, for instance, allows us to describe the $ D \left( G \right) $ models based on \emph{star-quantum groupoids} [\citen{chang,nikshych}]. After all, in addition to being possible to prove that, whenever a group $ G $ acts on a ring $ \mathcal{A} $ that can be interpreted as a star-algebra, there is a co-action $ \mathcal{F} : \mathcal{A} \rightarrow \mathds{C} \left( G \right) \otimes \mathcal{A} $, it is also possible to prove that this $ \mathcal{F} $ can be given by (\ref{co-action}) as long as $ \mathfrak{f} $ is a \emph{homomorphism}: i.e., this function $ \mathfrak{f} $, which we need to define the \textquotedblleft fake holonomy\textquotedblright \hspace*{0.01cm} (\ref{fake-holonomy}), exists and allows us to interpret (\ref{co-action}) as a \emph{co-action homomorphism}\footnote{\label{discussion}An excellent discussion of why such inductions exist can be found, for instance, in https://mathoverflow.net/questions/190812/coaction-of-a-group. And for the sake of completeness, Ref. [\citen{dan}] shows some examples that make it very clear that such co-action homomorphism can be defined.}.
			
				Another interesting fact, which also points to the convenience of taking (\ref{b-prime}) as the $ D^{K} \left( G \right) $ face operators, is that all of them commute with themselves and, according to Figure \ref{dual-comut-ab},
				\begin{figure}[!t]
					\begin{flushleft}
						\begin{tikzpicture}[
							scale=0.3,
							equation/.style={thin},
							trans/.style={thin,shorten >=0.5pt,shorten <=0.5pt,>=stealth},
							flecha/.style={thin,->,shorten >=0.5pt,shorten <=0.5pt,>=stealth}
							]
							\draw[equation] (-14.6,0.7) -- (-14.6,0.7) node[midway,right] {$ A^{\prime } _{v} \circ B^{\prime \left( h^{\prime } \right) } _{f} $};
							\draw[trans] (-7.9,3.9) -- (-7.9,-2.3) node[above=2pt,right=-1pt] {};
							\draw[trans] (-0.1,3.9) -- (1.3,0.68) node[above=2pt,right=-1pt] {};
							\draw[trans] (-0.1,-2.3) -- (1.3,0.80) node[above=2pt,right=-1pt] {};
							\draw[flecha] (-6.4,1.2) -- (-5.1,1.2) node[above=2pt,right=-1pt] {};
							\draw[flecha] (-5.3,1.2) -- (-2.0,1.2) node[above=2pt,right=-1pt] {};
							\draw[trans] (-2.2,1.2) -- (-0.1,1.2) node[above=2pt,right=-1pt] {};
							\draw[flecha] (-6.4,-1.2) -- (-5.1,-1.2) node[above=2pt,right=-1pt] {};
							\draw[flecha] (-5.3,-1.2) -- (-2.0,-1.2) node[above=2pt,right=-1pt] {};
							\draw[trans] (-2.2,-1.2) -- (-0.1,-1.2) node[above=2pt,right=-1pt] {};
							\draw[flecha] (-3.8,-2.0) -- (-3.8,0.0) node[above=2pt,right=-1pt] {};
							\draw[flecha] (-3.8,-0.2) -- (-3.8,2.9) node[above=2pt,right=-1pt] {};
							\draw[trans] (-3.8,2.7) -- (-3.8,3.6) node[above=2pt,right=-1pt] {};
							\draw[flecha] (-1.0,-2.0) -- (-1.0,0.0) node[above=2pt,right=-1pt] {};
							\draw[flecha] (-1.0,-0.2) -- (-1.0,2.9) node[above=2pt,right=-1pt] {};
							\draw[trans] (-1.0,2.7) -- (-1.0,3.6) node[above=2pt,right=-1pt] {};
							\draw[trans,fill=white] (-2.4,0.0) circle (0.8);
							\draw[equation] (-7.8,1.2) -- (-7.8,1.2) node[midway,right] {$ d $};
							\draw[equation] (-3.7,-0.2) -- (-3.7,-0.2) node[midway,left] {$ c $};
							\draw[equation] (-5.3,2.6) -- (-5.3,2.6) node[midway,right] {$ a $};
							\draw[equation] (-2.3,-2.0) -- (-2.3,-2.0) node[midway] {$ m $};
							\draw[equation] (0.6,-0.2) -- (0.6,-0.2) node[midway,left] {$ r $};
							\draw[equation] (-2.3,2.0) -- (-2.3,2.0) node[midway] {$ b $};
							\draw[equation] (-2.4,0.0) -- (-2.4,0.0) node[midway] {$ \tilde{\alpha } $};
							\draw[equation] (10.2,0.8) -- (10.2,0.8) node[midway] {$ =  \ \delta \left( h^{\prime } , \mathfrak{f} \left( \tilde{\alpha } \right) \cdot b^{-1} c^{-1} mr \right) \cdot A^{\prime } _{v} $};
							\draw[trans] (19.0,3.9) -- (19.0,-2.3) node[above=2pt,right=-1pt] {};
							\draw[trans] (26.8,3.9) -- (28.2,0.68) node[above=2pt,right=-1pt] {};
							\draw[trans] (26.8,-2.3) -- (28.2,0.80) node[above=2pt,right=-1pt] {};
							\draw[flecha] (20.5,1.2) -- (22.8,1.2) node[above=2pt,right=-1pt] {};
							\draw[flecha] (21.6,1.2) -- (24.9,1.2) node[above=2pt,right=-1pt] {};
							\draw[trans] (24.7,1.2) -- (26.8,1.2) node[above=2pt,right=-1pt] {};
							\draw[flecha] (20.5,-1.2) -- (21.8,-1.2) node[above=2pt,right=-1pt] {};
							\draw[flecha] (21.6,-1.2) -- (24.9,-1.2) node[above=2pt,right=-1pt] {};
							\draw[trans] (24.7,-1.2) -- (26.8,-1.2) node[above=2pt,right=-1pt] {};
							\draw[flecha] (23.1,-2.0) -- (23.1,0.0) node[above=2pt,right=-1pt] {};
							\draw[flecha] (23.1,-0.2) -- (23.1,2.9) node[above=2pt,right=-1pt] {};
							\draw[trans] (23.1,2.7) -- (23.1,3.6) node[above=2pt,right=-1pt] {};
							\draw[flecha] (25.9,-2.0) -- (25.9,0.0) node[above=2pt,right=-1pt] {};
							\draw[flecha] (25.9,-0.2) -- (25.9,2.9) node[above=2pt,right=-1pt] {};
							\draw[trans] (25.9,2.7) -- (25.9,3.6) node[above=2pt,right=-1pt] {};
							\draw[trans,fill=white] (24.5,0.0) circle (0.8);
							\draw[equation] (19.0,1.2) -- (19.0,1.2) node[midway,right] {$ d $};
							\draw[equation] (23.1,-0.2) -- (23.1,-0.2) node[midway,left] {$ c $};
							\draw[equation] (21.6,2.6) -- (21.6,2.6) node[midway,right] {$ a $};
							\draw[equation] (24.6,-2.0) -- (24.6,-2.0) node[midway] {$ m $};
							\draw[equation] (27.5,-0.2) -- (27.5,-0.2) node[midway,left] {$ r $};
							\draw[equation] (24.6,2.0) -- (24.6,2.0) node[midway] {$ b $};
							\draw[equation] (24.5,0.0) -- (24.5,0.0) node[midway] {$ \tilde{\alpha } $};
						\end{tikzpicture} \\ \hspace*{3.5cm}
						\begin{tikzpicture}[
							scale=0.3,
							equation/.style={thin},
							trans/.style={thin,shorten >=0.5pt,shorten <=0.5pt,>=stealth},
							flecha/.style={thin,->,shorten >=0.5pt,shorten <=0.5pt,>=stealth}
							]					
							\draw[equation] (-1.7,0.6) -- (-1.7,0.6) node[midway] {$ = \ $};
							\draw[equation] (0.1,1.5) -- (0.1,1.5) node[midway] {$ 1 $};
							\draw[trans] (-1.0,0.67) -- (1.2,0.67) node[above=2pt,right=-1pt] {};
							\draw[equation] (0.1,-0.3) -- (0.1,-0.3) node[midway] {$ \left\vert G \right\vert $};
							\draw[equation] (2.2,0.9) -- (2.2,0.9) node[midway] {$ \sum $};
							\draw[equation] (2.2,-0.45) -- (2.2,-0.45) node[midway] {$ _{g \in G} $};
							\draw[equation] (9.1,0.8) -- (9.1,0.8) node[midway] {$ \delta \left( h^{\prime } , \mathfrak{f} \left( \tilde{\alpha } \right) \cdot b^{-1} c^{-1} mr \right) $};
							\draw[trans] (15.5,3.9) -- (15.5,-2.3) node[above=2pt,right=-1pt] {};
							\draw[trans] (24.6,3.9) -- (26.0,0.68) node[above=2pt,right=-1pt] {};
							\draw[trans] (24.6,-2.3) -- (26.0,0.80) node[above=2pt,right=-1pt] {};
							\draw[flecha] (18.3,1.2) -- (20.6,1.2) node[above=2pt,right=-1pt] {};
							\draw[flecha] (19.4,1.2) -- (22.7,1.2) node[above=2pt,right=-1pt] {};
							\draw[trans] (22.5,1.2) -- (24.6,1.2) node[above=2pt,right=-1pt] {};
							\draw[flecha] (18.3,-1.2) -- (19.6,-1.2) node[above=2pt,right=-1pt] {};
							\draw[flecha] (19.4,-1.2) -- (22.7,-1.2) node[above=2pt,right=-1pt] {};
							\draw[trans] (22.5,-1.2) -- (24.6,-1.2) node[above=2pt,right=-1pt] {};
							\draw[flecha] (20.9,-2.0) -- (20.9,0.0) node[above=2pt,right=-1pt] {};
							\draw[flecha] (20.9,-0.2) -- (20.9,2.9) node[above=2pt,right=-1pt] {};
							\draw[trans] (20.9,2.7) -- (20.9,3.6) node[above=2pt,right=-1pt] {};
							\draw[flecha] (23.7,-2.0) -- (23.7,0.0) node[above=2pt,right=-1pt] {};
							\draw[flecha] (23.7,-0.2) -- (23.7,2.9) node[above=2pt,right=-1pt] {};
							\draw[trans] (23.7,2.7) -- (23.7,3.6) node[above=2pt,right=-1pt] {};
							\draw[trans,fill=white] (22.3,0.0) circle (0.8);
							\draw[equation] (15.6,1.4) -- (15.6,1.4) node[midway,right] {$ dg^{-1} $};
							\draw[equation] (21.0,0.0) -- (21.0,0.0) node[midway,left] {$ cg^{-1} $};
							\draw[equation] (18.8,2.6) -- (18.8,2.6) node[midway,right] {$ ga $};
							\draw[equation] (22.4,-2.0) -- (22.4,-2.0) node[midway] {$ m $};
							\draw[equation] (25.3,-0.2) -- (25.3,-0.2) node[midway,left] {$ r $};
							\draw[equation] (22.4,2.0) -- (22.4,2.0) node[midway] {$ gb $};
							\draw[equation] (22.3,0.0) -- (22.3,0.0) node[midway] {$ \tilde{\alpha } $};
						\end{tikzpicture} \\ \vspace*{0.8cm}
						\begin{tikzpicture}[
							scale=0.3,
							equation/.style={thin},
							trans/.style={thin,shorten >=0.5pt,shorten <=0.5pt,>=stealth},
							flecha/.style={thin,->,shorten >=0.5pt,shorten <=0.5pt,>=stealth}
							]
							\draw[equation] (-14.6,0.47) -- (-14.6,0.47) node[midway,right] {$ B^{\prime \left( h^{\prime } \right) } _{f} \circ A^{\prime } _{v} $};
							\draw[trans] (-7.9,3.9) -- (-7.9,-2.3) node[above=2pt,right=-1pt] {};
							\draw[trans] (-0.1,3.9) -- (1.3,0.68) node[above=2pt,right=-1pt] {};
							\draw[trans] (-0.1,-2.3) -- (1.3,0.80) node[above=2pt,right=-1pt] {};
							\draw[flecha] (-6.4,1.2) -- (-5.1,1.2) node[above=2pt,right=-1pt] {};
							\draw[flecha] (-5.3,1.2) -- (-2.0,1.2) node[above=2pt,right=-1pt] {};
							\draw[trans] (-2.2,1.2) -- (-0.1,1.2) node[above=2pt,right=-1pt] {};
							\draw[flecha] (-6.4,-1.2) -- (-5.1,-1.2) node[above=2pt,right=-1pt] {};
							\draw[flecha] (-5.3,-1.2) -- (-2.0,-1.2) node[above=2pt,right=-1pt] {};
							\draw[trans] (-2.2,-1.2) -- (-0.1,-1.2) node[above=2pt,right=-1pt] {};
							\draw[flecha] (-3.8,-2.0) -- (-3.8,0.0) node[above=2pt,right=-1pt] {};
							\draw[flecha] (-3.8,-0.2) -- (-3.8,2.9) node[above=2pt,right=-1pt] {};
							\draw[trans] (-3.8,2.7) -- (-3.8,3.6) node[above=2pt,right=-1pt] {};
							\draw[flecha] (-1.0,-2.0) -- (-1.0,0.0) node[above=2pt,right=-1pt] {};
							\draw[flecha] (-1.0,-0.2) -- (-1.0,2.9) node[above=2pt,right=-1pt] {};
							\draw[trans] (-1.0,2.7) -- (-1.0,3.6) node[above=2pt,right=-1pt] {};
							\draw[trans,fill=white] (-2.4,0.0) circle (0.8);
							\draw[equation] (-7.8,1.2) -- (-7.8,1.2) node[midway,right] {$ d $};
							\draw[equation] (-3.7,-0.2) -- (-3.7,-0.2) node[midway,left] {$ c $};
							\draw[equation] (-5.3,2.6) -- (-5.3,2.6) node[midway,right] {$ a $};
							\draw[equation] (-2.3,-2.0) -- (-2.3,-2.0) node[midway] {$ m $};
							\draw[equation] (0.6,-0.2) -- (0.6,-0.2) node[midway,left] {$ r $};
							\draw[equation] (-2.3,2.0) -- (-2.3,2.0) node[midway] {$ b $};
							\draw[equation] (-2.4,0.0) -- (-2.4,0.0) node[midway] {$ \tilde{\alpha } $};
							\draw[equation] (2.2,0.64) -- (2.2,0.64) node[midway] {$ = $};
							\draw[equation] (4.2,1.5) -- (4.2,1.5) node[midway] {$ 1 $};
							\draw[trans] (3.1,0.67) -- (5.3,0.67) node[above=2pt,right=-1pt] {};
							\draw[equation] (4.2,-0.3) -- (4.2,-0.3) node[midway] {$ \left\vert G \right\vert $};
							\draw[equation] (6.3,0.5) -- (6.3,0.5) node[midway] {$ \sum $};
							\draw[equation] (6.3,-0.85) -- (6.3,-0.85) node[midway] {$ _{g \in G} $};
							\draw[equation] (8.9,0.7) -- (8.9,0.7) node[midway] {$ B^{\prime \left( h^{\prime } \right) } _{f} $};
							\draw[trans] (10.7,3.9) -- (10.7,-2.3) node[above=2pt,right=-1pt] {};
							\draw[trans] (19.8,3.9) -- (21.2,0.68) node[above=2pt,right=-1pt] {};
							\draw[trans] (19.8,-2.3) -- (21.2,0.80) node[above=2pt,right=-1pt] {};
							\draw[flecha] (13.5,1.2) -- (15.8,1.2) node[above=2pt,right=-1pt] {};
							\draw[flecha] (14.6,1.2) -- (17.9,1.2) node[above=2pt,right=-1pt] {};
							\draw[trans] (17.7,1.2) -- (19.8,1.2) node[above=2pt,right=-1pt] {};
							\draw[flecha] (13.5,-1.2) -- (14.8,-1.2) node[above=2pt,right=-1pt] {};
							\draw[flecha] (14.6,-1.2) -- (17.9,-1.2) node[above=2pt,right=-1pt] {};
							\draw[trans] (17.7,-1.2) -- (19.8,-1.2) node[above=2pt,right=-1pt] {};
							\draw[flecha] (16.1,-2.0) -- (16.1,0.0) node[above=2pt,right=-1pt] {};
							\draw[flecha] (16.1,-0.2) -- (16.1,2.9) node[above=2pt,right=-1pt] {};
							\draw[trans] (16.1,2.7) -- (16.1,3.6) node[above=2pt,right=-1pt] {};
							\draw[flecha] (18.9,-2.0) -- (18.9,0.0) node[above=2pt,right=-1pt] {};
							\draw[flecha] (18.9,-0.2) -- (18.9,2.9) node[above=2pt,right=-1pt] {};
							\draw[trans] (18.9,2.7) -- (18.9,3.6) node[above=2pt,right=-1pt] {};
							\draw[trans,fill=white] (17.5,0.0) circle (0.8);
							\draw[equation] (10.8,1.4) -- (10.8,1.4) node[midway,right] {$ dg^{-1} $};
							\draw[equation] (16.2,0.0) -- (16.2,0.0) node[midway,left] {$ cg^{-1} $};
							\draw[equation] (14.0,2.6) -- (14.0,2.6) node[midway,right] {$ ga $};
							\draw[equation] (17.6,-2.0) -- (17.6,-2.0) node[midway] {$ m $};
							\draw[equation] (20.5,-0.2) -- (20.5,-0.2) node[midway,left] {$ r $};
							\draw[equation] (17.6,2.0) -- (17.6,2.0) node[midway] {$ gb $};
							\draw[equation] (17.5,0.0) -- (17.5,0.0) node[midway] {$ \tilde{\alpha } $};
						\end{tikzpicture} \\ \hspace*{2.5cm}
						\begin{tikzpicture}[
							scale=0.3,
							equation/.style={thin},
							trans/.style={thin,shorten >=0.5pt,shorten <=0.5pt,>=stealth},
							flecha/.style={thin,->,shorten >=0.5pt,shorten <=0.5pt,>=stealth}
							]					
							\draw[equation] (-7.1,0.6) -- (-7.1,0.6) node[midway] {$ = \ $};
							\draw[equation] (-5.3,1.5) -- (-5.3,1.5) node[midway] {$ 1 $};
							\draw[trans] (-6.4,0.67) -- (-4.2,0.67) node[above=2pt,right=-1pt] {};
							\draw[equation] (-5.3,-0.3) -- (-5.3,-0.3) node[midway] {$ \left\vert G \right\vert $};
							\draw[equation] (-3.2,0.8) -- (-3.2,0.8) node[midway] {$ \sum $};
							\draw[equation] (-3.2,-0.55) -- (-3.2,-0.55) node[midway] {$ _{g \in G} $};
							\draw[equation] (6.0,0.8) -- (6.0,0.8) node[midway] {$ \delta \left( h^{\prime } , \mathfrak{f} \bigl( \tilde{\alpha } \right) \cdot \left( gb \right) ^{-1} \left( cg^{-1} \right) ^{-1} mr \bigr) $};
							\draw[trans] (14.5,3.9) -- (14.5,-2.3) node[above=2pt,right=-1pt] {};
							\draw[trans] (23.6,3.9) -- (25.0,0.68) node[above=2pt,right=-1pt] {};
							\draw[trans] (23.6,-2.3) -- (25.0,0.80) node[above=2pt,right=-1pt] {};
							\draw[flecha] (17.3,1.2) -- (19.6,1.2) node[above=2pt,right=-1pt] {};
							\draw[flecha] (18.4,1.2) -- (21.7,1.2) node[above=2pt,right=-1pt] {};
							\draw[trans] (21.5,1.2) -- (23.6,1.2) node[above=2pt,right=-1pt] {};
							\draw[flecha] (17.3,-1.2) -- (18.6,-1.2) node[above=2pt,right=-1pt] {};
							\draw[flecha] (18.4,-1.2) -- (21.7,-1.2) node[above=2pt,right=-1pt] {};
							\draw[trans] (21.5,-1.2) -- (23.6,-1.2) node[above=2pt,right=-1pt] {};
							\draw[flecha] (19.9,-2.0) -- (19.9,0.0) node[above=2pt,right=-1pt] {};
							\draw[flecha] (19.9,-0.2) -- (19.9,2.9) node[above=2pt,right=-1pt] {};
							\draw[trans] (19.9,2.7) -- (19.9,3.6) node[above=2pt,right=-1pt] {};
							\draw[flecha] (22.7,-2.0) -- (22.7,0.0) node[above=2pt,right=-1pt] {};
							\draw[flecha] (22.7,-0.2) -- (22.7,2.9) node[above=2pt,right=-1pt] {};
							\draw[trans] (22.7,2.7) -- (22.7,3.6) node[above=2pt,right=-1pt] {};
							\draw[trans,fill=white] (21.3,0.0) circle (0.8);
							\draw[equation] (14.6,1.4) -- (14.6,1.4) node[midway,right] {$ dg^{-1} $};
							\draw[equation] (20.0,0.0) -- (20.0,0.0) node[midway,left] {$ cg^{-1} $};
							\draw[equation] (17.8,2.6) -- (17.8,2.6) node[midway,right] {$ ga $};
							\draw[equation] (21.4,-2.0) -- (21.4,-2.0) node[midway] {$ m $};
							\draw[equation] (24.3,-0.2) -- (24.3,-0.2) node[midway,left] {$ r $};
							\draw[equation] (21.4,2.0) -- (21.4,2.0) node[midway] {$ gb $};
							\draw[equation] (21.3,0.0) -- (21.3,0.0) node[midway] {$ \tilde{\alpha } $};
						\end{tikzpicture}
					\end{flushleft}
					\caption{\label{dual-comut-ab} Proof that the operators $ A^{\prime } _{v} $ and $ B^{\prime } _{f,h} \equiv B^{\prime \left( h^{\prime } \right) } _{f} $ commute because the elements of $ G $ are such that $ \left( gb \right) ^{-1} \left( cg^{-1} \right) ^{-1} = b^{-1} \left( g^{-1} g \right) c^{-1} = b^{-1} c^{-1} $. Here, $ A^{\prime } _{v} $ and $ B^{\prime } _{f,h} $ act only on the vertex and face sectors whose intersection is not empty because, when this intersection is empty, these operators commute by definition.} 
				\end{figure}
				with all the vertex operators inherited from the $ D \left( G \right) $ models. And undoubtedly this fact is extremely relevant because, in order to make these $ D^{K} \left( G \right) $ models exactly solvable, it is essential that all these operators are interpreted as projectors onto
				\begin{equation*}
					\mathfrak{H} _{D^{K} \left( G \right) } = \underbrace{\mathfrak{H} _{\left\vert G \right\vert } \otimes \ldots \otimes \mathfrak{H} _{\left\vert G \right\vert }} _{N_{\ell } \ \textnormal{\tiny{times}}} \ \otimes \ \underbrace{\mathfrak{H} _{K} \otimes \ldots \otimes \mathfrak{H} _{K}} _{N_{f} \ \textnormal{\tiny{times}}} \ .
				\end{equation*}
				However, as (\ref{H-qdmf}) shows us that the $ D^{K} \left( G \right) $ Hamiltonian is also defined by an operator $ C^{\prime } _{\ell } $, which acts only on the dual link sectors of $ \mathcal{L} _{2} $, it becomes clear that all the $ D^{K} \left( G \right) $ link operators also need to commute with themselves and these other operators for the same reason.
				
				Note that, although we have not yet presented the exact expressions of these $ D^{K} \left( G \right) $ link operators, one thing we already know about them is: they must be interpreted as the duals of the $ D_{M} \left( G \right) $ link operators, and vice versa, both from a geometric and algebraic point of view. But even though Figure \ref{QMDf-rede} already shows the need for this geometric point of view, what does it mean to say that the $ D_{M} \left( G \right) $ and $ D^{K} \left( G \right) $ link operators are the duals of each other from an algebraic point of view? Based on the dual relationship between the $ D_{M} \left( G \right) $ ($ D^{K} \left( G \right) $) vertex and $ D^{K} \left( G \right) $ ($ D_{M} \left( G \right) $) face operators, it is correct to say that this means that, while the $ D_{M} \left( G \right) $ link operators just compares two matter fields without performing any transformation on these fields, the $ D^{K} \left( G \right) $ link operators must
				\begin{itemize}
					\item do this same kind of comparison, in some way, with the help of $ \mathfrak{f} $, and
					\item necessarily perform some kind of transformation in the gauge and dual matter fields on which it acts.
				\end{itemize}
				And given this scenario, the expression that best fits the needs of the link operator that make up the $ D^{K} \left( G \right) $ Hamiltonian (\ref{H-qdmf}) is
				\begin{equation}
					C^{\prime } _{\ell } = \frac{1}{\vert \tilde{S} \vert } \sum _{\tilde{\lambda } \in \tilde{S}} C^{\prime \left( \tilde{\lambda } \right) } _{\ell } \ , \label{qdmf-edge-operator}
				\end{equation}
				whose components are defined in Figure \ref{QMDf-edge-operator-components}.
				\begin{figure}[!t]
					\begin{center}
						\begin{tikzpicture}[
							scale=0.3,
							equation/.style={thin},
							trans/.style={thin,shorten >=0.5pt,shorten <=0.5pt,>=stealth},
							flecha/.style={thin,->,shorten >=0.5pt,shorten <=0.5pt,>=stealth},
							dual/.style={dotted,-,shorten >=0.5pt,shorten <=0.5pt}
							]
							\draw[equation] (-9.1,0.1) -- (-9.1,0.1) node[midway,right] {$ C^{\prime \tilde{\left( \lambda \right) }} _{\ell } $};
							\draw[trans] (-5.6,2.3) -- (-5.6,-2.3) node[above=2pt,right=-1pt] {};
							\draw[trans] (0.1,2.3) -- (1.3,-0.06) node[above=2pt,right=-1pt] {};
							\draw[trans] (0.1,-2.3) -- (1.3,0.06) node[above=2pt,right=-1pt] {};
							\draw[dual] (-4.4,0.0) -- (-0.2,0.0) node[above=2pt,right=-1pt] {};
							\draw[flecha] (-2.3,-1.0) -- (-2.3,0.3) node[above=2pt,right=-1pt] {};
							\draw[trans] (-2.3,0.0) -- (-2.3,1.0) node[above=2pt,right=-1pt] {};
							\draw[trans,fill=white] (-4.4,0.0) circle (0.9);
							\draw[trans,fill=white] (-0.2,0.0) circle (0.9);
							\draw[equation] (-2.3,-0.8) -- (-2.3,-0.8) node[midway,below] {$ a $};
							\draw[equation] (-4.4,0.0) -- (-4.4,0.0) node[midway] {$ \tilde{\alpha } $};
							\draw[equation] (-0.2,-0.02) -- (-0.2,-0.02) node[midway] {$ \tilde{\beta } $};	
							\draw[equation] (2.6,-0.07) -- (2.6,-0.07) node[midway] {$ = $};
							\draw[trans] (4.0,2.3) -- (4.0,-2.3) node[above=2pt,right=-1pt] {};
							\draw[trans] (9.7,2.3) -- (10.9,-0.06) node[above=2pt,right=-1pt] {};
							\draw[trans] (9.7,-2.3) -- (10.9,0.06) node[above=2pt,right=-1pt] {};
							\draw[dual] (5.2,0.0) -- (9.4,0.0) node[above=2pt,right=-1pt] {};
							\draw[flecha] (7.2,-1.0) -- (7.2,0.3) node[above=2pt,right=-1pt] {};
							\draw[trans] (7.2,0.0) -- (7.2,1.0) node[above=2pt,right=-1pt] {};
							\draw[trans,fill=white] (5.2,0.0) circle (0.9);
							\draw[trans,fill=white] (9.4,0.0) circle (0.9);
							\draw[equation] (7.3,-0.8) -- (7.3,-0.8) node[midway,below] {$ a^{\prime } $};
							\draw[equation] (5.2,0.0) -- (5.2,0.0) node[midway] {$ \tilde{\alpha } ^{\prime } $};
							\draw[equation] (9.4,-0.02) -- (9.4,-0.02) node[midway] {$ \tilde{\beta } ^{\prime } $};
						\end{tikzpicture}
					\end{center}
					\caption{\label{QMDf-edge-operator-components} Definition of the components $ C^{\prime \left( \tilde{\lambda } \right) } _{\ell } $ that define the link operator (\ref{qdmf-edge-operator}). Note that, since $ C^{\prime \left( \tilde{\lambda } \right) } _{\ell } $ is defined by taking $ a^{\prime } = \mathfrak{f} \bigl( \tilde{\lambda } \bigr) \cdot a $, $ \tilde{\alpha } ^{\prime } = \tilde{\lambda } ^{-1} \ast \tilde{\alpha } $ and $ \tilde{\beta } ^{\prime } = \tilde{\beta } \ast \tilde{\lambda } $, this shows that $ C^{\prime }_{\ell } $ actually performs transformations in the gauge and matter fields on which it acts. Here, in the same way that the symbol \textquotedblleft $ \cdot $\textquotedblright \hspace*{0.01cm} is used, when necessary, to indicate a product between the elements of the gauge group $ G $, the symbol \textquotedblleft $ \ast $\textquotedblright \hspace*{0.01cm} is used to indicate a product between the elements of $ \tilde{S} $.}
				\end{figure} 
			
			\subsubsection{What are the requirements for $ A^{\prime } _{v} $, $ B^{\prime } _{f} $ and $ C^{\prime } _{\ell } $ to be projectors?}
			
				It is obvious that there are several questions that still need to be answered about all these $ D^{K} \left( G \right) $ operators. And one of these questions refers, for instance, to the reason why we have presented a definition for only the link operator $ C^{\prime } _{\ell } $ and not for all the $ D^{K} \left( G \right) $ link operators. Note that a good answer to this question requires us to remember that, since all the $ D^{K} \left( G \right) $ vertex, face and link operators must define three sets
				\begin{eqnarray*}
					\mathfrak{A} ^{\prime } & = & \bigl\{ A^{\prime } _{v,0} \ , \ A^{\prime } _{v,1} \ , \ \ldots \ , \ A^{\prime } _{v, R-1} \bigr\} \ , \\
					\mathfrak{B} ^{\prime } & = & \bigl\{ B^{\prime } _{f,0} \ , \ B^{\prime } _{f,1} \ , \ \ldots \ , \ B^{\prime } _{f, R-1} \bigr\} \quad \textnormal{and} \\
					\mathfrak{C} ^{\prime } & = & \bigl\{ C^{\prime } _{\ell ,0} \ , \ C^{\prime } _{\ell ,1} \ , \ \ldots \ , \ C^{\prime } _{\ell ,K-1} \bigr\}
				\end{eqnarray*}
				of orthogonal projectors onto $ \mathfrak{H} _{D^{K} \left( G \right) } $, they must also satisfy the same properties \textbf{(a)}, \textbf{(b)} and \textbf{(c)} as their dual counterparts. After all, as the previous Section already made it clear that the orthogonality of these operators can be delegated, for instance, to the characters of a group, all the commutation relations that are satisfied by $ A^{\prime } _{v} $, $ B^{\prime } _{f} $ and $ C^{\prime } _{\ell } $ will apply to those operators that complete $ \mathfrak{A} ^{\prime } $, $ \mathfrak{B} ^{\prime } $ and $ \mathfrak{C} ^{\prime } $\label{ortho-comment}. We will return to this point later on.
				
				As a consequence of this good answer, it is correct to say that, if we want to evaluate whether $ C^{\prime } _{\ell } $ qualifies as a projector, we need to evaluate the commutation relations between it and all the operators that make up (\ref{H-qdmf}). And because Figure \ref{dual-comut-ac}
				\begin{figure}[!t]
					\begin{flushleft}
						\begin{tikzpicture}[
							scale=0.3,
							equation/.style={thin},
							trans/.style={thin,shorten >=0.5pt,shorten <=0.5pt,>=stealth},
							flecha/.style={thin,->,shorten >=0.5pt,shorten <=0.5pt,>=stealth}
							]
							\draw[equation] (-13.5,-0.1) -- (-13.5,-0.1) node[midway,right] {$ A^{\prime } _{v} \circ C^{\prime } _{\ell } $};
							\draw[trans] (-7.9,2.7) -- (-7.9,-2.7) node[above=2pt,right=-1pt] {};
							\draw[trans] (-0.1,2.7) -- (1.1,-0.06) node[above=2pt,right=-1pt] {};
							\draw[trans] (-0.1,-2.7) -- (1.1,0.06) node[above=2pt,right=-1pt] {};
							\draw[flecha] (-6.4,0.0) -- (-5.1,0.0) node[above=2pt,right=-1pt] {};
							\draw[flecha] (-5.3,0.0) -- (-2.0,0.0) node[above=2pt,right=-1pt] {};
							\draw[trans] (-2.2,0.0) -- (-0.9,0.0) node[above=2pt,right=-1pt] {};
							\draw[flecha] (-3.8,-2.4) -- (-3.8,-1.2) node[above=2pt,right=-1pt] {};
							\draw[flecha] (-3.8,-1.4) -- (-3.8,1.7) node[above=2pt,right=-1pt] {};
							\draw[trans] (-3.8,1.5) -- (-3.8,2.4) node[above=2pt,right=-1pt] {};
							\draw[trans,fill=white] (-2.4,-1.2) circle (0.8);
							\draw[trans,fill=white] (-2.4,1.2) circle (0.8);
							\draw[equation] (-7.8,0.0) -- (-7.8,0.0) node[midway,right] {$ d $};
							\draw[equation] (-3.7,-1.4) -- (-3.7,-1.4) node[midway,left] {$ c $};
							\draw[equation] (-5.3,1.4) -- (-5.3,1.4) node[midway,right] {$ a $};
							\draw[equation] (-0.3,0.0) -- (-0.3,0.0) node[midway] {$ b $};
							\draw[equation] (-2.4,-1.2) -- (-2.4,-1.2) node[midway] {$ \tilde{\beta } $};
							\draw[equation] (-2.4,1.2) -- (-2.4,1.2) node[midway] {$ \tilde{\alpha } $};
							\draw[equation] (2.5,0.0) -- (2.5,0.0) node[midway] {$ = $};
							\draw[equation] (4.9,0.8) -- (4.9,0.8) node[midway] {$ 1 $};
							\draw[trans] (3.8,0.0) -- (6.0,0.0) node[above=2pt,right=-1pt] {};
							\draw[equation] (4.9,-1.1) -- (4.9,-1.1) node[midway] {$ \bigl\vert \tilde{S} \bigr\vert $};
							\draw[equation] (7.2,0.0) -- (7.2,0.0) node[midway] {$ \sum $};
							\draw[equation] (7.2,-1.35) -- (7.2,-1.35) node[midway] {$ _{\tilde{\lambda } \in \tilde{S}} $};
							\draw[equation] (7.7,-0.1) -- (7.7,-0.1) node[midway,right] {$ A^{\prime } _{v} $};
							\draw[trans] (10.3,2.7) -- (10.3,-2.7) node[above=2pt,right=-1pt] {};
							\draw[trans] (22.0,2.7) -- (23.1,-0.06) node[above=2pt,right=-1pt] {};
							\draw[trans] (22.0,-2.7) -- (23.1,0.06) node[above=2pt,right=-1pt] {};
							\draw[flecha] (11.8,0.0) -- (13.6,0.0) node[above=2pt,right=-1pt] {};
							\draw[flecha] (13.4,0.0) -- (16.7,0.0) node[above=2pt,right=-1pt] {};
							\draw[trans] (16.5,0.0) -- (17.8,0.0) node[above=2pt,right=-1pt] {};
							\draw[flecha] (14.9,-2.4) -- (14.9,-1.2) node[above=2pt,right=-1pt] {};
							\draw[flecha] (14.9,-1.4) -- (14.9,1.7) node[above=2pt,right=-1pt] {};
							\draw[trans] (14.9,1.5) -- (14.9,2.4) node[above=2pt,right=-1pt] {};
							\draw[trans,fill=white] (16.3,-1.2) circle (0.8);
							\draw[trans,fill=white] (16.3,1.2) circle (0.8);
							\draw[equation] (10.4,0.0) -- (10.4,0.0) node[midway,right] {$ d $};
							\draw[equation] (15.0,-1.4) -- (15.0,-1.4) node[midway,left] {$ c $};
							\draw[equation] (13.5,1.3) -- (13.5,1.3) node[midway,right] {$ a $};
							\draw[equation] (20.3,0.0) -- (20.3,0.0) node[midway] {$ \mathfrak{f} \bigl( \tilde{\lambda } \bigr) \cdot b $};
							\draw[equation] (16.3,-1.2) -- (16.3,-1.2) node[midway] {$ \tilde{\beta } ^{\prime } $};
							\draw[equation] (16.3,1.2) -- (16.3,1.2) node[midway] {$ \tilde{\alpha } ^{\prime } $};
						\end{tikzpicture} \\ \hspace*{4.0cm}
						\begin{tikzpicture}[
							scale=0.3,
							equation/.style={thin},
							trans/.style={thin,shorten >=0.5pt,shorten <=0.5pt,>=stealth},
							flecha/.style={thin,->,shorten >=0.5pt,shorten <=0.5pt,>=stealth}
							]
							\draw[equation] (-5.3,0.0) -- (-5.3,0.0) node[midway] {$ = $};
							\draw[equation] (-2.9,0.8) -- (-2.9,0.8) node[midway] {$ 1 $};
							\draw[trans] (-4.0,0.0) -- (-1.8,0.0) node[above=2pt,right=-1pt] {};
							\draw[equation] (-2.9,-1.1) -- (-2.9,-1.1) node[midway] {$ \left\vert G \right\vert $};
							\draw[equation] (-0.4,0.8) -- (-0.4,0.8) node[midway] {$ 1 $};
							\draw[trans] (-1.5,0.0) -- (0.7,0.0) node[above=2pt,right=-1pt] {};
							\draw[equation] (-0.4,-1.1) -- (-0.4,-1.1) node[midway] {$ \bigl\vert \tilde{S} \bigr\vert $};
							\draw[equation] (2.1,0.0) -- (2.1,0.0) node[midway] {$ \sum $};
							\draw[equation] (2.1,-1.35) -- (2.1,-1.35) node[midway] {$ _{\tilde{\lambda } \in \tilde{S}} $};
							\draw[equation] (4.5,0.0) -- (4.5,0.0) node[midway] {$ \sum $};
							\draw[equation] (4.5,-1.35) -- (4.5,-1.35) node[midway] {$ _{g \in G} $};
							\draw[trans] (5.7,2.7) -- (5.7,-2.7) node[above=2pt,right=-1pt] {};
							\draw[trans] (21.1,2.7) -- (22.2,-0.06) node[above=2pt,right=-1pt] {};
							\draw[trans] (21.1,-2.7) -- (22.2,0.06) node[above=2pt,right=-1pt] {};
							\draw[flecha] (9.1,0.0) -- (10.9,0.0) node[above=2pt,right=-1pt] {};
							\draw[flecha] (10.7,0.0) -- (14.0,0.0) node[above=2pt,right=-1pt] {};
							\draw[trans] (13.8,0.0) -- (15.1,0.0) node[above=2pt,right=-1pt] {};
							\draw[flecha] (12.2,-2.4) -- (12.2,-1.2) node[above=2pt,right=-1pt] {};
							\draw[flecha] (12.2,-1.4) -- (12.2,1.7) node[above=2pt,right=-1pt] {};
							\draw[trans] (12.2,1.5) -- (12.2,2.4) node[above=2pt,right=-1pt] {};
							\draw[trans,fill=white] (13.6,-1.2) circle (0.8);
							\draw[trans,fill=white] (13.6,1.2) circle (0.8);
							\draw[equation] (5.7,0.05) -- (5.7,0.05) node[midway,right] {$ dg^{-1} $};
							\draw[equation] (12.4,-1.4) -- (12.4,-1.4) node[midway,left] {$ cg^{-1} $};
							\draw[equation] (9.9,1.4) -- (9.9,1.4) node[midway,right] {$ ga $};
							\draw[equation] (18.3,0.0) -- (18.3,0.0) node[midway] {$ g \cdot \mathfrak{f} \bigl( \tilde{\lambda } \bigr) \cdot b $};
							\draw[equation] (13.6,-1.2) -- (13.6,-1.2) node[midway] {$ \tilde{\beta } ^{\prime } $};
							\draw[equation] (13.6,1.2) -- (13.6,1.2) node[midway] {$ \tilde{\alpha } ^{\prime } $};
						\end{tikzpicture} \\ \vspace*{0.8cm}
						\begin{tikzpicture}[
							scale=0.3,
							equation/.style={thin},
							trans/.style={thin,shorten >=0.5pt,shorten <=0.5pt,>=stealth},
							flecha/.style={thin,->,shorten >=0.5pt,shorten <=0.5pt,>=stealth}
							]
							\draw[equation] (-13.5,-0.1) -- (-13.5,-0.1) node[midway,right] {$ C^{\prime } _{\ell } \circ A^{\prime } _{v} $};
							\draw[trans] (-7.9,2.7) -- (-7.9,-2.7) node[above=2pt,right=-1pt] {};
							\draw[trans] (-0.1,2.7) -- (1.1,-0.06) node[above=2pt,right=-1pt] {};
							\draw[trans] (-0.1,-2.7) -- (1.1,0.06) node[above=2pt,right=-1pt] {};
							\draw[flecha] (-6.4,0.0) -- (-5.1,0.0) node[above=2pt,right=-1pt] {};
							\draw[flecha] (-5.3,0.0) -- (-2.0,0.0) node[above=2pt,right=-1pt] {};
							\draw[trans] (-2.2,0.0) -- (-0.9,0.0) node[above=2pt,right=-1pt] {};
							\draw[flecha] (-3.8,-2.4) -- (-3.8,-1.2) node[above=2pt,right=-1pt] {};
							\draw[flecha] (-3.8,-1.4) -- (-3.8,1.7) node[above=2pt,right=-1pt] {};
							\draw[trans] (-3.8,1.5) -- (-3.8,2.4) node[above=2pt,right=-1pt] {};
							\draw[trans,fill=white] (-2.4,-1.2) circle (0.8);
							\draw[trans,fill=white] (-2.4,1.2) circle (0.8);
							\draw[equation] (-7.8,0.0) -- (-7.8,0.0) node[midway,right] {$ d $};
							\draw[equation] (-3.7,-1.4) -- (-3.7,-1.4) node[midway,left] {$ c $};
							\draw[equation] (-5.3,1.4) -- (-5.3,1.4) node[midway,right] {$ a $};
							\draw[equation] (-0.3,0.0) -- (-0.3,0.0) node[midway] {$ b $};
							\draw[equation] (-2.4,-1.2) -- (-2.4,-1.2) node[midway] {$ \tilde{\beta } $};
							\draw[equation] (-2.4,1.2) -- (-2.4,1.2) node[midway] {$ \tilde{\alpha } $};
							\draw[equation] (2.4,-0.1) -- (2.4,-0.1) node[midway] {$ = $};
							\draw[equation] (4.9,0.8) -- (4.9,0.8) node[midway] {$ 1 $};
							\draw[trans] (3.8,0.0) -- (6.0,0.0) node[above=2pt,right=-1pt] {};
							\draw[equation] (4.9,-1.1) -- (4.9,-1.1) node[midway] {$ \left\vert G \right\vert $};
							\draw[equation] (7.2,0.0) -- (7.2,0.0) node[midway] {$ \sum $};
							\draw[equation] (7.2,-1.35) -- (7.2,-1.35) node[midway] {$ _{g \in G} $};
							\draw[equation] (7.7,-0.1) -- (7.7,-0.1) node[midway,right] {$ C^{\prime } _{\ell } $};
							\draw[trans] (10.1,2.7) -- (10.1,-2.7) node[above=2pt,right=-1pt] {};
							\draw[trans] (20.5,2.7) -- (21.6,-0.06) node[above=2pt,right=-1pt] {};
							\draw[trans] (20.5,-2.7) -- (21.6,0.06) node[above=2pt,right=-1pt] {};
							\draw[flecha] (13.3,0.0) -- (15.1,0.0) node[above=2pt,right=-1pt] {};
							\draw[flecha] (14.9,0.0) -- (18.2,0.0) node[above=2pt,right=-1pt] {};
							\draw[trans] (18.0,0.0) -- (19.3,0.0) node[above=2pt,right=-1pt] {};
							\draw[flecha] (16.4,-2.4) -- (16.4,-1.2) node[above=2pt,right=-1pt] {};
							\draw[flecha] (16.4,-1.4) -- (16.4,1.7) node[above=2pt,right=-1pt] {};
							\draw[trans] (16.4,1.5) -- (16.4,2.4) node[above=2pt,right=-1pt] {};
							\draw[trans,fill=white] (17.8,-1.2) circle (0.8);
							\draw[trans,fill=white] (17.8,1.2) circle (0.8);
							\draw[equation] (10.0,0.1) -- (10.0,0.1) node[midway,right] {$ dg^{-1} $};
							\draw[equation] (16.6,-1.4) -- (16.6,-1.4) node[midway,left] {$ cg^{-1} $};
							\draw[equation] (14.2,1.4) -- (14.2,1.4) node[midway,right] {$ ga $};
							\draw[equation] (20.3,0.0) -- (20.3,0.0) node[midway] {$ gb $};
							\draw[equation] (17.8,-1.2) -- (17.8,-1.2) node[midway] {$ \tilde{\beta } $};
							\draw[equation] (17.8,1.2) -- (17.8,1.2) node[midway] {$ \tilde{\alpha } $};
						\end{tikzpicture} \\ \hspace*{4.0cm}
						\begin{tikzpicture}[
							scale=0.3,
							equation/.style={thin},
							trans/.style={thin,shorten >=0.5pt,shorten <=0.5pt,>=stealth},
							flecha/.style={thin,->,shorten >=0.5pt,shorten <=0.5pt,>=stealth}
							]
							\draw[equation] (-5.3,0.0) -- (-5.3,0.0) node[midway] {$ = $};
							\draw[equation] (-2.9,0.8) -- (-2.9,0.8) node[midway] {$ 1 $};
							\draw[trans] (-4.0,0.0) -- (-1.8,0.0) node[above=2pt,right=-1pt] {};
							\draw[equation] (-2.9,-1.1) -- (-2.9,-1.1) node[midway] {$ \bigl\vert \tilde{S} \bigr\vert $};
							\draw[equation] (-0.4,0.8) -- (-0.4,0.8) node[midway] {$ 1 $};
							\draw[trans] (-1.5,0.0) -- (0.7,0.0) node[above=2pt,right=-1pt] {};
							\draw[equation] (-0.4,-1.1) -- (-0.4,-1.1) node[midway] {$ \left\vert G \right\vert $};
							\draw[equation] (2.1,0.0) -- (2.1,0.0) node[midway] {$ \sum $};
							\draw[equation] (2.1,-1.35) -- (2.1,-1.35) node[midway] {$ _{g \in G} $};
							\draw[equation] (4.5,0.0) -- (4.5,0.0) node[midway] {$ \sum $};
							\draw[equation] (4.5,-1.35) -- (4.5,-1.35) node[midway] {$ _{\tilde{\lambda } \in \tilde{S}} $};
							\draw[trans] (5.7,2.7) -- (5.7,-2.7) node[above=2pt,right=-1pt] {};
							\draw[trans] (20.4,2.7) -- (21.5,-0.06) node[above=2pt,right=-1pt] {};
							\draw[trans] (20.4,-2.7) -- (21.5,0.06) node[above=2pt,right=-1pt] {};
							\draw[flecha] (9.1,0.0) -- (10.9,0.0) node[above=2pt,right=-1pt] {};
							\draw[flecha] (10.7,0.0) -- (14.0,0.0) node[above=2pt,right=-1pt] {};
							\draw[trans] (13.8,0.0) -- (15.1,0.0) node[above=2pt,right=-1pt] {};
							\draw[flecha] (12.2,-2.4) -- (12.2,-1.2) node[above=2pt,right=-1pt] {};
							\draw[flecha] (12.2,-1.4) -- (12.2,1.7) node[above=2pt,right=-1pt] {};
							\draw[trans] (12.2,1.5) -- (12.2,2.4) node[above=2pt,right=-1pt] {};
							\draw[trans,fill=white] (13.6,-1.2) circle (0.8);
							\draw[trans,fill=white] (13.6,1.2) circle (0.8);
							\draw[equation] (5.7,0.05) -- (5.7,0.05) node[midway,right] {$ dg^{-1} $};
							\draw[equation] (12.4,-1.4) -- (12.4,-1.4) node[midway,left] {$ cg^{-1} $};
							\draw[equation] (9.9,1.4) -- (9.9,1.4) node[midway,right] {$ ga $};
							\draw[equation] (18.0,0.0) -- (18.0,0.0) node[midway] {$ \mathfrak{f} \bigl( \tilde{\lambda } \bigr) \cdot gb $};
							\draw[equation] (13.6,-1.2) -- (13.6,-1.2) node[midway] {$ \tilde{\beta } ^{\prime } $};
							\draw[equation] (13.6,1.2) -- (13.6,1.2) node[midway] {$ \tilde{\alpha } ^{\prime } $};
						\end{tikzpicture}
					\end{flushleft}
					\caption{\label{dual-comut-ac} Results of the action of the operators $ A^{\prime } _{v} \circ C^{\prime } _{\ell } $ and $ C^{\prime } _{\ell } \circ A^{\prime } _{v} $ on the lattice $ \mathcal{L} _{2} $, from which it is clear that $ \bigl[ A^{\prime } _{v} , C^{\prime } _{\ell } \bigr] $ will only be equal to zero if $ \mathfrak{f} \bigl( \tilde{\lambda } \bigr) $ belongs to $ \mathcal{Z} \left( G \right) $ (i.e., if $ \mathfrak{f} \bigl( \tilde{\lambda } \bigr) $ belongs to the centre of $ G $). Analogous to what has already been observed in Figure \ref{dual-comut-ab}, $ A^{\prime } _{v} $ and $ C^{\prime } _{\ell } $ act only on the vertex and edge sectors whose intersection is not empty because, when this intersection is empty, these operators commute by definition. Note that the order in which the summations are performed is irrelevant.}
				\end{figure}
				 shows us that the only way to cancel $ \bigl[ A^{\prime } _{v} , C^{\prime } _{\ell } \bigr] $ is by taking
				\begin{equation*}
					\mathfrak{f} \bigl( \tilde{\gamma } \bigr) \cdot g = g \cdot \mathfrak{f} \bigl( \tilde{\gamma } \bigr) \ ,
				\end{equation*}
				it turns out to be quite clear that, for $ C^{\prime } _{\ell } $ to be interpreted as a projector, $ \mathsf{Im} \left( \mathfrak{f} \right) \subseteq \mathcal{Z} \left( G \right) $ (i.e., $ \mathfrak{f} \bigl( \tilde{\gamma } \bigr) $ must belong to the centre of $ G $) [\citen{james}]. Note that this need is also reinforced by Figure \ref{dual-comut-bc},
				\begin{figure}[!t]
					\begin{flushleft}
						\begin{tikzpicture}[
							scale=0.3,
							equation/.style={thin},
							trans/.style={thin,shorten >=0.5pt,shorten <=0.5pt,>=stealth},
							flecha/.style={thin,->,shorten >=0.5pt,shorten <=0.5pt,>=stealth}
							]
							\draw[equation] (-13.0,-0.1) -- (-13.0,-0.1) node[midway,right] {$ B^{\prime \left( h^{\prime } \right) } _{f} \circ C^{\prime } _{\ell } $};
							\draw[trans] (-6.1,2.3) -- (-6.1,-2.3) node[above=2pt,right=-1pt] {};
							\draw[trans] (4.4,2.3) -- (5.6,-0.06) node[above=2pt,right=-1pt] {};
							\draw[trans] (4.4,-2.3) -- (5.6,0.06) node[above=2pt,right=-1pt] {};
							\draw[flecha] (-5.2,1.2) -- (-2.6,1.2) node[above=2pt,right=-1pt] {};
							\draw[flecha] (-2.8,1.2) -- (2.0,1.2) node[above=2pt,right=-1pt] {};
							\draw[trans] (1.8,1.2) -- (4.0,1.2) node[above=2pt,right=-1pt] {};
							\draw[flecha] (-5.2,-1.2) -- (-2.6,-1.2) node[above=2pt,right=-1pt] {};
							\draw[flecha] (-2.8,-1.2) -- (2.0,-1.2) node[above=2pt,right=-1pt] {};
							\draw[trans] (1.5,-1.2) -- (4.0,-1.2) node[above=2pt,right=-1pt] {};
							\draw[flecha] (-4.3,-2.0) -- (-4.3,0.3) node[above=2pt,right=-1pt] {};
							\draw[trans] (-4.3,0.0) -- (-4.3,2.0) node[above=2pt,right=-1pt] {};
							\draw[flecha] (0.3,-2.0) -- (0.3,0.3) node[above=2pt,right=-1pt] {};
							\draw[trans] (0.3,0.0) -- (0.3,2.0) node[above=2pt,right=-1pt] {};
							\draw[flecha] (3.1,-2.0) -- (3.1,0.3) node[above=2pt,right=-1pt] {};
							\draw[trans] (3.1,0.0) -- (3.1,2.0) node[above=2pt,right=-1pt] {};
							\draw[trans,fill=white] (-2.85,0.0) circle (0.9);
							\draw[trans,fill=white] (1.75,0.0) circle (0.9);
							\draw[equation] (1.8,2.0) -- (1.8,2.0) node[midway] {$ a $};
							\draw[equation] (4.8,0.0) -- (4.8,0.0) node[midway,left] {$ d $};
							\draw[equation] (1.8,-2.0) -- (1.8,-2.0) node[midway] {$ c $};
							\draw[equation] (-1.4,0.0) -- (-1.4,0.0) node[midway,right] {$ b $};
							\draw[equation] (-2.85,0.0) -- (-2.85,0.0) node[midway] {$ \tilde{\alpha } $};
							\draw[equation] (1.75,0.0) -- (1.75,0.0) node[midway] {$ \tilde{\beta } $};
							\draw[equation] (7.0,0.0) -- (7.0,0.0) node[midway] {$ = $};
							\draw[equation] (9.4,0.8) -- (9.4,0.8) node[midway] {$ 1 $};
							\draw[trans] (8.3,0.0) -- (10.5,0.0) node[above=2pt,right=-1pt] {};
							\draw[equation] (9.4,-1.1) -- (9.4,-1.1) node[midway] {$ \bigl\vert \tilde{S} \bigr\vert $};
							\draw[equation] (11.6,-0.2) -- (11.6,-0.2) node[midway] {$ \sum $};
							\draw[equation] (11.6,-1.55) -- (11.6,-1.55) node[midway] {$ _{\tilde{\lambda } \in \tilde{S}} $};
							\draw[equation] (12.1,-0.1) -- (12.1,-0.1) node[midway,right] {$ B^{\prime \left( h^{\prime } \right) } _{f} $};
							\draw[trans] (16.0,2.3) -- (16.0,-2.3) node[above=2pt,right=-1pt] {};
							\draw[trans] (29.9,2.3) -- (31.1,-0.06) node[above=2pt,right=-1pt] {};
							\draw[trans] (29.9,-2.3) -- (31.1,0.06) node[above=2pt,right=-1pt] {};
							\draw[flecha] (16.9,1.2) -- (22.9,1.2) node[above=2pt,right=-1pt] {};
							\draw[flecha] (22.7,1.2) -- (27.5,1.2) node[above=2pt,right=-1pt] {};
							\draw[trans] (27.3,1.2) -- (29.5,1.2) node[above=2pt,right=-1pt] {};
							\draw[flecha] (16.9,-1.2) -- (22.9,-1.2) node[above=2pt,right=-1pt] {};
							\draw[flecha] (22.7,-1.2) -- (27.5,-1.2) node[above=2pt,right=-1pt] {};
							\draw[trans] (27.3,-1.2) -- (29.5,-1.2) node[above=2pt,right=-1pt] {};
							\draw[flecha] (17.8,-2.0) -- (17.8,0.3) node[above=2pt,right=-1pt] {};
							\draw[trans] (17.8,0.0) -- (17.8,2.0) node[above=2pt,right=-1pt] {};
							\draw[flecha] (25.8,-2.0) -- (25.8,0.3) node[above=2pt,right=-1pt] {};
							\draw[trans] (25.8,0.0) -- (25.8,2.0) node[above=2pt,right=-1pt] {};
							\draw[flecha] (28.6,-2.0) -- (28.6,0.3) node[above=2pt,right=-1pt] {};
							\draw[trans] (28.6,0.0) -- (28.6,2.0) node[above=2pt,right=-1pt] {};
							\draw[trans,fill=white] (19.25,0.0) circle (0.9);
							\draw[trans,fill=white] (27.25,0.0) circle (0.9);
							\draw[equation] (27.3,2.0) -- (27.3,2.0) node[midway] {$ a $};
							\draw[equation] (30.3,0.0) -- (30.3,0.0) node[midway,left] {$ d $};
							\draw[equation] (27.3,-2.0) -- (27.3,-2.0) node[midway] {$ c $};
							\draw[equation] (20.8,0.0) -- (20.8,0.0) node[midway,right] {$ \mathfrak{f} \bigl( \tilde{\lambda } \bigr) \cdot b $};
							\draw[equation] (19.25,0.0) -- (19.25,0.0) node[midway] {$ \tilde{\alpha } ^{\prime } $};
							\draw[equation] (27.25,0.0) -- (27.25,0.0) node[midway] {$ \tilde{\beta } ^{\prime } $};
						\end{tikzpicture} \\ \hspace*{1.0cm}
						\begin{tikzpicture}[
							scale=0.3,
							equation/.style={thin},
							trans/.style={thin,shorten >=0.5pt,shorten <=0.5pt,>=stealth},
							flecha/.style={thin,->,shorten >=0.5pt,shorten <=0.5pt,>=stealth}
							]
							\draw[equation] (-8.6,0.0) -- (-8.6,0.0) node[midway] {$ = $};
							\draw[equation] (-6.2,0.8) -- (-6.2,0.8) node[midway] {$ 1 $};
							\draw[trans] (-7.3,0.0) -- (-5.1,0.0) node[above=2pt,right=-1pt] {};
							\draw[equation] (-6.2,-1.1) -- (-6.2,-1.1) node[midway] {$ \bigl\vert \tilde{S} \bigr\vert $};
							\draw[equation] (-4.0,0.0) -- (-4.0,0.0) node[midway] {$ \sum $};
							\draw[equation] (-4.0,-1.35) -- (-4.0,-1.35) node[midway] {$ _{\tilde{\lambda } \in \tilde{S}} $};
							\draw[equation] (6.3,-0.1) -- (6.3,-0.1) node[midway] {$ \delta \bigl( h^{\prime } , \mathfrak{f} \bigl( \tilde{\beta } \ast \tilde{\lambda } \bigr) \cdot a^{-1} \cdot \bigl[ \mathfrak{f} \bigl( \tilde{\lambda } \bigr) \cdot b \bigr] ^{-1} \cdot cd \bigr) $};
							\draw[trans] (16.0,2.3) -- (16.0,-2.3) node[above=2pt,right=-1pt] {};
							\draw[trans] (29.9,2.3) -- (31.1,-0.06) node[above=2pt,right=-1pt] {};
							\draw[trans] (29.9,-2.3) -- (31.1,0.06) node[above=2pt,right=-1pt] {};
							\draw[flecha] (16.9,1.2) -- (22.9,1.2) node[above=2pt,right=-1pt] {};
							\draw[flecha] (22.7,1.2) -- (27.5,1.2) node[above=2pt,right=-1pt] {};
							\draw[trans] (27.3,1.2) -- (29.5,1.2) node[above=2pt,right=-1pt] {};
							\draw[flecha] (16.9,-1.2) -- (22.9,-1.2) node[above=2pt,right=-1pt] {};
							\draw[flecha] (22.7,-1.2) -- (27.5,-1.2) node[above=2pt,right=-1pt] {};
							\draw[trans] (27.3,-1.2) -- (29.5,-1.2) node[above=2pt,right=-1pt] {};
							\draw[flecha] (17.8,-2.0) -- (17.8,0.3) node[above=2pt,right=-1pt] {};
							\draw[trans] (17.8,0.0) -- (17.8,2.0) node[above=2pt,right=-1pt] {};
							\draw[flecha] (25.8,-2.0) -- (25.8,0.3) node[above=2pt,right=-1pt] {};
							\draw[trans] (25.8,0.0) -- (25.8,2.0) node[above=2pt,right=-1pt] {};
							\draw[flecha] (28.6,-2.0) -- (28.6,0.3) node[above=2pt,right=-1pt] {};
							\draw[trans] (28.6,0.0) -- (28.6,2.0) node[above=2pt,right=-1pt] {};
							\draw[trans,fill=white] (19.25,0.0) circle (0.9);
							\draw[trans,fill=white] (27.25,0.0) circle (0.9);
							\draw[equation] (27.3,2.0) -- (27.3,2.0) node[midway] {$ a $};
							\draw[equation] (30.3,0.0) -- (30.3,0.0) node[midway,left] {$ d $};
							\draw[equation] (27.3,-2.0) -- (27.3,-2.0) node[midway] {$ c $};
							\draw[equation] (20.8,0.0) -- (20.8,0.0) node[midway,right] {$ \mathfrak{f} \bigl( \tilde{\lambda } \bigr) \cdot b $};
							\draw[equation] (19.25,0.0) -- (19.25,0.0) node[midway] {$ \tilde{\alpha } ^{\prime } $};
							\draw[equation] (27.25,0.0) -- (27.25,0.0) node[midway] {$ \tilde{\beta } ^{\prime } $};
						\end{tikzpicture} \\ \vspace*{0.8cm}
						\begin{tikzpicture}[
							scale=0.3,
							equation/.style={thin},
							trans/.style={thin,shorten >=0.5pt,shorten <=0.5pt,>=stealth},
							flecha/.style={thin,->,shorten >=0.5pt,shorten <=0.5pt,>=stealth}
							]
							\draw[equation] (-12.6,-0.1) -- (-12.6,-0.1) node[midway,right] {$ C^{\prime } _{\ell } \circ B^{\prime \left( h^{\prime } \right) } _{f} $};
							\draw[trans] (-6.1,2.3) -- (-6.1,-2.3) node[above=2pt,right=-1pt] {};
							\draw[trans] (4.4,2.3) -- (5.6,-0.06) node[above=2pt,right=-1pt] {};
							\draw[trans] (4.4,-2.3) -- (5.6,0.06) node[above=2pt,right=-1pt] {};
							\draw[flecha] (-5.2,1.2) -- (-2.6,1.2) node[above=2pt,right=-1pt] {};
							\draw[flecha] (-2.8,1.2) -- (2.0,1.2) node[above=2pt,right=-1pt] {};
							\draw[trans] (1.8,1.2) -- (4.0,1.2) node[above=2pt,right=-1pt] {};
							\draw[flecha] (-5.2,-1.2) -- (-2.6,-1.2) node[above=2pt,right=-1pt] {};
							\draw[flecha] (-2.8,-1.2) -- (2.0,-1.2) node[above=2pt,right=-1pt] {};
							\draw[trans] (1.5,-1.2) -- (4.0,-1.2) node[above=2pt,right=-1pt] {};
							\draw[flecha] (-4.3,-2.0) -- (-4.3,0.3) node[above=2pt,right=-1pt] {};
							\draw[trans] (-4.3,0.0) -- (-4.3,2.0) node[above=2pt,right=-1pt] {};
							\draw[flecha] (0.3,-2.0) -- (0.3,0.3) node[above=2pt,right=-1pt] {};
							\draw[trans] (0.3,0.0) -- (0.3,2.0) node[above=2pt,right=-1pt] {};
							\draw[flecha] (3.1,-2.0) -- (3.1,0.3) node[above=2pt,right=-1pt] {};
							\draw[trans] (3.1,0.0) -- (3.1,2.0) node[above=2pt,right=-1pt] {};
							\draw[trans,fill=white] (-2.85,0.0) circle (0.9);
							\draw[trans,fill=white] (1.75,0.0) circle (0.9);
							\draw[equation] (1.8,2.0) -- (1.8,2.0) node[midway] {$ a $};
							\draw[equation] (4.8,0.0) -- (4.8,0.0) node[midway,left] {$ d $};
							\draw[equation] (1.8,-2.0) -- (1.8,-2.0) node[midway] {$ c $};
							\draw[equation] (-1.4,0.0) -- (-1.4,0.0) node[midway,right] {$ b $};
							\draw[equation] (-2.85,0.0) -- (-2.85,0.0) node[midway] {$ \tilde{\alpha } $};
							\draw[equation] (1.75,0.0) -- (1.75,0.0) node[midway] {$ \tilde{\beta } $};
							\draw[equation] (6.1,-0.1) -- (6.1,-0.1) node[midway,right] {$ = \ \delta \bigl( h^{\prime } , \mathfrak{f} \bigl( \tilde{\beta } \bigr) \cdot a^{-1} b^{-1} cd \bigr) \cdot C^{\prime } _{\ell } $}; 
							\draw[trans] (22.1,2.3) -- (22.1,-2.3) node[above=2pt,right=-1pt] {};
							\draw[trans] (32.4,2.3) -- (33.6,-0.06) node[above=2pt,right=-1pt] {};
							\draw[trans] (32.4,-2.3) -- (33.6,0.06) node[above=2pt,right=-1pt] {};
							\draw[flecha] (22.8,1.2) -- (25.4,1.2) node[above=2pt,right=-1pt] {};
							\draw[flecha] (25.2,1.2) -- (30.0,1.2) node[above=2pt,right=-1pt] {};
							\draw[trans] (29.8,1.2) -- (32.0,1.2) node[above=2pt,right=-1pt] {};
							\draw[flecha] (22.8,-1.2) -- (25.4,-1.2) node[above=2pt,right=-1pt] {};
							\draw[flecha] (25.2,-1.2) -- (30.0,-1.2) node[above=2pt,right=-1pt] {};
							\draw[trans] (29.5,-1.2) -- (32.0,-1.2) node[above=2pt,right=-1pt] {};
							\draw[flecha] (23.7,-2.0) -- (23.7,0.3) node[above=2pt,right=-1pt] {};
							\draw[trans] (23.7,0.0) -- (23.7,2.0) node[above=2pt,right=-1pt] {};
							\draw[flecha] (28.3,-2.0) -- (28.3,0.3) node[above=2pt,right=-1pt] {};
							\draw[trans] (28.3,0.0) -- (28.3,2.0) node[above=2pt,right=-1pt] {};
							\draw[flecha] (31.1,-2.0) -- (31.1,0.3) node[above=2pt,right=-1pt] {};
							\draw[trans] (31.1,0.0) -- (31.1,2.0) node[above=2pt,right=-1pt] {};
							\draw[trans,fill=white] (25.15,0.0) circle (0.9);
							\draw[trans,fill=white] (29.75,0.0) circle (0.9);
							\draw[equation] (29.8,2.0) -- (29.8,2.0) node[midway] {$ a $};
							\draw[equation] (32.8,0.0) -- (32.8,0.0) node[midway,left] {$ d $};
							\draw[equation] (29.8,-2.0) -- (29.8,-2.0) node[midway] {$ c $};
							\draw[equation] (26.6,0.0) -- (26.6,0.0) node[midway,right] {$ b $};
							\draw[equation] (25.15,0.0) -- (25.15,0.0) node[midway] {$ \tilde{\alpha } $};
							\draw[equation] (29.75,0.0) -- (29.75,0.0) node[midway] {$ \tilde{\beta } $};
						\end{tikzpicture} \\ \hspace*{2.0cm}
						\begin{tikzpicture}[
							scale=0.3,
							equation/.style={thin},
							trans/.style={thin,shorten >=0.5pt,shorten <=0.5pt,>=stealth},
							flecha/.style={thin,->,shorten >=0.5pt,shorten <=0.5pt,>=stealth}
							]
							\draw[equation] (-1.6,0.0) -- (-1.6,0.0) node[midway] {$ = $};
							\draw[equation] (0.8,0.8) -- (0.8,0.8) node[midway] {$ 1 $};
							\draw[trans] (-0.3,0.0) -- (1.9,0.0) node[above=2pt,right=-1pt] {};
							\draw[equation] (0.8,-1.1) -- (0.8,-1.1) node[midway] {$ \bigl\vert \tilde{S} \bigr\vert $};
							\draw[equation] (3.0,0.0) -- (3.0,0.0) node[midway] {$ \sum $};
							\draw[equation] (3.0,-1.35) -- (3.0,-1.35) node[midway] {$ _{\tilde{\lambda } \in \tilde{S}} $};
							\draw[equation] (9.6,-0.1) -- (9.6,-0.1) node[midway] {$ \delta \bigl( h^{\prime } , \mathfrak{f} \bigl( \tilde{\beta } \bigr) \cdot a^{-1} b^{-1} cd \bigr) $};
							\draw[trans] (16.0,2.3) -- (16.0,-2.3) node[above=2pt,right=-1pt] {};
							\draw[trans] (29.9,2.3) -- (31.1,-0.06) node[above=2pt,right=-1pt] {};
							\draw[trans] (29.9,-2.3) -- (31.1,0.06) node[above=2pt,right=-1pt] {};
							\draw[flecha] (16.9,1.2) -- (22.9,1.2) node[above=2pt,right=-1pt] {};
							\draw[flecha] (22.7,1.2) -- (27.5,1.2) node[above=2pt,right=-1pt] {};
							\draw[trans] (27.3,1.2) -- (29.5,1.2) node[above=2pt,right=-1pt] {};
							\draw[flecha] (16.9,-1.2) -- (22.9,-1.2) node[above=2pt,right=-1pt] {};
							\draw[flecha] (22.7,-1.2) -- (27.5,-1.2) node[above=2pt,right=-1pt] {};
							\draw[trans] (27.3,-1.2) -- (29.5,-1.2) node[above=2pt,right=-1pt] {};
							\draw[flecha] (17.8,-2.0) -- (17.8,0.3) node[above=2pt,right=-1pt] {};
							\draw[trans] (17.8,0.0) -- (17.8,2.0) node[above=2pt,right=-1pt] {};
							\draw[flecha] (25.8,-2.0) -- (25.8,0.3) node[above=2pt,right=-1pt] {};
							\draw[trans] (25.8,0.0) -- (25.8,2.0) node[above=2pt,right=-1pt] {};
							\draw[flecha] (28.6,-2.0) -- (28.6,0.3) node[above=2pt,right=-1pt] {};
							\draw[trans] (28.6,0.0) -- (28.6,2.0) node[above=2pt,right=-1pt] {};
							\draw[trans,fill=white] (19.25,0.0) circle (0.9);
							\draw[trans,fill=white] (27.25,0.0) circle (0.9);
							\draw[equation] (27.3,2.0) -- (27.3,2.0) node[midway] {$ a $};
							\draw[equation] (30.3,0.0) -- (30.3,0.0) node[midway,left] {$ d $};
							\draw[equation] (27.3,-2.0) -- (27.3,-2.0) node[midway] {$ c $};
							\draw[equation] (20.8,0.0) -- (20.8,0.0) node[midway,right] {$ \mathfrak{f} \bigl( \tilde{\lambda } \bigr) \cdot b $};
							\draw[equation] (19.25,0.0) -- (19.25,0.0) node[midway] {$ \tilde{\alpha } ^{\prime } $};
							\draw[equation] (27.25,0.0) -- (27.25,0.0) node[midway] {$ \tilde{\beta } ^{\prime } $};
						\end{tikzpicture}
					\end{flushleft}
					\caption{\label{dual-comut-bc} Results of the action of the operators $ B^{\prime \left( h^{\prime } \right) } _{f} \circ C^{\prime } _{\ell } $ and $ C^{\prime } _{\ell } \circ B^{\prime \left( h^{\prime } \right) } _{f} $ on the lattice $ \mathcal{L} _{2} $, which not only reinforces that $ \mathfrak{f} \bigl( \tilde{\lambda } \bigr) $ must belong to $ \mathcal{Z} \left( G \right) $, but also indicates that $ \mathfrak{f} $ must be a group homomorphism. Just as we did in Figures \ref{dual-comut-ab} and \ref{dual-comut-ac}, it is also worth noting that, here, $ B^{\prime } _{f,h^{\prime }} $ and $ C^{\prime } _{\ell } $ act only on the face and edge sectors whose intersection is not empty because, when this intersection is empty, these operators commute by definition.}
				\end{figure}
				since it shows us that
				\begin{equation}
					\mathfrak{f} \bigl( \tilde{\beta } \ast \tilde{\lambda } \bigr) \cdot a^{-1} \cdot \bigl[ \mathfrak{f} \bigl( \tilde{\lambda } \bigr) \cdot b \bigr] ^{-1} = \mathfrak{f} \bigl( \tilde{\beta } \ast \tilde{\lambda } \bigr) \cdot ab^{-1} \cdot \bigl[ \mathfrak{f} \bigl( \tilde{\lambda } \bigr) \bigr] ^{-1} = \mathfrak{f} \bigl( \tilde{\beta } \bigr) \cdot a^{-1} b^{-1} \label{condition-2}
				\end{equation}
				needs to also be satisfied for $ \bigl[ B^{\prime } _{f} , C^{\prime } _{\ell } \bigr] $ to vanish. Here, $ \bigl[ \mathfrak{f} \bigl( \tilde{\beta} \bigr) \bigr] ^{-1} $ is the inverse of the (group) element $ \mathfrak{f} \bigl( \tilde{\beta} \bigr) $. After all, since $ \mathsf{Im} \left( \mathfrak{f} \right) \subseteq \mathcal{Z} \left( G \right) $ allows us to conclude that (\ref{condition-2}) is equivalent to
				\begin{equation*}
					\mathfrak{f} \bigl( \tilde{\beta } \ast \tilde{\lambda } \bigr) \cdot \bigl[ \mathfrak{f} \bigl( \tilde{\lambda } \bigr) \bigr] ^{-1} = \mathfrak{f} \bigl( \tilde{\beta } \bigr) \ ,
				\end{equation*} 
				this result is in full agreement with the fact that $ \mathfrak{f} $ is a homomorphism.
				
				By the way, since $ \mathfrak{f} $ is a homomorphism whose codomain is the gauge group $ G $, it is of paramount importance to point out that this allows us to conclude that $ \tilde{S} $ is also a group (whose neutral element will be denoted by $ \tilde{\varepsilon } $) because every homomorphism is a structure-preserving map between two algebraic structures of the same type [\citen{fraleigh}]. And the importance of pointing this out is that, in addition to the group homomorphism properties
				\begin{equation*}
					\mathfrak{f} \bigl( \tilde{\varepsilon } \bigr) = e \ , \ \ \bigl[ \mathfrak{f} \bigl( \tilde{\alpha } \bigr) \bigr] ^{\dagger } = \mathfrak{f} \bigl( \tilde{\alpha } ^{-1} \bigr) = \bigl[ \mathfrak{f} \bigl( \tilde{\alpha } \bigr) \bigr] ^{-1} \ \ \textnormal{and} \ \ \mathfrak{f} \bigl( \tilde{\alpha } _{1} \bigr) \cdot \mathfrak{f} \bigl( \tilde{\alpha } _{2} \bigr) = \mathfrak{f} \bigl( \tilde{\alpha } _{1} \ast \tilde{\alpha } _{2} \bigr)
				\end{equation*} 
				ensure that $ \bigl[ A^{\prime } _{v} , C^{\prime } _{\ell } \bigr] = \bigl[ B^{\prime } _{f} , C^{\prime } _{\ell } \bigr] = 0 $, they also ensure that the requirements
				\begin{subequations} \label{requirements-double-c}
					\begin{align}
						\tilde{\alpha } ^{\prime \prime } = \bigl( \tilde{\lambda } ^{\prime } \bigr) ^{-1} \ast \tilde{\alpha } ^{\prime } & = \bigl( \tilde{\lambda } ^{\prime } \bigr) ^{-1} \ast \tilde{\lambda } ^{-1} \ast \tilde{\alpha } = \bigl( \tilde{\lambda } \ast \tilde{\lambda } ^{\prime } \bigr) ^{-1} \ast \tilde{\alpha } \ , \\
						\tilde{\beta } ^{\prime \prime } & = \tilde{\beta } ^{\prime } \ast \tilde{\lambda } ^{\prime } = \tilde{\beta } ^{\prime } \ast \bigl( \tilde{\lambda } \ast \tilde{\lambda } ^{\prime } \bigr) \quad \textnormal{and} \\
						a^{\prime \prime } = \mathfrak{f} \bigl( \tilde{\lambda } ^{\prime } & \bigr) \cdot a^{\prime } = \mathfrak{f} \bigl( \tilde{\lambda } ^{\prime } \bigr) \cdot \mathfrak{f} \bigl( \tilde{\lambda } \bigr) \cdot a = \mathfrak{f} \bigl( \tilde{\lambda } ^{\prime } \ast \tilde{\lambda } \bigr) \cdot a \ ,
					\end{align}
				\end{subequations}
				which need to be satisfied in the double action of $ C^{\left( \tilde{\lambda } \right) } _{\ell } $ that appears in Figure \ref{double-action},
				\begin{figure}[!t]
					\begin{flushleft}
						\begin{tikzpicture}[
							scale=0.3,
							equation/.style={thin},
							trans/.style={thin,shorten >=0.5pt,shorten <=0.5pt,>=stealth},
							flecha/.style={thin,->,shorten >=0.5pt,shorten <=0.5pt,>=stealth},
							dual/.style={dotted,-,shorten >=0.5pt,shorten <=0.5pt}
							]
							\draw[equation] (-8.4,-0.1) -- (-8.4,-0.1) node[midway,right] {$ C^{\prime }_{\ell } $};
							\draw[trans] (-5.6,2.3) -- (-5.6,-2.3) node[above=2pt,right=-1pt] {};
							\draw[trans] (0.1,2.3) -- (1.3,-0.06) node[above=2pt,right=-1pt] {};
							\draw[trans] (0.1,-2.3) -- (1.3,0.06) node[above=2pt,right=-1pt] {};
							\draw[dual] (-4.4,0.0) -- (-0.2,0.0) node[above=2pt,right=-1pt] {};
							\draw[flecha] (-2.3,-1.0) -- (-2.3,0.3) node[above=2pt,right=-1pt] {};
							\draw[trans] (-2.3,0.0) -- (-2.3,1.0) node[above=2pt,right=-1pt] {};
							\draw[trans,fill=white] (-4.4,0.0) circle (0.9);
							\draw[trans,fill=white] (-0.2,0.0) circle (0.9);
							\draw[equation] (-2.3,-0.8) -- (-2.3,-0.8) node[midway,below] {$ a $};
							\draw[equation] (-4.4,0.0) -- (-4.4,0.0) node[midway] {$ \tilde{\alpha } $};
							\draw[equation] (-0.2,-0.02) -- (-0.2,-0.02) node[midway] {$ \tilde{\beta } $};	
							\draw[equation] (2.6,-0.07) -- (2.6,-0.07) node[midway] {$ \ = \ $};
							\draw[equation] (4.9,0.8) -- (4.9,0.8) node[midway] {$ 1 $};
							\draw[trans] (3.8,0.0) -- (6.0,0.0) node[above=2pt,right=-1pt] {};
							\draw[equation] (4.9,-1.1) -- (4.9,-1.1) node[midway] {$ \bigl\vert \tilde{S} \bigr\vert $};
							\draw[equation] (7.2,0.0) -- (7.2,0.0) node[midway] {$ \sum $};
							\draw[equation] (7.2,-1.3) -- (7.2,-1.3) node[midway] {$ _{\tilde{\lambda } \in \tilde{S}} $};
							\draw[trans] (8.4,2.3) -- (8.4,-2.3) node[above=2pt,right=-1pt] {};
							\draw[trans] (14.1,2.3) -- (15.3,-0.06) node[above=2pt,right=-1pt] {};
							\draw[trans] (14.1,-2.3) -- (15.3,0.06) node[above=2pt,right=-1pt] {};
							\draw[dual] (9.6,0.0) -- (13.8,0.0) node[above=2pt,right=-1pt] {};
							\draw[flecha] (11.6,-1.0) -- (11.6,0.3) node[above=2pt,right=-1pt] {};
							\draw[trans] (11.6,0.0) -- (11.6,1.0) node[above=2pt,right=-1pt] {};
							\draw[trans,fill=white] (9.6,0.0) circle (0.9);
							\draw[trans,fill=white] (13.8,0.0) circle (0.9);
							\draw[equation] (11.7,-0.8) -- (11.7,-0.8) node[midway,below] {$ a^{\prime } $};
							\draw[equation] (9.6,0.0) -- (9.6,0.0) node[midway] {$ \tilde{\alpha } ^{\prime } $};
							\draw[equation] (13.8,-0.02) -- (13.8,-0.02) node[midway] {$ \tilde{\beta } ^{\prime } $};
						\end{tikzpicture} \\ \hspace*{3.0cm}
						\begin{tikzpicture}[
							scale=0.3,
							equation/.style={thin},
							trans/.style={thin,shorten >=0.5pt,shorten <=0.5pt,>=stealth},
							flecha/.style={thin,->,shorten >=0.5pt,shorten <=0.5pt,>=stealth},
							dual/.style={dotted,-,shorten >=0.5pt,shorten <=0.5pt}
							]
							\draw[equation] (15.3,-0.2) -- (15.3,-0.2) node[midway] {$ \Rightarrow \ \ C^{\prime }_{\ell } \circ C^{\prime }_{\ell } $};
							\draw[trans] (19.4,2.3) -- (19.4,-2.3) node[above=2pt,right=-1pt] {};
							\draw[trans] (25.1,2.3) -- (26.3,-0.06) node[above=2pt,right=-1pt] {};
							\draw[trans] (25.1,-2.3) -- (26.3,0.06) node[above=2pt,right=-1pt] {};
							\draw[dual] (20.6,0.0) -- (24.8,0.0) node[above=2pt,right=-1pt] {};
							\draw[flecha] (22.7,-1.0) -- (22.7,0.3) node[above=2pt,right=-1pt] {};
							\draw[trans] (22.7,0.0) -- (22.7,1.0) node[above=2pt,right=-1pt] {};
							\draw[trans,fill=white] (20.6,0.0) circle (0.9);
							\draw[trans,fill=white] (24.8,0.0) circle (0.9);
							\draw[equation] (22.7,-0.8) -- (22.7,-0.8) node[midway,below] {$ a $};
							\draw[equation] (20.6,0.0) -- (20.6,0.0) node[midway] {$ \tilde{\alpha } $};
							\draw[equation] (24.8,-0.02) -- (24.8,-0.02) node[midway] {$ \tilde{\beta } $};	
							\draw[equation] (27.6,-0.07) -- (27.6,-0.07) node[midway] {$ = $};
							\draw[equation] (29.9,0.8) -- (29.9,0.8) node[midway] {$ 1 $};
							\draw[trans] (28.8,0.0) -- (31.0,0.0) node[above=2pt,right=-1pt] {};
							\draw[equation] (29.9,-1.1) -- (29.9,-1.1) node[midway] {$ \bigl\vert \tilde{S} \bigr\vert ^{2} $};
							\draw[equation] (32.4,0.0) -- (32.4,0.0) node[midway] {$ \sum $};
							\draw[equation] (32.4,-1.3) -- (32.4,-1.3) node[midway] {$ _{\tilde{\lambda } ^{\prime } \in \tilde{S}} $};
							\draw[equation] (34.9,0.0) -- (34.9,0.0) node[midway] {$ \sum $};
							\draw[equation] (34.9,-1.3) -- (34.9,-1.3) node[midway] {$ _{\tilde{\lambda } \in \tilde{S}} $};
							\draw[trans] (36.1,2.3) -- (36.1,-2.3) node[above=2pt,right=-1pt] {};
							\draw[trans] (41.8,2.3) -- (43.0,-0.06) node[above=2pt,right=-1pt] {};
							\draw[trans] (41.8,-2.3) -- (43.0,0.06) node[above=2pt,right=-1pt] {};
							\draw[dual] (37.3,0.0) -- (41.5,0.0) node[above=2pt,right=-1pt] {};
							\draw[flecha] (39.3,-1.0) -- (39.3,0.3) node[above=2pt,right=-1pt] {};
							\draw[trans] (39.3,0.0) -- (39.3,1.0) node[above=2pt,right=-1pt] {};
							\draw[trans,fill=white] (37.3,0.0) circle (0.9);
							\draw[trans,fill=white] (41.5,0.0) circle (0.9);
							\draw[equation] (39.4,-0.8) -- (39.4,-0.8) node[midway,below] {$ a^{\prime \prime } $};	
							\draw[equation] (37.3,0.0) -- (37.3,0.0) node[midway] {$ \tilde{\alpha } ^{\prime \prime } $};
							\draw[equation] (41.5,-0.02) -- (41.5,-0.02) node[midway] {$ \tilde{\beta } ^{\prime \prime } $};
						\end{tikzpicture}
					\end{flushleft}
					\caption{\label{double-action} Scheme related to the double action of the link operator $ C^{\left( \tilde{\lambda } \right) } _{\ell } $ on the same edge sector of $ \mathcal{L} _{2} $. Here, $ \tilde{\alpha } ^{\prime \prime } $, $ \tilde{\beta } ^{\prime \prime } $ and $ a^{\prime \prime } $ are given by the expressions (\ref{requirements-double-c}), which also reinforce the need for $ \mathfrak{f} $ to be a group homomorphism.}
				\end{figure}
				are respected. In this way, by
				\begin{itemize}
					\item remembering that the $ D^{K} \left( G \right) $ vertex operators were inherited from the $ D \left( G \right) $ models, and
					\item noticing that the double action of $ B^{\prime } _{f,h} $ (on the same face sector of $ \mathcal{L} _{2} $) shows that it is, in fact, a projector because
					\begin{eqnarray*}
						\lefteqn{\delta \bigl( h^{\prime } , \mathfrak{f} \bigl( \tilde{\alpha } \bigr) \cdot a^{-1} b^{-1} cd \bigr) \cdot \delta \bigl( h^{\prime } , \mathfrak{f} \bigl( \tilde{\alpha } \bigr) \cdot a^{-1} b^{-1} cd \bigr) } \hspace*{5.5cm} \\
						& = & \delta \bigl( h^{\prime } , \mathfrak{f} \bigl( \tilde{\alpha } \bigr) \cdot a^{-1} b^{-1} cd \bigr) \ ,
					\end{eqnarray*}
				\end{itemize}
				we can conclude that
				\begin{itemize}
					\item it is actually reasonable that the vertex and face operators in (\ref{H-qdmf}) are defined as
					\begin{equation*}
						A^{\prime } _{v} = A^{\prime } _{v,0} \quad \textnormal{and} \quad B^{\prime } _{f} = B^{\prime } _{f,0}
					\end{equation*}
					respectively, with $ A^{\prime } _{v,J} $ and $ B^{\prime } _{f,L} $ being defined by (\ref{a-prime}) and (\ref{b-prime}) also respectively, and
					\item it is correct to assert that $ C^{\prime } _{\ell } $, in addition to making the $ D^{K} \left( G \right) $ models exactly solvable along with the other projectors in $ \mathfrak{A} ^{\prime } $ and $ \mathfrak{B} ^{\prime } $, can be interpreted as a special case of the operators
					\begin{equation}
						C^{\prime } _{\ell ,\Lambda } = \frac{1}{\vert \tilde{S} \vert } \sum _{\tilde{\lambda } \in \tilde{S}} \chi _{1 + \Lambda} \bigl( \tilde{\lambda } \bigr) \cdot C^{\bigl( \tilde{\lambda } \bigr) } _{\ell } \label{c-prime}
					\end{equation}
					(i.e., $ C^{\prime } _{\ell } = C^{\prime } _{\ell , 0} $) since, as $ \tilde{S} $ is a group, its characters $ \chi _{1 + \Lambda } \bigl( \tilde{\lambda } \bigr) $ confer the necessary orthonormality to the link operators that complete $ \mathfrak{C} ^{\prime } $.
				\end{itemize}
				
			\subsubsection{The dual behaviour of the link operator $ C^{\prime } _{\ell } $ as a comparator}
			
				Of course, even though it was clear that $ C^{\prime } _{\ell , \Lambda } $ are operators that make the $ D^{K} \left( G \right) $ models exactly solvable, we still need to evaluate them a little further. After all, despite them doing some transformations on the gauge and matter fields, we still need to evaluate if, in fact, $ C^{\prime } _{\ell } $ can be interpreted as the dual of $ C_{\ell } $. That is, we need to evaluated if $ C^{\prime } _{\ell } $ actually behaves as a comparator within some dual context.
				
				In order to make this evaluation, it is imperative to note that, since $ \mathsf{Im} \left( \mathfrak{f} \right) \subseteq \mathcal{Z} \left( G \right) $, it is highly recommended to assume that $ \tilde{S} $ is an Abelian group. And even though there are many good examples of group homomorphisms $ \mathfrak{f} : \tilde{S} \rightarrow G $ where $ \mathsf{Im} \left( \mathfrak{f} \right) \subseteq \mathcal{Z} \left( G \right) $ and $ \tilde{S} $ is a non-Abelian group [\citen{james}], one of the things that reinforces this recommendation is the fact that, when $ \tilde{S} $ and $ \mathsf{Im} \left( \mathfrak{f} \right) $ are two finite Abelian groups, there is a \emph{Fourier transform} [\citen{terras}] that allows us to observe that, in fact, $ C^{\prime } _{\ell } $ is endowed with the dual behaviour that it needs to portray. After all, note that, since $ \tilde{S} $ and $ \mathsf{Im} \left( \mathfrak{f} \right) $ are finite groups, this already allows us to rewrite
				\begin{equation*}
					C^{\prime } _{\ell } \bigl\vert \tilde{\alpha } , g , \tilde{\beta } \bigr\rangle = \frac{1}{\bigl\vert \tilde{S} \bigr\vert } \sum _{\tilde{\lambda } \in \tilde{S}} \bigl\vert \tilde{\lambda } ^{-1} \ast \tilde{\alpha } , \mathfrak{f} \bigl( \tilde{\lambda } \bigr) \cdot g , \tilde{\beta } \ast \tilde{\lambda } \bigr\rangle 
				\end{equation*} 
				as
				\begin{equation}
					C^{\prime } _{\ell } \bigl\vert \tilde{\alpha } ^{\prime } , g^{\prime } , \tilde{\beta } ^{\prime } \bigr\rangle = \frac{1}{\bigl\vert \tilde{S} \bigr\vert } \sum _{\tilde{\lambda } \in \tilde{S}} \bar{\chi } _{\tilde{\alpha } ^{\prime }} \bigl( \tilde{\lambda } \bigr) \ \omega _{g^{\prime }} \bigl( \mathfrak{f} \bigl( \tilde{\lambda } \bigr) \bigr) \ \chi _{\tilde{\beta } ^{\prime }} \bigl( \tilde{\lambda } \bigr) \bigl\vert \tilde{\alpha } , g , \tilde{\beta } \bigr\rangle \label{exp25}
				\end{equation}
				by using the unitary transformations
				\begin{equation*}
					\left\vert g^{\prime } \right\rangle = \frac{1}{\vert G \vert } \sum _{g \in G} \omega _{g^{\prime }} \bigl( g \bigr) \left\vert g \right\rangle \ \ \textnormal{and} \ \ \left\vert \tilde{\alpha } ^{\prime } \right\rangle = \frac{1}{\bigl\vert \tilde{S} \bigr\vert } \sum _{\tilde{\alpha } \in \tilde{S}} \bar{\chi } _{\tilde{\alpha } ^{\prime }} \left( \tilde{\alpha } \right) \left\vert \tilde{\alpha } \right\rangle \ ,
				\end{equation*} 
				where $ \omega _{g^{\prime }} \left( g \right) $ and $ \chi _{\tilde{\alpha } ^{\prime }} \left( \tilde{\alpha } \right) $ are characters of $ G $ and $ \tilde{S} $ respectively. Now if, in addition to $ \tilde{S} $ and $ \mathsf{Im} \left( \mathfrak{f} \right) $ being finite groups, they are also two Abelian groups, there will be a Fourier transform $ \hat{\mathfrak{f}} \in L \bigl( \tilde{S} ^{\ast } \bigr) $ such that
				\begin{equation}
					\hat{\mathfrak{f}} \left( \chi \right) = \sum _{\tilde{\lambda } \in \tilde{S}} \mathfrak{f} \bigl( \tilde{\lambda } \bigr) \chi \bigl( \tilde{\lambda } \bigr) \ \ \textnormal{and} \ \ \mathfrak{f} \bigl( \tilde{\lambda } \bigr) = \frac{1}{\bigl\vert \tilde{S} \bigr\vert } \sum _{\chi \in \tilde{S} ^{\ast }} \hat{\mathfrak{f}} \left( \chi \right) \chi \bigl( \tilde{\lambda } \bigr) \ , \label{fourier}
				\end{equation}
				where the dual group $ \tilde{S} ^{\ast } $ is isomorphic to $ \tilde{S} $ [\citen{james,rudin,barut,hall}]. And why is this Fourier transform important? Because, by noting that an expression of the sort $ \bar{\chi } _{\tilde{\alpha } ^{\prime }} \bigl( \tilde{\lambda } \bigr) \chi  _{\tilde{\beta } ^{\prime }} \bigl( \tilde{\lambda } \bigr) = \bar{\chi } _{\left\{ \tilde{\alpha } ^{\prime } , \tilde{\beta } ^{\prime } \right\}} \bigl( \tilde{\lambda } \bigr) $ is also a character, the substitution of (\ref{fourier}) into (\ref{exp25}) allows to see that
				\begin{eqnarray*}
					C^{\prime } _{\ell } \bigl\vert \tilde{\alpha } ^{\prime } , g^{\prime } , \tilde{\beta } ^{\prime } \bigr\rangle & = & \frac{1}{\bigl\vert \tilde{S} \bigr\vert } \sum _{\chi _{\tilde{\gamma }} \in \tilde{S} ^{\ast }} \widehat{ \left[ \omega _{g^{\prime }} \circ \mathfrak{f} \right] } \left( \chi _{\tilde{\gamma }} \right) \left( \frac{1}{\bigl\vert \tilde{S} \bigr\vert } \sum _{\tilde{\lambda } \in \tilde{S}} \bar{\chi } _{\left\{ \tilde{\alpha } ^{\prime } , \tilde{\beta } ^{\prime } \right\}} \bigl( \tilde{\lambda } \bigr) \chi  _{\tilde{\gamma }} \bigl( \tilde{\lambda } \bigr) \right) \bigl\vert \tilde{\alpha } , g , \tilde{\beta } \bigr\rangle \\
					& = & \negthickspace \frac{1}{\bigl\vert \tilde{S} \bigr\vert } \sum _{\chi _{\tilde{\gamma }} \in \tilde{S} ^{\ast }} \widehat{ \left[ \omega _{g^{\prime }} \circ \mathfrak{f} \right] } \left( \chi _{\tilde{\gamma }} \right) \cdot \delta \left( \chi _{\left\{ \tilde{\alpha } ^{\prime } , \tilde{\beta } ^{\prime } \right\}} , \chi _{\tilde{\gamma }} \right) \bigl\vert \tilde{\alpha } , g , \tilde{\beta } \bigr\rangle \\
					& = & \negthickspace \frac{1}{\bigl\vert \tilde{S} \bigr\vert } \ \widehat{ \left[ \omega _{g^{\prime }} \circ \mathfrak{f} \right] } \bigl( \chi _{\left\{ \tilde{\alpha } ^{\prime } , \tilde{\beta } ^{\prime } \right\}} \bigr) \bigl\vert \tilde{\alpha } , g , \tilde{\beta } \bigr\rangle \ .
				\end{eqnarray*}
				In other words, although the exact form of the index $ {\bigl\{ \tilde{\alpha } ^{\prime } , \tilde{\beta } ^{\prime } \bigr\}} $ depends on the nature of the group $ \tilde{S} $, it is undeniable that, when $ \tilde{S} $ and $ \mathsf{Im} \left( \mathfrak{f} \right) $ are two finite Abelian groups, $ C^{\prime } _{\ell } $ can actually be interpreted as an operator that compares two neighbouring matter fields (i.e., that compares two matter fields that belong to the same dual edge sector) differently, which only becomes clear when this operator acts on a diagonal basis
				\begin{equation*}
					\left\{ \bigl\vert \tilde{\alpha } , g , \tilde{\beta } \bigr\rangle : g \in G \ \right. \textnormal{and} \ \left. \tilde{\alpha } , \tilde{\beta } \in \tilde{S} \ \right\} \ .
				\end{equation*}
				This different way of comparing two neighbouring matter fields rests on the \emph{Pontryagin duality}, which ensures that there is a one-to-one correspondence between the characters $ \chi _{\tilde{\lambda }} $ and the elements of $ \tilde{S} $ [\citen{pontryagin}]. 
		
		\section{\label{QDMf-properties} General properties of these $ D^{K} \left( G \right) $ models}
			
			According to what we saw in the last Section, it is impossible not to recognize that, when $ \tilde{S} $ and $ \mathsf{Im} \left( \mathfrak{f} \right) $ are Abelian finite groups, all the operators (\ref{a-prime}), (\ref{b-prime}) and (\ref{c-prime}), in addition to being dual to $ B_{f,L} $, $ A_{v,J} $ and $ C_{\ell , \Lambda } $ respectively, are also projectors that make the $ D^{K} \left( G \right) $ models exactly solvable. And since all these properties are only achieved when $ \tilde{S} $ and $ \mathsf{Im} \left( \mathfrak{f} \right) $ are Abelian finite groups, it is correct to say that this duality, which we so wanted to see between the $ D_{M} \left( G \right) $ and $ D^{K} \left( G \right) $ models, only exists when $ G $ is an Abelian finite group. Of course, if $ \mathfrak{f} : \tilde{S} \rightarrow G $ were a group homomorphism without any commitment to the definition of the co-action (\ref{co-action}), the fact that $ \mathsf{Im} \left( \mathfrak{f} \right) \subseteq \mathcal{Z} \left( G \right) $ would be completely incapable of making $ G $ also an Abelian group. But given that
			\begin{itemize}
				\item this $ \mathfrak{f} $ defines (\ref{co-action}) as co-action homomorphism, and
				\item it is well known that, when $ G $ and $ \mathcal{Z} \left( G \right) $ are both Abelian groups, $ \mathcal{Z} \left( G \right) = G $ [\citen{james}],
			\end{itemize}
			this is precisely what allows us to assert that the duality between these $ D_{M} \left( G \right) $ and $ D^{K} \left( G \right) $ models only exists when $ G $ is an Abelian finite gauge group.
			
			But by speaking of the projectivity of these operators, it is worth mentioning that, just as the $ D_{M} \left( G \right) $ vertex, face and link operators are responsible for the decomposition of $ \mathfrak{H} _{D_{M} \left( G \right) } $ as (\ref{soma-direta}), all these $ D^{K} \left( G \right) $ vertex, face and link operators are also responsible for the decomposition of $ \mathfrak{H} _{D^{K} \left( G \right) } $ into the direct sum
			\begin{equation*}
				\mathfrak{H} _{D^{K} \left( G \right) } = \mathfrak{H} ^{\left( 0 \right) } _{D^{K} \left( G \right) } \oplus \mathfrak{H} ^{\perp } _{D^{K} \left( G \right) } \ .
			\end{equation*} 
			Here, $ \mathfrak{H} ^{\left( 0 \right) } _{D^{K} \left( G \right) } $ and $ \mathfrak{H} ^{\perp } _{D^{K} \left( G \right) } $ are the orthogonal subspaces that contain all the $ D^{K} \left( G \right) $ vacuum and non-vacuum states respectively.
			
			As a matter of fact, in the case of the operators $ A^{\prime } _{v,0} $, $ B^{\prime } _{f,0} $ and $ C^{\prime } _{\ell ,0} $ that make up the $ D^{K} \left( G \right) $ Hamiltonian, it is crucial to note that they are responsible for projecting any state onto $ \mathfrak{H} ^{\left( 0 \right) } _{D^{K} \left( G \right) } $. After all, this is what explains not only why the $ D^{K} \left( G \right) $ vacuum states are such that
			\begin{equation*}
				A^{\prime } _{v,0} \bigl\vert \tilde{\xi } _{0} \bigr\rangle = \bigl\vert \tilde{\xi } _{0} \bigr\rangle \ , \ \ B^{\prime } _{f,0} \bigl\vert \tilde{\xi } _{0} \bigr\rangle = \bigl\vert \tilde{\xi } _{0} \bigr\rangle \ \ \textnormal{and} \ \ C^{\prime } _{\ell ,0} \bigl\vert \tilde{\xi } _{0} \bigr\rangle = \bigl\vert \tilde{\xi } _{0} \bigr\rangle 
			\end{equation*} 
			hold for all the $ N_{v} $ vertices, $ N_{f} $ faces and $ N_{\ell } $ edges of $ \mathcal{L} _{2} $, but also why
			\begin{equation}
				\bigl\vert \tilde{\xi } ^{\left( 0 \right) } _{0} \bigr\rangle = \prod _{\ell } C^{\prime } _{\ell } \prod _{v} A^{\prime } _{v} \underbrace{\ \left\vert e \right\rangle \otimes \ldots \otimes \left\vert e \right\rangle \ } _{N_{\ell } \ \textnormal{times}} \otimes \underbrace{\ \left\vert 0 \right\rangle \otimes \ldots \otimes \left\vert 0 \right\rangle \ } _{N_{f} \ \textnormal{times}} \label{dk-vacuo-1} 
			\end{equation}
			is a vacuum state that is common to all the $ D^{K} \left( G \right) $ models. Alongside this, it is also very crucial to note that it is the projectivity of the other vertex, face and link operators on $ \mathfrak{H} ^{\perp } _{D^{K} \left( G \right) } $ that explains why all the operators $ \tilde{W} ^{\left( J , L , \Lambda \right) } _{\ell } $ and $ \tilde{W} ^{\left( J , \Lambda \right) } _{f} $, which are such that
			\begin{subequations} \label{DK-quasiparticles-creation}
				\begin{align}
					\tilde{W} ^{\left( J , L , \Lambda \right) } _{\ell } \circ A^{\prime } _{v,0} & = A^{\prime } _{v,J} \circ \tilde{W} ^{\left( J , L , \Lambda \right) } _{\ell } \ , \label{DK-quasiparticles-creation-a} \\
					\tilde{W} ^{\left( J , L , \Lambda \right) } _{\ell } \circ B^{\prime } _{f,0} & = B^{\prime } _{f,L} \circ \tilde{W} ^{\left( J , L , \Lambda \right) } _{\ell } \ , \label{DK-quasiparticles-creation-b} \\
					\tilde{W} ^{\left( J , L , \Lambda \right) } _{\ell } \circ C^{\prime } _{\ell ,0} & = C^{\prime } _{\ell , \Lambda } \circ \tilde{W} ^{\left( J , L , \Lambda \right) } _{\ell } \ , \label{DK-quasiparticles-creation-c}
				\end{align}
			\end{subequations}
			\begin{subequations} \label{dual-matter-quasiparticles-creation}
				\begin{align}
					\tilde{W} ^{\left( L , \Lambda \right) } _{f} \circ B^{\prime } _{f,0} & = B^{\prime } _{f, L} \circ \tilde{W} ^{\left( L , \Lambda \right) } _{f} \quad \textnormal{and} \label{dual-matter-quasiparticles-creation-a} \\
					\tilde{W} ^{\left( L , \Lambda \right) } _{f} \circ C^{\prime } _{\ell ,0} & = C^{\prime } _{\ell , \Lambda } \circ \tilde{W} ^{\left( L , \Lambda \right) } _{f} \label{dual-matter-quasiparticles-creation-c}
				\end{align}
			\end{subequations}
			respectively, can remove these $ D^{K} \left( G \right) $ models from their ground states by producing energy excitations when $ \left( J , L , \Lambda \right) \neq \left( 0 , 0 , 0 \right) $ and $ \left( J , \Lambda \right) \neq \left( 0 , 0 \right) $. Once again, note that, since these $ D^{K} \left( G \right) $ models are also quantum-computational models that try/need to model some reality that can be physically implemented, all the energy excitations $ \tilde{q} ^{\left( J , L , \Lambda \right) } $ and $ \tilde{Q} ^{\left( J , \Lambda \right) } $, which are locally produced by the action of $ \tilde{W} ^{\left( J , L , \Lambda \right) } _{\ell } $ and $ \tilde{W} ^{\left( J , \Lambda \right) } _{f} $ respectively, need to be, at least, such that
			\begin{eqnarray*}
				\tilde{q} ^{\left( J^{\prime } , L^{\prime } , \Lambda ^{\prime } \right) } \times \tilde{q} ^{\left( J^{\prime \prime } , L^{\prime \prime } , \Lambda ^{\prime \prime } \right) } & = & \tilde{q} ^{\left( J^{\prime \prime } , L^{\prime \prime } , \Lambda ^{\prime \prime } \right) } \times \tilde{q} ^{\left( J^{\prime } , L^{\prime } , \Lambda ^{\prime } \right) } \ , \\
				\tilde{q} ^{\left( J^{\prime } , L^{\prime } , \Lambda ^{\prime } \right) } \times \tilde{Q} ^{\left( J^{\prime \prime } , \Lambda ^{\prime } \right) } & = & \tilde{Q} ^{\left( J^{\prime \prime } , \Lambda ^{\prime } \right) } \times \tilde{q} ^{\left( J^{\prime } , L^{\prime } , \Lambda ^{\prime } \right) } \quad \textnormal{and} \\
				\tilde{Q} ^{\left( J^{\prime } , \Lambda ^{\prime } \right) } \times \tilde{Q} ^{\left( J^{\prime \prime } , \Lambda ^{\prime \prime } \right) } & = & \tilde{Q} ^{\left( J^{\prime \prime } , \Lambda ^{\prime \prime } \right) } \times \tilde{Q} ^{\left( J^{\prime } , \Lambda ^{\prime } \right) } \ .
			\end{eqnarray*}
			in order to $ \tilde{q} ^{\left( J , L , \Lambda \right) } $ and $ \tilde{Q} ^{\left( J , \Lambda \right) } $ can be interpreted as quasiparticles. 
			
			\subsection{The matrix representation of the $ D^{K} \left( \mathds{Z} _{N} \right) $ vertex, face and edge operators}
				
				Given that we paid more attention to the $ D_{M} \left( G \right) $ models where $ G = \mathds{Z} _{N} $, it makes sense that we turn our attention to the $ D^{K} \left( G \right) $ ones where we have this same gauge group. After all, in addition to allowing us to better compare these $ D^{K} \left( \mathds{Z} _{N} \right) $ models with the $ D \left( \mathds{Z} _{N} \right) $ and $ D_{M} \left( \mathds{Z} _{N} \right) $ ones, it also seems reasonable to take $ G = \mathds{Z} _{N} $ for two other reasons. And the first one is that, by remembering that the fact that $ \mathfrak{f} $ is a group homomorphism suggests that we also take $ \tilde{S} $ as another cyclic Abelian group (because every homomorphism is a structure-preserving map between two algebraic structures of the same type [\citen{fraleigh}]), this allows to deal with a well-defined matrix representation
				\begin{subequations} \label{abc-dk}
					\begin{align}
						A^{\prime } _{v,J} & = \frac{1}{\left\vert G \right\vert } \sum _{g \in \mathds{Z} _{N}} \chi _{1 + J} \left( g^{-1} \right) \cdot \left( \prod _{\ell ^{\prime } \in S^{\uparrow } _{v}} X^{g} _{\ell ^{\prime }} \right) \left( \prod _{\ell ^{\prime \prime } \in S^{\downarrow } _{v}} X^{-g} _{\ell ^{\prime \prime }} \right) \ , \label{a-dk} \\
						B^{\prime } _{f,L} = \frac{1}{\left\vert G \right\vert } & \sum _{g \in \mathds{Z} _{N}} \chi _{1 + L} \left( g \right) \cdot F_{f} \left( g \right) \left( \prod _{\ell ^{\prime } \in S^{\circlearrowleft } _{f}} Z^{g} _{\ell ^{\prime }} \right) \left( \prod _{\ell ^{\prime \prime } \in S^{\circlearrowright } _{f}} Z^{-g} _{\ell ^{\prime \prime }} \right) \quad \textnormal{and} \label{b-dk} \\
						C^{\prime } _{\ell ,\Lambda } & = \frac{1}{\bigl\vert \tilde{S} \bigr\vert } \sum _{\tilde{\gamma } \in \tilde{S}} \tilde{\chi } _{1 + \Lambda } \left( \tilde{\gamma } \right) \cdot \bigl( \tilde{X} ^{\dagger } _{f_{1}} \bigr) ^{\tilde{\gamma }} \otimes F_{\ell } \left( \tilde{\gamma } \right) \otimes \bigl( \tilde{X} _{f_{2}} \bigr) ^{\tilde{\gamma }} \ , \label{c-dk}
					\end{align}
				\end{subequations}
				for the $ D^{K} \left( \mathds{Z} _{N} \right) $ vertex, face and link operators. Here,
				\begin{itemize}
					\item[(i)] $ S^{\uparrow } _{v} $ and $ S^{\downarrow } _{v} $ are disjoint edge subsets of $ S_{v} $, whose edge orientations pointing in and out of the $ v $-th vertex respectively,
					\item[(ii)] $ S^{\circlearrowleft } _{f} $ and $ S^{\circlearrowright } _{f} $ are disjoint edge subsets of $ S_{f} $, whose edges have counterclockwise and clockwise orientations respectively, and
					\item[(iii)] $ \chi $ and $ \tilde{\chi } $ are characters of the matrix representations of $ G = \mathds{Z} _{N} $ and $ \tilde{S} = \mathds{Z} _{K} $ respectively.
				\end{itemize}
				Note also that, in the case of this matrix representation (\ref{abc-dk}), it leads us to
				\begin{subequations} \label{matrices} 
					\begin{align}
						X = \sum _{h \in \mathds{Z} _{N}} \left\vert \left( h + 1 \right) \textnormal{mod} \ N \right\rangle \left\langle h \right\vert \ \ & , \ \ Z = \sum _{h \in \mathds{Z} _{N}} \omega ^{h} \left\vert h \right\rangle \left\langle h \right\vert \ , \label{matrices-xz} \\
						\tilde{X} = \sum _{\tilde{\alpha } \in \mathds{Z} _{K}} \left\vert \left( \tilde{\alpha } + 1 \right) \textnormal{mod} \ K \right\rangle \left\langle \tilde{\alpha } \right\vert \quad & \textnormal{and} \quad \tilde{Z} = \sum _{\tilde{\alpha } \in \mathds{Z} _{K}} \tilde{\omega } ^{\tilde{\alpha }} \left\vert \tilde{\alpha } \right\rangle \left\langle \tilde{\alpha } \right\vert \ , \label{tilde-matrices-xz} 
					\end{align}
				\end{subequations}
				since the correspondence principle dictates that these $ D^{K} \left( \mathds{Z} _{N} \right) $ models must be reduced to the $ D \left( \mathds{Z} _{N} \right) $ ones in some special cases. Here, $ \omega = e^{i \left( 2 \pi / N \right) } $ and $ \tilde{\omega } = e^{i \left( 2 \pi / K \right) } $ are the generators of the gauge ($ G = \mathds{Z} _{N} $) and matter ($ \tilde{S} = \mathds{Z} _{K} $) groups\footnote{Although we have not evaluated the commutation relations when $ J, L, \Lambda \neq 0 $, the expressions (\ref{abc-dk}) justify the comment that we made on page \pageref{ortho-comment} because the only difference that exists among them concerns the characters that multiply each of the components $ A^{\prime \left( g \right) } _{v} $, $ B^{\prime \left( g \right) } _{f} $ and $ C^{\prime \left( \tilde{\lambda } \right) } _{\ell } $. And as these characters are constants that commute with each other, there is no way not to conclude that
				\begin{equation*}
					\bigl[ A^{\prime } _{v,J} , B^{\prime } _{f,L} \bigr] = \bigl[ A^{\prime } _{v,J} , C^{\prime } _{\ell , \Lambda } \bigr] = \bigl[ B^{\prime } _{f,L} , C^{\prime } _{\ell , \Lambda } \bigr] = 0
				\end{equation*}
				holds not only for all the values of $ J^{\prime \left( \prime \right) } , L^{\prime \left( \prime \right) } = 0 , 1 , \ldots , N-1 $ and $ \Lambda ^{\prime \left( \prime \right) } = 0 , 1 , \ldots , K-1 $, but also for all the vertices, faces and edges of $ \mathcal{L} _{2} $.}.
				
				Now, with respect to the matrices $ F_{f} \left( g \right) $ and $ F_{\ell } \left( \tilde{\gamma } \right) $ that appear in (\ref{abc-dk}), it is important to say that they represent how $ \mathfrak{f} $ couples these $ D \left( \mathds{Z} _{N} \right) $ models to the matter fields. And in order to understand not only how these matrices make this coupling, but also the second reason why it is reasonable to take $ G = \mathds{Z} _{N} $ and $ \tilde{S} = \mathds{Z} _{K} $, it is of paramount importance to pay attention to the statements of Theorems \ref{gcd-gallian} and \ref{proposition} below, whose proofs are in Refs. [\citen{gallian}] and [\citen{beachy}] respectively. \bigskip
				
				\noindent\fbox{
					\parbox[center]{4.8in}{
						\begin{theorem}
							The number of group homomorphisms from $ \mathds{Z} _{K} $ into $ \mathds{Z} _{N} $ is $ \gcd \left( K , N \right) $ (i.e., this number is the greatest common divisor of $ K $ and $ N $). \label{gcd-gallian}
						\end{theorem}
						\begin{theorem}
							Every group homomorphism $ \mathfrak{f} : \mathds{Z} _{K} \rightarrow \mathds{Z} _{N} $ can be completely determined by
							\begin{equation}
								\mathfrak{f} \bigl( \bigl[ \tilde{\alpha } \bigr] \bigr) \ = \ \bigl[ \mathsf{n} \tilde{\alpha } \bigr] \quad , \label{regulator-homomorphism}
							\end{equation}
							where $ \mathsf{n} $ is a natural number that assumes values other than zero if, and only if, $ \mathsf{n} K $ is a natural number divisible by $ N $. \label{proposition}
						\end{theorem}
					}
				} \bigskip
				
				\noindent That is, when we deal with $ D^{K} \left( \mathds{Z} _{N} \right) $ models where $ \tilde{S} = \mathds{Z} _{K} $, we can rely on these two theorems and, according to what these theorems claim, all these $ D^{K} \left( \mathds{Z} _{N} \right) $ models have, at least, a description where $ F_{f} \left( g \right) $ and $ F_{\ell } \left( \tilde{\gamma } \right) $ are identity matrices: after all, for all the values of $ N $ and $ K $, there will always be a group homomorphism
				\begin{equation}
					\mathfrak{f} \bigl( \bigl[ \tilde{\alpha } \bigr] \bigr) \ = \ \bigl[ e \bigr] \label{trivial-via-proposition}
				\end{equation}
				that maps all the elements of $ \mathds{Z} _{K} $ to the identity element of $ \mathds{Z} _{N} $. Note that, when these $ D^{K} \left( \mathds{Z} _{N} \right) $ models are such that $ N $ and $ K $ are \emph{coprime numbers}, the only way to define these models is by using this trivial group homomorphism (\ref{trivial-via-proposition}).
				
				\subsubsection{\label{first-degeneracy-comment}A first comment on the $ D^{K} \left( \mathds{Z} _{N} \right) $ ground state degeneracy} 
				
					When we come across this description, where all these models are defined by using (\ref{trivial-via-proposition}), one of the things that we can say about them is that their \textquotedblleft fake holonomies\textquotedblright \hspace*{0.01cm} (\ref{fake-holonomy}) reduce to the true holonomies (\ref{holonomy}).  And since this reduction allows us to identify $ B^{\prime } _{f,L} $ as the same face operator $ B_{f,L} $ of the $ D \left( \mathds{Z} _{N} \right) $ models, there is no way not to conclude that all the $ D^{K} \left( \mathds{Z} _{N} \right) $ models support the same quasiparticles as the $ D \left( \mathds{Z} _{N} \right) $ models.
					
					Observe that this conclusion is not surprising because, similar to what was discussed in Section \ref{QDM-review}, the correspondence principle already requires that all the $ D^{K} \left( \mathds{Z} _{N} \right) $ models support these quasiparticles in some way. And, in fact, this is reinforced by the fact that the face and link operators of the trivial $ D^{K} \left( \mathds{Z} _{N} \right) $ models (i.e., of the $ D^{K} \left( \mathds{Z} _{N} \right) $ models that are defined by using (\ref{trivial-via-proposition})) are given by
					\begin{subequations} \label{bc-dk-trivial}
						\begin{align}
							B^{\prime } _{f,L} = \frac{1}{\left\vert G \right\vert } & \sum _{g \in \mathds{Z} _{N}} \chi _{1 + L} \left( g \right) \cdot \mathds{1} _{f} \left( \prod _{\ell ^{\prime } \in S^{\circlearrowleft } _{f}} Z^{g} _{\ell ^{\prime }} \right) \left( \prod _{\ell ^{\prime \prime } \in S^{\circlearrowright } _{f}} Z^{-g} _{\ell ^{\prime \prime }} \right) \quad \textnormal{and} \label{b-dk-trivial} \\
							C^{\prime } _{\ell ,\Lambda } & = \frac{1}{\bigl\vert \tilde{S} \bigr\vert } \sum _{\tilde{\gamma } \in \tilde{S}} \tilde{\chi } _{1 + \Lambda } \left( \tilde{\gamma } \right) \cdot \bigl( \tilde{X} ^{\dagger } _{f_{1}} \bigr) ^{\tilde{\gamma }} \otimes \mathds{1} _{\ell } \otimes \bigl( \tilde{X} _{f_{2}} \bigr) ^{\tilde{\gamma }} \label{c-dk-trivial}
						\end{align}
					\end{subequations}
					respectively because this allows us to conclude, for instance, that these operators cannot detect any matter and gauge quasiparticles also respectively. In other words, this allows us to conclude that, in the same way as with the trivial $ D_{M} \left( \mathds{Z} _{N} \right) $ models, the trivial $ D^{K} \left( \mathds{Z} _{N} \right) $ ones do not couple the gauge fields with those of matter since $ B^{\prime } _{f,L} $ and $ C^{\prime } _{\ell ,\Lambda } $ are \textquotedblleft blind\textquotedblright \hspace*{0.01cm} to the presence of the quasiparticles $ \tilde{Q} ^{\left( J , \Lambda \right) } $ and $ \tilde{q} ^{\left( J , L , \Lambda \right) } $ respectively. As a consequence, it is correct to say that, due to this \textquotedblleft blindness\textquotedblright , all these trivial $ D^{K} \left( \mathds{Z} _{N} \right) $ models support the same electric and magnetic quasiparticles, with the same properties, as the $ D \left( \mathds{Z} _{N} \right) $ ones.
					
					In light of these comments, it is also interesting to note that, since
					\begin{itemize}
						\item none of these operators is able to detect any change $ \left\vert \tilde{\alpha } ^{\prime } \right\rangle _{f} \leftrightarrow \left\vert \tilde{\alpha } ^{\prime \prime } \right\rangle _{f} $ \footnote{By paraphrasing the footnote on page \pageref{comment-v-index}: here, we are using the index $ f $ only to emphasize that $ \left\vert \tilde{\alpha } \right\rangle $ is an element associated with a face of $ \mathcal{L} _{2} $.}, and
						\item none of the operators $ \bigl( \tilde{X} _{f} \bigr) ^{\tilde{\gamma }} $ (which make all these changes $ \left\vert \tilde{\alpha } ^{\prime } \right\rangle _{f} \leftrightarrow \left\vert \tilde{\alpha } ^{\prime \prime } \right\rangle _{f} $) can be expressed as a product involving the vertex, face and link operators,
					\end{itemize}
					all the vacuum states
					\begin{equation}
						\bigl\vert \tilde{\xi } ^{\left( \tilde{\alpha } \right) } _{0} \bigr\rangle = \prod _{\ell ^{\prime }} C_{\ell ^{\prime }} \prod _{v^{\prime }} A_{v^{\prime }} \left( \bigotimes _{\ell \in \mathcal{L} _{2}} \left\vert e \right\rangle \right) \otimes \left( \bigotimes _{f \in \mathcal{L} _{2}} \left\vert 0 \right\rangle \right) _{f \neq f^{\prime \prime }} \hspace*{-0.8cm} \otimes \left\vert \tilde{\alpha } \right\rangle _{f^{\prime \prime }} \ , \label{dk-trivial-vacuum-states}
					\end{equation}
					which are defined by taking $ \tilde{\alpha } = 0 , 1 , \ldots , K-1 $, are independent of each other. Here, the rationale for why all the vacuum states, where the $ \tilde{\alpha } \neq 0 $, have only a single $ f^{\prime \prime } $-th lattice face filled with $ \left\vert \tilde{\alpha } \right\rangle \neq \left\vert 0 \right\rangle $ is due to the simple fact that, just in this case, there are no transformations, which can be expressed as a product of the vertex, face and link operators, that can connect these $ K $ vacuum states (\ref{dk-trivial-vacuum-states}). Note that, since the inability of (\ref{c-dk-trivial}) to detect the quasiparticles $ \tilde{q} ^{\left( J , L , \Lambda \right) } $ implies, for instance, that all of these quasiparticles can be transported without increasing/decreasing the energy of the system, it is not difficult to conclude that the action of an operator
					\begin{equation*}
						\tilde{O} _{L} \left( \overline{\gamma } ^{\ast } \right) = \prod _{\ell ^{\prime } \in \Gamma ^{\ast } _{\circlearrowleft }} \tilde{W} ^{\left( 0 , L , 0 \right) } _{\ell ^{\prime }} \prod _{\ell ^{\prime } \in \Gamma ^{\ast } _{\circlearrowright }} \left( \tilde{W} ^{\left( 0 , L , 0 \right) } _{\ell ^{\prime }} \right) ^{\dagger }
					\end{equation*}
					does not lead to an excited state when it acts on any of the vacuum states (\ref{dk-trivial-vacuum-states}). And since this allows us to recognize that all the vacuum states
					\begin{equation}
						\bigl\vert \xi ^{\left( \tilde{\alpha } , \vec{\lambda } , \vec{L} \right) } _{0} \bigr\rangle  = \prod ^{s} _{p=1} \left[ \tilde{O} _{L_{p}} \left( \overline{\gamma } ^{\ast } _{p} \right) \right] ^{\lambda _{p}} \bigl\vert \tilde{\xi } ^{\left( \tilde{\alpha } \right) } _{0} \bigr\rangle \label{trivial-dk-vacuum-states}
					\end{equation}
					are topologically independent of each other due to the non-contractility of $ \overline{\gamma } ^{\ast } _{p} $ [\citen{mf-pedagogical}], it is also not difficult to conclude that all these
					\begin{equation}
						\tilde{\mathfrak{n}} = \left\vert \ \ker \left( \mathfrak{f} \right) \ \right\vert \cdot \mathfrak{d} _{D \left( \mathds{Z} _{N} \right) } \label{formula-trivial}
					\end{equation}
					vacuum states (\ref{trivial-dk-vacuum-states}) are mere replicas of the $ D \left( \mathds{Z} _{N} \right) $ vacuum states\footnote{Here, we are using the same notation used on page \pageref{ddg-notation}, now to refer to the number $ \mathfrak{d} _{D \left( \mathds{Z} _{N} \right) } $ of vacuum states that define the $ D \left( \mathds{Z} _{N} \right) $ ground state.}. 
		
		\subsection{But what happens when $ \mathfrak{f} $ is not a trivial group homomorphism?}
		
			From the point of view of the duality that we want to identify between the $ D_{M} \left( \mathds{Z} _{N} \right) $ and $ D^{K} \left( \mathds{Z} _{N} \right) $ models, all these last conclusions/observations about the trivial $ D^{K} \left( \mathds{Z} _{N} \right) $ models are very welcome. After all, note that, since the co-action that (\ref{trivial-via-proposition}) defines can be always induced by a trivial (sub)group action
			\begin{equation}
				\tilde{\mu } _{\mathfrak{f}} \bigl( \mathfrak{f} \bigl( \tilde{\alpha } \bigr) , \tilde{\gamma } \bigr) = \tilde{\gamma } \ , \label{trivial-dual-group-action}
			\end{equation} \label{duality-comment-formula}
			which can be represented by the same matrix
			\begin{equation*}
				F_{f} \left( g \right) = \mathds{1} _{f}
			\end{equation*}
			that composes (\ref{b-dk-trivial}), this trivial (sub)group action defines the set
			\begin{equation*}
				\mathfrak{Fix} _{\tilde{\mu }}  = \left\{ \bigl\vert \tilde{\alpha } \bigr\rangle _{f} \in \mathfrak{H} _{K} : F_{f} \left( g \right) \bigl\vert \tilde{\alpha } \bigr\rangle _{f} = \bigl\vert \tilde{\alpha } \bigr\rangle _{f} \right. \textnormal{for all} \left. g \in \mathds{Z} _{N} \right\} 
			\end{equation*}
			of points of $ \mathfrak{H} _{K} $ that are fixed by (\ref{trivial-dual-group-action}). As a consequence, as the cardinality of this set is precisely equal to $ \left\vert \ \ker \left( \mathfrak{f} \right) \ \right\vert $, it is not difficult to conclude that the result (\ref{formula-trivial}) corresponds to the same expression (\ref{degeneracy-trivial-dmn}) that defines the degree of degeneracy of the ground states of all the trivial $ D_{M} \left( \mathds{Z} _{N} \right) $ models.
		
			But given that we already know a lot about these trivial $ D^{K} \left( \mathds{Z} _{N} \right) $ models, it is time to analyse the main properties of the non-trivial $ D^{K} \left( \mathds{Z} _{N} \right) $ ones: i.e., of the $ D^{K} \left( \mathds{Z} _{N} \right) $ models where $ \mathfrak{f} $ is a non-trivial group homomorphism. And in order to start this analysis, it is interesting to take the $ D^{2} \left( \mathds{Z} _{2} \right) $ model as an example. After all, in view of what was stated by Theorem \ref{gcd-gallian}, there are two ways to define this model:
			\begin{itemize}
				\item[\textbf{[1st]}] one, which we presented in Subsubsection \ref{first-degeneracy-comment} by using a trivial group homomorphism, that has the same quasiparticles, with the same properties, as the $ D \left( \mathds{Z} _{2} \right) $ model; and
				\item[\textbf{[2nd]}] another that, because it needs to be defined by using a non-trivial group homomorphism, has vertex, face and edge operators represented by
				\begin{subequations} \label{abc-tck}
					\begin{align}
						A^{\prime } _{v,J} = \frac{1}{\left\vert G \right\vert } & \sum _{g \in \mathds{Z} _{N}} \left( -1 \right) ^{Jg} \cdot \prod _{\ell \in S_{v}} \left( \sigma ^{x} _{\ell } \right) ^{g} \ , \label{a-tck} \\
						B^{\prime } _{f,L} = \frac{1}{\left\vert G \right\vert } \sum _{g \in \mathds{Z} _{N}} & \left( -1 \right) ^{Lg} \cdot F_{f} \left( g \right) \prod _{\ell \in S_{f}} \left( \sigma ^{z} _{\ell } \right) ^{g} \quad \textnormal{and} \label{b-tck} \\
						C^{\prime } _{\ell ,\Lambda } = \frac{1}{\bigl\vert \tilde{S} \bigr\vert } \sum _{\tilde{\gamma } \in \tilde{S}} & \left( -1 \right) ^{\Lambda g} \cdot F_{\ell } \left( \tilde{\gamma } \right) \prod _{f \in S_{\ell }} \left( \sigma ^{x} _{f} \right) ^{\tilde{\gamma }} \label{c-tck}
					\end{align}
				\end{subequations}
				respectively, where $ F_{f} \left( g \right) $ and $ F_{\ell } \left( \tilde{\gamma } \right) $ cannot be identity matrices.
			\end{itemize}
			In this fashion, by noticing that Theorem \ref{proposition} guarantees that the only non-trivial group homomorphism $ \mathfrak{f} : \mathds{Z} _{2} \rightarrow \mathds{Z} _{2} $ that exists is 
			\begin{equation}
				\mathfrak{f} \left( 0 \right) = 0 \quad \textnormal{and} \quad \mathfrak{f} \left( 1 \right) = 1 \ , \label{group-iso-22}
			\end{equation}
			the fact that $ \mathsf{Im} \left( \mathfrak{f} \right) = G $ allows us to conclude that, in this \textbf{[2nd]} way, we have
			\begin{equation*}
				F_{f} \left( g \right) = \left( \sigma ^{z} _{f} \right) ^{g} \quad \textnormal{and} \quad  F_{\ell } \left( \tilde{\gamma } \right) = \left( \sigma ^{x} _{\ell } \right) ^{\tilde{\gamma }} \ .
			\end{equation*} 
			And by according to this picture, it is not difficult to recognize that, in this \textbf{[2nd]} way, the state
			\begin{equation*}
				\bigl\vert \tilde{\xi } ^{\left( \tilde{\alpha } \right) } _{0} \bigr\rangle = \prod _{\ell ^{\prime }} C_{\ell ^{\prime }} \prod _{v^{\prime }} A_{v^{\prime }} \left( \bigotimes _{\ell \in \mathcal{L} _{2}} \left\vert e \right\rangle \right) \otimes \left( \bigotimes _{f \in \mathcal{L} _{2}} \left\vert 0 \right\rangle \right) _{f \neq f^{\prime \prime }} \hspace*{-0.8cm} \otimes \left\vert \tilde{\alpha } \right\rangle _{f^{\prime \prime }}
			\end{equation*}
			with $ \tilde{\alpha } \neq 0 $ (i.e., with $ \tilde{\alpha } = 1 $) cannot be interpreted as a vacuum state. After all, as all the operators
			\begin{equation*}
				\left( \sigma ^{x,z} _{f} \right) ^{g} \quad \textnormal{and} \quad  \left( \sigma ^{x,z} _{\ell } \right) ^{\tilde{\gamma }} \ ,
			\end{equation*}				
			which compose the vertex, face and link operators (\ref{abc-tck}) (and, consequently, the Hamiltonian (\ref{H-qdmf})), produce quasiparticles in this model\footnote{See the comments made in the footnote on page \pageref{footnote-qft}.}, it is not difficult to recognize that
			\begin{equation*}
				\tilde{W} ^{\left( 1 , 0 \right) } _{f} = \sigma ^{x} _{f}
			\end{equation*}
			(which satisfies (\ref{dual-matter-quasiparticles-creation}) with $ L=1 $ and $ \Lambda = 0 $) produces a quasiparticle $ \tilde{Q} ^{\left( 1 , 0 \right) } $, throughout a permutation $ \left\vert 0 \right\rangle _{f} \leftrightarrow \left\vert 1 \right\rangle _{f} $, that can be detected by $ B^{\prime } _{f} $.
			
			\subsubsection{Are there \textquotedblleft confined\textquotedblright \hspace*{0.01cm} quasiparticles in the $ D^{K} \left( \mathds{Z} _{N} \right) $ models?}
			
				Another important point that deserves to be mentioned here is that, in addition to the \emph{group isomorphism} (\ref{group-iso-22}) defines a $ D^{2} \left( \mathds{Z} _{2} \right) $ model that does not have an algebraically degenerate ground state\footnote{That is, this group isomorphism defines a $ D^{2} \left( \mathds{Z} _{2} \right) $ model that has a set of vacuum states that are indexed only by $ \tilde{\alpha } = 0 $.}, it also makes $ C^{\prime } _{\ell } $ able to detect the pairs of quasiparticles $ \tilde{q} ^{\left( 1 , L , 1 \right) } $ that are produced by
				\begin{equation*}
					\tilde{W} ^{\left( 1 , L , 1 \right) } _{\ell } = \sigma ^{z} _{\ell } \circ \left( \sigma ^{x} _{\ell } \right) ^{L} \quad \textnormal{or} \quad \tilde{W} ^{\left( 1 , L , 1 \right) } _{\ell } = \left( \sigma ^{x} _{\ell } \right) ^{L} \circ \sigma ^{z} _{\ell } \ .
				\end{equation*} 
				That is, (\ref{group-iso-22}) causes $ C^{\prime } _{\ell } $ to be able to detect the same pair of quasiparticles that are detected individually by the operator $ A^{\prime } _{v} $. And why does this deserve to be mentioned here? Because this situation is entirely analogous, for instance, to that of the $ D_{2} \left( \mathds{Z} _{2} \right) $ model. After all, contrary to what happens in the $ D \left( \mathds{Z} _{2} \right) $ model, where it is possible to transport the electric quasiparticles without changing the energy of the system, this is not possible in this $ D^{2} \left( \mathds{Z} _{2} \right) $ model: whenever the transport of these electric quasiparticles occurs, the energy of the system increases when $ \mathfrak{f} $ is defined by (\ref{group-iso-22}). And since this increase is not welcome for the same reasons as outlined in Section \ref{QDM-review}, we need to do the same thing we did before: i.e., we need to ignore that the transport of these quasiparticles $ \tilde{q} ^{\left( 1 , L , 1 \right) } $ is mathematically possible and consider all of them to be confined.
				\begin{figure}[!t]
					\begin{center}
						\tikzstyle myBG=[line width=3pt,opacity=1.0]
						\newcommand{\drawLatticeLine}[2]
						{
							\draw[gray,very thick] (#1) -- (#2);
						}
						\newcommand{\drawLatticeLineFlex}[2]
						{
							\draw[->,gray,very thick,>=stealth] (#1) -- (#2);
						}
						\newcommand{\drawExcitedLine}[2]{
							\draw[black,ultra thick] (#1) -- (#2);
						}
						\newcommand{\graphLinesHorizontal}
						{
							\drawLatticeLineFlex{2,1}{2,3.1};
							\drawLatticeLine{2,3}{2,5};
							\drawLatticeLineFlex{4,1}{4,3.1};
							\drawLatticeLine{4,3}{4,5};
							\drawLatticeLineFlex{6,1}{6,3.1};
							\drawLatticeLine{6,3}{6,5};
							\drawLatticeLineFlex{8,1}{8,3.1};
							\drawLatticeLine{8,3}{8,5};
							\drawLatticeLineFlex{10,1}{10,3.1};
							\drawLatticeLine{10,3}{10,5};
							\drawLatticeLineFlex{12,1}{12,3.1};
							\drawLatticeLine{12,3}{12,5};
							\drawLatticeLineFlex{1,2}{3.1,2};
							\drawLatticeLineFlex{3,2}{5.1,2};
							\drawLatticeLineFlex{5,2}{7.1,2};
							\drawLatticeLineFlex{7,2}{9.1,2};
							\drawLatticeLineFlex{9,2}{11.1,2};
							\drawLatticeLine{11,2}{13,2};
							\drawLatticeLineFlex{1,4}{3.1,4};
							\drawLatticeLineFlex{3,4}{5.1,4};
							\drawLatticeLineFlex{5,4}{7.1,4};
							\drawLatticeLineFlex{7,4}{9.1,4};
							\drawLatticeLineFlex{9,4}{11.1,4};
							\drawLatticeLine{11,4}{13,4};
							\drawExcitedLine{2,4}{4,4};
						}
						\begin{tikzpicture}
							\graphLinesHorizontal;
							\draw[color=red!70,fill=yellow!70] (2,4) circle (1.4ex);
							\draw[color=red!70,fill=red!70] (2,4) circle (1.2ex);
							\node [] (2,4) at (2,4) {$ {\textcolor{white}{\boldsymbol{-}}} $};
							\draw[color=red!70,fill=yellow!70] (4,4) circle (1.4ex);
							\draw[color=red!70,fill=red!70] (4,4) circle (1.2ex);
							\node [] (4,4) at (4,4) {$ {\textcolor{white}{\boldsymbol{+}}} $};
							\draw[color=black,fill=green!70] (3,4) circle (0.7ex);
						\end{tikzpicture}
					\end{center} \bigskip
					\begin{center}
						\tikzstyle myBG=[line width=3pt,opacity=1.0]
						\newcommand{\drawLatticeLine}[2]
						{
							\draw[gray,very thick] (#1) -- (#2);
						}
						\newcommand{\drawLatticeLineFlex}[2]
						{
							\draw[->,gray,very thick,>=stealth] (#1) -- (#2);
						}
						\newcommand{\drawExcitedLine}[2]{
							\draw[black,ultra thick] (#1) -- (#2);
						}
						\newcommand{\drawExcitedPath}[2]{
							\draw[black,dashed,ultra thick] (#1) -- (#2);
						}
						\newcommand{\graphLinesHorizontal}
						{
							\drawLatticeLineFlex{2,1}{2,3.1};
							\drawLatticeLine{2,3}{2,5};
							\drawLatticeLineFlex{4,1}{4,3.1};
							\drawLatticeLine{4,3}{4,5};
							\drawLatticeLineFlex{6,1}{6,3.1};
							\drawLatticeLine{6,3}{6,5};
							\drawLatticeLineFlex{8,1}{8,3.1};
							\drawLatticeLine{8,3}{8,5};
							\drawLatticeLineFlex{10,1}{10,3.1};
							\drawLatticeLine{10,3}{10,5};
							\drawLatticeLineFlex{12,1}{12,3.1};
							\drawLatticeLine{12,3}{12,5};
							\drawLatticeLineFlex{1,2}{3.1,2};
							\drawLatticeLineFlex{3,2}{5.1,2};
							\drawLatticeLineFlex{5,2}{7.1,2};
							\drawLatticeLineFlex{7,2}{9.1,2};
							\drawLatticeLineFlex{9,2}{11.1,2};
							\drawLatticeLine{11,2}{13,2};
							\drawLatticeLineFlex{1,4}{3.1,4};
							\drawLatticeLineFlex{3,4}{5.1,4};
							\drawLatticeLineFlex{5,4}{7.1,4};
							\drawLatticeLineFlex{7,4}{9.1,4};
							\drawLatticeLineFlex{9,4}{11.1,4};
							\drawLatticeLine{11,4}{13,4};
							\drawExcitedLine{4,4}{4,2};
							\drawExcitedLine{4,2}{12,2};
							\drawExcitedLine{12,2}{12,4};
							\drawExcitedLine{12,4}{8,4};
						}
						\begin{tikzpicture}
							\graphLinesHorizontal;
							\draw[color=red!70,fill=yellow!70] (2,4) circle (1.4ex);
							\draw[color=red!70,fill=red!70] (2,4) circle (1.2ex);
							\node [] (2,4) at (2,4) {$ {\textcolor{white}{\boldsymbol{-}}} $};
							\draw[color=red!70,fill=white] (4,4) circle (1.4ex);
							\draw[color=red!70,fill=yellow!70] (8,4) circle (1.4ex);
							\draw[color=red!70,fill=red!70] (8,4) circle (1.2ex);
							\node [] (8,4) at (8,4) {$ {\textcolor{white}{\boldsymbol{+}}} $};
							\draw[color=black,fill=green!70] (3,4) circle (0.7ex);
							\draw[color=black,fill=green!70] (4,3) circle (0.7ex);
							\draw[color=black,fill=green!70] (5,2) circle (0.7ex);
							\draw[color=black,fill=green!70] (7,2) circle (0.7ex);
							\draw[color=black,fill=green!70] (9,2) circle (0.7ex);
							\draw[color=black,fill=green!70] (11,2) circle (0.7ex);
							\draw[color=black,fill=green!70] (12,3) circle (0.7ex);
							\draw[color=black,fill=green!70] (11,4) circle (0.7ex);
							\draw[color=black,fill=green!70] (9,4) circle (0.7ex);
						\end{tikzpicture}
					\end{center}
					\caption{\label{confinamento-eletrons} Here, we see a kind of replica of Figures \ref{e-transporte} and \ref{e-general-transporte} highlighting the situation of the same electric quasiparticles, at two different times, but now in the $ D^{2} \left( \mathds{Z} _{2} \right) $ model. In the first instant $ t_{1} $ (above) we have a pair of quasiparticles $ \tilde{q} ^{\left( 1 , 0 , 1 \right) } _{+} $ and $ \tilde{q} ^{\left( 1 , 0 , 1 \right) } _{-} $ (red outlined and purposely indexed with the \textquotedblleft $ + $\textquotedblright \hspace*{0.01cm} and \textquotedblleft $ - $\textquotedblright \hspace*{0.01cm} symbols respectively), which were produced by the action of a single operator $ \tilde{W} ^{\left( 1 , 0 , 1 \right) } _{\ell } $. Note that, since the production of this pair is detected by $ C^{\prime } _{\ell , 0} $, the green dot corresponds to the unique vacuum violation detected by this vertex operator. Now, in the second instant $ t_{2} > t_{1} $ (below) we have these same quasiparticles after one of them has been transported away from the other due to the action of operators $ \tilde{W} ^{\left( 1 , 0 , 1 \right) } _{\ell ^{\prime }} $ on all the edges highlighted in black colour. In this latter case, we have new eight green dots: one for each edge involved in this transport, making clear the linearity related to the growth of the system energy in this transport.}
				\end{figure}
				
				In view of this \textquotedblleft confinement\textquotedblright , it is not wrong to say that this $ D^{2} \left( \mathds{Z} _{2} \right) $ model, where $ \mathfrak{f} $ is a group isomorphism, has properties that are dual to those of the $ D_{2} \left( \mathds{Z} _{2} \right) $ model. After all, it is quite clear, for instance, that
				\begin{itemize}
					\item while, in the $ D_{2} \left( \mathds{Z} _{2} \right) $ model, the \textquotedblleft confined\textquotedblright \hspace*{0.01cm} quasiparticles are detected by the face operator $ B_{f} $,
					\item here, in the $ D^{2} \left( \mathds{Z} _{2} \right) $ model, the \textquotedblleft confined\textquotedblright \hspace*{0.01cm} quasiparticles are detected by the vertex operator $ A^{\prime } _{v} $, which is dual to $ B_{f} $.
				\end{itemize}
				However, there is, at least, one aspect of this $ D^{2} \left( \mathds{Z} _{2} \right) $ model that seems to spoil this duality. What is this aspect? It is the aspect that is related precisely to the fact that these \textquotedblleft confined\textquotedblright \hspace*{0.01cm} quasiparticles $ \tilde{q} ^{\left( 1 , 0 , 1 \right) } $ are not detected by $ B^{\prime } _{f,L} $. And why does this seem to spoil the duality between the $ D_{2} \left( \mathds{Z} _{2} \right) $ and $ D^{2} \left( \mathds{Z} _{2} \right) $ models? Because, as these quasiparticles are not detected by any of the operators that measure the (\textquotedblleft fake\textquotedblright ) holonomies around the lattice faces, this means that their production cannot be associated with any type of local deformation of $ \mathcal{L} _{2} $. In this way, by noting that the action of an operator
				\begin{equation*}
					O_{1} \left( \overline{\gamma } \right) = \prod _{\ell ^{\prime } \in \overline{\gamma }} \tilde{W} ^{\left( 1 , 0 , 1 \right) } _{\ell ^{\prime }} \ ,
				\end{equation*} 
				on a set of edges that form a non-contractile closed path $ \overline{\gamma } $, does not have the slightest importance for the determination of vacuum states that are topologically independent of\footnote{Observe that (\ref{dk22-ground-state}) is just a more streamlined way of writing the same vacuum state (\ref{dk-vacuo-1}).} [\citen{mf-pedagogical}]
				\begin{equation}
					\bigl\vert \tilde{\xi } ^{\left( 0 \right) } _{0} \bigr\rangle = \prod _{\ell ^{\prime }} C_{\ell ^{\prime }} \prod _{v^{\prime }} A_{v^{\prime }} \left( \bigotimes _{\ell \in \mathcal{L} _{2}} \left\vert e \right\rangle \right) \otimes \left( \bigotimes _{f \in \mathcal{L} _{2}} \left\vert 0 \right\rangle \right) \ , \label{dk22-ground-state}
				\end{equation}
				the fact that the quasiparticles $ \tilde{q} ^{\left( 1 , 0 , 1 \right) } $ are \textquotedblleft confined\textquotedblright \hspace*{0.01cm} does not prevent the $ D^{2} \left( \mathds{Z} _{2} \right) $ ground state from depending on the first homotopy group $ \pi _{1} \left( \mathcal{M} _{2} \right) $. In other words, the ground state of this $ D^{2} \left( \mathds{Z} _{2} \right) $ model, where $ \mathfrak{f} $ is given by (\ref{group-iso-22}), is made up of all the vacuum states
				\begin{equation*}
					\bigl\vert \xi ^{\left( 0 , \vec{\lambda } \right) } _{0} \bigr\rangle  = \prod ^{s} _{p=1} \left[ \tilde{O} _{1} \left( \overline{\gamma } ^{\ast } _{p} \right) \right] ^{\lambda _{p}} \bigl\vert \tilde{\xi } ^{\left( 0 \right) } _{0} \bigr\rangle \ ,
				\end{equation*}
				which are topologically independent of each other due to the non-contractility of $ \overline{\gamma } ^{\ast } _{p} $. 
			
			\subsubsection{The $ D^{N} \left( \mathds{Z} _{N} \right) $ models as other examples}
			
				Given all that we have just understood about the $ D^{2} \left( \mathds{Z} _{2} \right) $ model, it is also not difficult to conclude that all the other $ D^{N} \left( \mathds{Z} _{N} \right) $ models where $ \mathfrak{f} $ is a group isomorphism (i.e., where $ \mathfrak{f} $ is a group homomorphism (\ref{regulator-homomorphism}) with $ N = K $ and $ \mathsf{n} = 1 $), have the same properties listed in the last two Subsubsections. After all, since this group isomorphism requires that
				\begin{equation*}
					F_{f} \left( g \right) = \bigl( \tilde{Z} _{f} \bigr) ^{g} \quad \textnormal{and} \quad F_{\ell } \left( \tilde{\gamma } \right) = \left( X_{\ell } \right) ^{\tilde{\gamma }} \ ,
				\end{equation*} 
				we can conclude that:
				\begin{itemize}
					\item[\textbf{I.}] All the operators
					\begin{equation*}
						\left( X_{\ell } \right) ^{g} \ , \ \ \left( Z_{\ell } \right) ^{g} \ , \ \ \bigl( \tilde{X} _{\ell } \bigr) ^{g} \quad \textnormal{and} \quad \bigl( \tilde{Z} _{\ell } \bigr) ^{g} \ ,
					\end{equation*}
					which compose the vertex, face and link operators (\ref{abc-tc}) (and, consequently, the Hamiltonian (\ref{H-qdmf})), produce quasiparticles in this model.
					
					\item[\textbf{II.}] Since the operators
					\begin{equation*}
						\tilde{W} ^{\left( g , 0 \right) } _{f} = \bigl( \tilde{X} _{\ell } \bigr) ^{g}
					\end{equation*}
					(which satisfy (\ref{dual-matter-quasiparticles-creation}) with $ L=g $ and $ \Lambda = 0 $) can make all the changes $ \left\vert \tilde{\alpha } ^{\prime } \right\rangle _{f} \leftrightarrow \left\vert \tilde{\alpha } ^{\prime \prime } \right\rangle _{f} $ that are allowed between the elements of $ \mathcal{B} _{f} $, a state
					\begin{equation*}
						\bigl\vert \tilde{\xi } ^{\left( \tilde{\alpha } \right) } _{0} \bigr\rangle = \prod _{\ell ^{\prime }} C_{\ell ^{\prime }} \prod _{v^{\prime }} A_{v^{\prime }} \left( \bigotimes _{\ell \in \mathcal{L} _{2}} \left\vert e \right\rangle \right) \otimes \left( \bigotimes _{f \in \mathcal{L} _{2}} \left\vert 0 \right\rangle \right) _{f \neq f^{\prime \prime }} \hspace*{-0.8cm} \otimes \left\vert \tilde{\alpha } \right\rangle _{f^{\prime \prime }} \ , 
					\end{equation*}
					with $ \tilde{\alpha } \neq 0 $, cannot be interpreted as a vacuum state.
					
					\item[\textbf{III.}] All the quasiparticles $ \tilde{q} ^{\left( 0 , L , 0 \right) } $, which are produced by an operator
					\begin{equation*}
						\tilde{W} ^{\left( 0 , L , 0 \right) } _{\ell } = \left( X_{\ell } \right) ^{L} \ ,
					\end{equation*} 
					can be transported without increasing/decreasing the energy of the system, while the others $ \tilde{q} ^{\left( J , L , \Lambda \right) } $, which are produced by any operator
					\begin{equation*}
						\tilde{W} ^{\left( J , L , \Lambda \right) } _{\ell } = \left( Z_{\ell } \right) ^{J} \circ \left( X_{\ell } \right) ^{L} \quad \textnormal{or} \quad \tilde{W} ^{\left( J , L , \Lambda \right) } _{\ell } = \left( X_{\ell } \right) ^{L} \circ \left( Z_{\ell } \right) ^{J} 
					\end{equation*} 
					with $ J \neq 0 $, should be regarded as \textquotedblleft confined\textquotedblright .
					
					\item[\textbf{IV.}] As a consequence of items \textbf{II} and \textbf{III}, the ground state of these $ D^{N} \left( \mathds{Z} _{N} \right) $ models are made up of
					\begin{equation*}
						\bigl\vert \xi ^{\left( 0 , \vec{\lambda } , \vec{L} \right) } _{0} \bigr\rangle  = \prod ^{s} _{p=1} \left[ \tilde{O} _{L_{p}} \left( \overline{\gamma } ^{\ast } _{p} \right) \right] ^{\lambda _{p}} \prod _{\ell ^{\prime }} C_{\ell ^{\prime }} \prod _{v^{\prime }} A_{v^{\prime }} \left( \bigotimes _{\ell \in \mathcal{L} _{2}} \left\vert e \right\rangle \right) \otimes \left( \bigotimes _{f \in \mathcal{L} _{2}} \left\vert 0 \right\rangle \right) 
					\end{equation*}
					since all these vacuum states are topologically independent of each other due.
				\end{itemize}
				
				However, something that becomes quite clear from Theorems \ref{gcd-gallian} and \ref{proposition} is that, except that $ N $ is a prime number, all these $ D^{N} \left( \mathds{Z} _{N} \right) $ models can also be defined by using an $ \mathfrak{f} $ that is neither a trivial group homomorphism nor a group isomorphism. This is, for example, the case of the $ D^{4} \left( \mathds{Z} _{4} \right) $ model that, in addition to being able to be defined by using these two group homomorphisms, can also be defined by using
				\begin{equation}
					\mathfrak{f} \left( 0 \right) = \mathfrak{f} \left( 2 \right) = 0 \ \ \textnormal{and} \ \ \mathfrak{f} \left( 1 \right) = \mathfrak{f} \left( 3 \right) = 2 \ . \label{non-trivial-44}
				\end{equation}
				
			\subsubsection{And what may happen when $ \mathfrak{f} $ is not a group isomorphism?}
			
				Although Theorem \ref{proposition} shows us that
				\begin{equation}
					\mathfrak{f} \left( 0 \right) = 0 \ , \ \ \mathfrak{f} \left( 1 \right) = 3 \ , \ \ \mathfrak{f} \left( 2 \right) = 2 \ \ \textnormal{and} \ \ \mathfrak{f} \left( 3 \right) = 1 \label{outro}
				\end{equation}
				is another non-trivial group homomorphism that can also be used to define this $ D^{4} \left( \mathds{Z} _{4} \right) $ model, the group homomorphism (\ref{non-trivial-44}) seems to be more interesting because
				\begin{equation*}
					\left\vert \ \ker \left( \mathfrak{f} \right) \ \right\vert > 1 \quad \textnormal{and} \quad \left\vert \ \textnormal{Im} \left( \mathfrak{f} \right) \ \right\vert = 2 \ .
				\end{equation*}
				After all, besides (\ref{outro}) is not very different from the group isomorphism $ \mathfrak{f} : \mathds{Z} _{4} \rightarrow \mathds{Z} _{4} $\footnote{Note that, as this group isomorphism is defined by
				\begin{equation}
					\mathfrak{f} \left( 0 \right) = 0 \ , \ \ \mathfrak{f} \left( 1 \right) = 1 \ , \ \ \mathfrak{f} \left( 2 \right) = 2 \ \ \textnormal{and} \ \ \mathfrak{f} \left( 3 \right) = 3 \ , \label{iso-44}
				\end{equation}
				the only difference between it and (\ref{outro}) can be justified in terms of a permutation.}, one of the things that this (\ref{non-trivial-44}), where $ \ker \left( \mathfrak{f} \right) = \left\{ 0 , 2 \right\} $, allows us to see is that the ground state of this model is defined by
				\begin{equation*}
					\bigl\vert \xi ^{\left( \tilde{\alpha } , \vec{\lambda } , \vec{L} \right) } _{0} \bigr\rangle  = \prod ^{s} _{p=1} \left[ \tilde{O} _{L_{p}} \left( \overline{\gamma } ^{\ast } _{p} \right) \right] ^{\lambda _{p}} \prod _{\ell ^{\prime }} C_{\ell ^{\prime }} \prod _{v^{\prime }} A_{v^{\prime }} \left( \bigotimes _{\ell \in \mathcal{L} _{2}} \left\vert e \right\rangle \right) \otimes \left( \bigotimes _{f \in \mathcal{L} _{2}} \left\vert 0 \right\rangle \right) _{f \neq f^{\prime \prime }} \hspace*{-0.8cm} \otimes \left\vert \tilde{\alpha } \right\rangle _{f^{\prime \prime }} \ ,
				\end{equation*} 
				where $ \tilde{\alpha } \in \ker \left( \mathfrak{f} \right) $. In other words, we are faced with a $ D^{4} \left( \mathds{Z} _{4} \right) $ model that has an algebraically degenerate ground state, but where this algebraic degeneracy is neither a maximum nor a minimum.
				
				Another interesting aspect of this $ D^{4} \left( \mathds{Z} _{4} \right) $ model, which is defined by using (\ref{non-trivial-44}), is related to the fact that				
				\begin{equation*}
					F_{f} \left( g \right) = \bigl( \tilde{Z} ^{2} _{f} \bigr) ^{g} \quad \textnormal{and} \quad  F_{\ell } \left( \tilde{\gamma } \right) = \left( X^{2} _{\ell } \right) ^{\tilde{\gamma }} \ . 
				\end{equation*} 
				And why is this interesting? Because, when we substitute these matrices into (\ref{abc-dk}), it is not difficult to see that not all quasiparticles, which are detected individually by the operator $ A^{\prime } _{v} $, can be considered as \textquotedblleft confined\textquotedblright . And how can we see it? By noting that
				\begin{itemize}
					\item the quasiparticles, which are detected by the operator $ A^{\prime } _{v} $, are produced by
					\begin{equation*}
						\tilde{W} ^{\left( g , 0 , \Lambda \right) } _{\ell } = \left( Z_{\ell } \right) ^{g} \ , \ \ \textnormal{and}
					\end{equation*}
					\item the link operators can be represented by
					\begin{equation*}
						C^{\prime } _{\ell ,\Lambda } = \frac{1}{4} \sum _{\tilde{\gamma } \in \tilde{S}} \left( i \right) ^{\Lambda \tilde{\gamma }} \cdot \bigl( \tilde{X} ^{\dagger } _{f_{1}} \bigr) ^{\tilde{\gamma }} \otimes \left( X^{2} _{\ell } \right) ^{\tilde{\gamma }} \otimes \bigl( \tilde{X} _{f_{2}} \bigr) ^{\tilde{\gamma }} \ ,
					\end{equation*}
					where $ i = e^{i \left( 2 \pi / 4 \right) } $ is the generator of the matter group.
				\end{itemize}
				After all, as the generator of the gauge group is also equal to $ i $ in this case where $ N = 4 $ and, therefore, the operators (\ref{matrices-xz}) are such that 
				\begin{equation*}
					Z^{g} X^{h} = i^{ \ \left[ \left( g + h \right) \hspace*{0.04cm} \textnormal{mod} \left( 4 \right) \right] } X^{h} Z^{g} \ ,
				\end{equation*}
				it is not difficult to conclude that all the quasiparticles $ \tilde{q} ^{\left( 2 , 0 , 0 \right) } $, which are produced by an operator $ \tilde{W} ^{\left( 2 , 0 , 0 \right) } _{\ell } $, are \textquotedblleft unconfined\textquotedblright \hspace*{0.01cm} (i.e., these quasiparticles can be transported without increasing/decreasing the energy of the system) because
				\begin{equation*}
					Z^{2} X^{2} = i^{\ \left[ 4 \hspace*{0.04cm} \textnormal{mod} \left( 4 \right) \right] } X^{2} Z^{2} = X^{2} Z^{2} \ .
				\end{equation*}
				Consequently, as there are $ N-1 $ quasiparticles $ \tilde{q} ^{\left( 2 , L , 0 \right) } $ that are produced by the operators 
				\begin{equation*}
					\tilde{W} ^{\left( g , L , 0 \right) } _{\ell } = \left( Z_{\ell } \right) ^{g} \circ \left( X_{\ell } \right) ^{L} \quad \textnormal{or} \quad \tilde{W} ^{\left( g , L , 0 \right) } _{\ell } = \left( X_{\ell } \right) ^{L} \circ \left( Z_{\ell } \right) ^{g}
				\end{equation*}
				through a fusion
				\begin{equation*}
					\tilde{q} ^{\left( 2 , 2 , 0 \right) } = \tilde{q} ^{\left( 2 , 0 , 0 \right) } \times \tilde{q} ^{\left( 0 , 2 , 0 \right) } = \tilde{q} ^{\left( 0 , 2 , 0 \right) } \times \tilde{q} ^{\left( 2 , 0 , 0 \right) }
				\end{equation*}				
				between the quasiparticles $ \tilde{q} ^{\left( 2 , 0 , 0 \right) } $ and $ \tilde{q} ^{\left( 0 , L , 0 \right) } $, all these quasiparticles $ \tilde{q} ^{\left( 2 , L , 0 \right) } $ are also interpreted as \textquotedblleft unconfined\textquotedblright \hspace*{0.01cm} since all $ \tilde{q} ^{\left( 0 , L , 0 \right) } $ are also \textquotedblleft unconfined\textquotedblright . 
		
		\subsection{The ground state degeneracy and the classifiability of the $ D^{K} \left( \mathds{Z} _{N} \right) $ models}
		
			In view of what we have just seen in this last Subsubsection, one thing that you, the reader, might be wondering is: \emph{is there some rule to determine when the $ D^{K} \left( \mathds{Z} _{N} \right) $ models have quasiparticles $ \tilde{q} ^{\left( J , L , 0 \right) } $ that are \textquotedblleft unconfined\textquotedblright ?} And in order for us to answer this question, it is very interesting to pay attention, for instance, to the trivial $ D^{K} \left( \mathds{Z} _{N} \right) $ models, because the trivial group homomorphisms $ \mathfrak{f} : \mathds{Z} _{K} \rightarrow \mathds{Z} _{N} $ always map every element of $ \mathds{Z} _{K} $ to the identity element of $ \mathds{Z} _{N} $. And since the definition of link operators makes it clear that it is precisely the result of this mapping that needs to change the gauge fields on which these operators act, it is also very clear that, when $ \mathfrak{f} \left( \tilde{\alpha } \right) = e $,
			\begin{itemize}
				\item these operators become \textquotedblleft blind\textquotedblright \hspace*{0.01cm} to the presence of the electric quasiparticles, and (therefore)
				\item the electric quasiparticles become \textquotedblleft unconfined\textquotedblright .
			\end{itemize}
			
			Note that, although the $ D^{4} \left( \mathds{Z} _{4} \right) $ model discussed above was not defined by using a trivial group homomorphism, (\ref{non-trivial-44}) places this model in a situation that, in some way, is comparable to this one. After all, unlike the (\ref{outro}) and (\ref{iso-44}), this group homomorphism (\ref{non-trivial-44}) defines two distinct equivalence classes: videlicet,
			\begin{equation*}
				\left[ 0 \right] = \left\{ a \in \mathds{Z} _{N} : a \equiv 0 \ \textnormal{mod} \left( 4 \right) \right\} \quad \textnormal{and} \quad \left[ 2 \right] = \left\{ a \in \mathds{Z} _{N} : a \equiv 2 \ \textnormal{mod} \left( 4 \right) \right\}
			\end{equation*}
			since (\ref{non-trivial-44}) is nothing more than the same group homomorphism (\ref{regulator-homomorphism}) where $ \mathsf{n} = 2 $. And why is it important to pay attention to the fact that (\ref{non-trivial-44}) defines these two distinct equivalence classes? Because (\ref{non-trivial-44}) is just one example of a group homomorphism that can do this: other functions (\ref{regulator-homomorphism}), which can also do this, can be identified whenever $ K $ and $ N $ are two even numbers. How? By considering that $ N = 2 \mathsf{n} $: after all, as $ K $ is also an even number and, therefore, $ \mathsf{n}K $ will always be divisible by $ N $, the Theorem \ref{proposition} guarantees the existence of the group homomorphism
			\begin{equation}
				\mathfrak{f} \bigl( \bigl[ \tilde{\alpha } \bigr] \bigr) \ = \ \bigl[ \mathsf{n} \tilde{\alpha } \bigr] \quad , \label{two-classes}
			\end{equation}
			which can be used to define two distinct equivalence classes
			\begin{equation*}
				\left[ 0 \right] = \left\{ a \in \mathds{Z} _{2 \mathsf{n}} : a \equiv 0 \ \textnormal{mod} \left( 2 \mathsf{n} \right) \right\} \quad \textnormal{and} \quad \left[ \mathsf{n} \right] = \left\{ a \in \mathds{Z} _{2 \mathsf{n}} : a \equiv \mathsf{n} \ \textnormal{mod} \left( 2 \mathsf{n} \right) \right\} \ .
			\end{equation*}
			The main consequence of this is that, whenever we define a $ D^{K} \left( \mathds{Z} _{2 \mathsf{n}} \right) $ model, where $ K $ is an even number, by using this group homomorphism (\ref{two-classes}), it leads us to
			\begin{equation*}
				F_{f} \left( g \right) = \bigl( \tilde{Z} ^{\mathsf{n}} _{f} \bigr) ^{g} \quad \textnormal{and} \quad F_{\ell } \left( \tilde{\gamma } \right) = \left( X^{\mathsf{n}} _{\ell } \right) ^{\tilde{\gamma }} \ , 
			\end{equation*} 
			and, therefore, all the quasiparticles produced by
			\begin{equation*}
				\tilde{W} ^{\left( \mathsf{n} , L , 0 \right) } _{\ell } = \left( Z_{\ell } \right) ^{\mathsf{n}} \circ \left( X_{\ell } \right) ^{L} \quad \textnormal{or} \quad \tilde{W} ^{\left( \mathsf{n} , L , 0 \right) } _{\ell } = \left( X_{\ell } \right) ^{L} \circ \left( Z_{\ell } \right) ^{\mathsf{n}}
			\end{equation*}
			will never be detected by the $ D^{K} \left( \mathds{Z} _{2 \mathsf{n}} \right) $ link operators as long as
			\begin{equation*}
				Z^{\mathsf{n}} X^{\mathsf{n}} = i^{ \ \left[ \left( \mathsf{n} + \mathsf{n} \right) \hspace*{0.04cm} \textnormal{mod} \left( 2 \mathsf{n} \right) \right] } X^{\mathsf{n}} Z^{\mathsf{n}} = X^{\mathsf{n}} Z^{\mathsf{n}} \ .
			\end{equation*}
			In this fashion, as $ \left\vert \ \textnormal{Im} \left( \mathfrak{f} \right) \ \right\vert $ is equal to the number of equivalence classes that $ \mathfrak{f} $ defines, this explains why we take, as an example, this $ D^{4} \left( \mathds{Z} _{4} \right) $ model where $ \left\vert \ \textnormal{Im} \left( \mathfrak{f} \right) \ \right\vert = 2 $. That is, as much as we have highlighted the fact that $ \left\vert \ \ker \left( \mathfrak{f} \right) \ \right\vert > 1 $, the necessary condition for the existence of \textquotedblleft unconfined\textquotedblright \hspace*{0.01cm} quasiparticles in the $ D^{K} \left( \mathds{Z} _{N} \right) $ models is that $ \left\vert \ \textnormal{Im} \left( \mathfrak{f} \right) \ \right\vert \leqslant 2 $.
			
			\subsubsection{\label{classification}What can we say about the quasiparticles that are produced by manipulating matter fields?}
			
				Notwithstanding, the information that $ \left\vert \ \ker \left( \mathfrak{f} \right) \ \right\vert > 1 $ is still relevant because it is precisely this $ \left\vert \ \ker \left( \mathfrak{f} \right) \ \right\vert $ that computes the number of quasiparticles $ \tilde{Q} ^{\left( J , \Lambda \right) } $ , which are produced by manipulating matter fields, that are not able to locally deform $ \mathcal{L} _{2} $. And although we still have not said a word about all these (matter) quasiparticles $ \tilde{Q} ^{\left( J , \Lambda \right) } $ , they are not as surprising as the quasiparticles $ Q^{\left( J , \Lambda \right) } $ of the $ D_{M} \left( \mathds{Z} _{N} \right) $ models: after all, as all the operators
				\begin{equation}
					\tilde{W} ^{\left( L , \Lambda \right) } _{f} = \bigl( \tilde{X} _{f} \bigr) ^{L} \circ \bigl[ F_{f} \left( g \right) \bigr] ^{\Lambda } \quad \textnormal{and} \quad \tilde{W} ^{\left( L , \Lambda \right) } _{f} = \bigl[ F_{f} \left( g \right) \bigr] ^{\Lambda } \circ \bigl( \tilde{X} _{f} \bigr) ^{L} \label{dual-matter-operators-excitations}
				\end{equation}
				that produce them can be identified in the expressions of the $ D^{K} \left( \mathds{Z} _{N} \right) $ face and link operators, it is not difficult to conclude that
				\begin{itemize}
					\item $ \tilde{Q} ^{\left( J , \Lambda \right) } $ have Abelian fusion rules, because we always have that
					\begin{equation*}
						F_{f} \left( g \right) = \bigl( \tilde{Z} ^{\mathsf{n}} _{f} \bigr) ^{g}
					\end{equation*}
					where $ \mathsf{n} $ takes the values that satisfy Theorem \ref{proposition}, and
					\item the action of these operators (\ref{dual-matter-operators-excitations}), with $ \Lambda = 0 $, is sufficient to perform transitions between/among the $ D^{K} \left( \mathds{Z} _{N} \right) $ vacuum states.
				\end{itemize}
				Note that, just by looking at the $ D^{K} \left( \mathds{Z} _{N} \right) $ vacuum states
				\begin{equation*}
					\bigl\vert \xi ^{\left( \tilde{\alpha } , \vec{\lambda } , \vec{L} \right) } _{0} \bigr\rangle  = \prod ^{s} _{p=1} \left[ \tilde{O} _{L_{p}} \left( \overline{\gamma } ^{\ast } _{p} \right) \right] ^{\lambda _{p}} \prod _{\ell ^{\prime }} C_{\ell ^{\prime }} \prod _{v^{\prime }} A_{v^{\prime }} \left( \bigotimes _{\ell \in \mathcal{L} _{2}} \left\vert e \right\rangle \right) \otimes \left( \bigotimes _{f \in \mathcal{L} _{2}} \left\vert 0 \right\rangle \right) _{f \neq f^{\prime \prime }} \hspace*{-0.8cm} \otimes \left\vert \tilde{\alpha } \right\rangle _{f^{\prime \prime }} \ ,
				\end{equation*} 
				this sufficient condition makes it very clear that all the (matter) quasiparticles, which are not able to locally deform $ \mathcal{L} _{2} $, are produced by the operators $ \tilde{W} ^{\left( L , 0 \right) } _{f} $ that reduce the \textquotedblleft fake holonomy\textquotedblright \hspace*{0.01cm} (\ref{fake-holonomy}) to the true holonomy (\ref{holonomy}).
				
				Based on these findings, it is not difficult to conclude that all these $ D^{K} \left( \mathds{Z} _{N} \right) $ models can be completely classified in terms of an ordered $ 3 $-tuple $ \left( N , K , \mathsf{n} \right) $, as follows: \label{classification-page}
				\begin{itemize}
					\item[\textbf{A.}] $ \left( N , K , 0 \right) $
					
					All the $ D^{K} \left( \mathds{Z} _{N} \right) $ models, which are characterized by an ordered $ 3 $-tuple where $ \mathsf{n} = 0 $, have ground states with an algebraic degeneracy $ \left\vert \ \ker \left( \mathfrak{f} \right) \ \right\vert $ that is maximal (i.e., it is equal to $ K $). As a consequence of this maximality, all the manipulations, which can be done on the matter fields by using only the operators that make up the $ D^{K} \left( \mathds{Z} _{N} \right) $ Hamiltonian, do not (locally) deform $ \mathcal{L} _{2} $ and, therefore, do not change the energy of the system. In this way, it is valid to affirm that all these models, with $ \left( N , K , 0 \right) $, have the same quasiparticles, with the same properties, as the $ D \left( \mathds{Z} _{N} \right) $ models.
					
					\item[\textbf{B.}] $ \left( N , N , N \right) $
					
					When $ \mathsf{n} = N $, all the $ D^{K} \left( \mathds{Z} _{N} \right) $ models have an algebraic degeneracy $ \left\vert \ \ker \left( \mathfrak{f} \right) \ \right\vert $ that is minimal (i.e., it is equal to $ 1 $). And as one of the consequence of this minimality is that $ \left\vert \ \textnormal{Im} \left( \mathfrak{f} \right) \ \right\vert = N $, we can affirm that, although these models house all the $ D \left( \mathds{Z} _{N} \right) $ quasiparticles among their energy excitations, all the quasiparticles that are detected by the $ D^{K} \left( \mathds{Z} _{N} \right) $ vertex operators are \textquotedblleft confined\textquotedblright . Observe that, since this minimality also implies that all the quasiparticles $ \tilde{Q} ^{\left( L , \Lambda \right) } _{f} $, where $ \left( L , \Lambda \right) \neq \left( 0 , 0 \right) $, are detectable by the $ D^{K} \left( \mathds{Z} _{N} \right) $ face and link operators, the ground state of all these models can only be indexed by $ \tilde{\alpha } = 0 $.
					
					\item[\textbf{C.}] $ \left( N , K , \mathsf{n} \right) $
					
					In this case, where this ordered $ 3 $-tuple is different from $ \left( N , K , 0 \right) $ or $ \left( N , N , N \right) $, it is possible to affirm that the $ D^{K} \left( \mathds{Z} _{N} \right) $ models may have intermediate properties between those of \textbf{A} and \textbf{B}. After all, although these models may be perfectly defined by using group homomorphisms that, for instance, confine all the quasiparticles $ \tilde{q} ^{\left( J , L , \Lambda \right) } $ with $ J \neq 0 $, whenever $ K $ is an even number and $ N = 2 \mathsf{n} $ we can also define such models by using (\ref{two-classes}): i.e., whenever $ K $ is an even number and $ N = 2 \mathsf{n} $, we can define the $ D^{K} \left( \mathds{Z} _{2 \mathsf{n}} \right) $ models where all the quasiparticles $ \tilde{q} ^{\left( J , L , 0 \right) } $, with $ J \in \left[ 0 \right] $, are unconfined. As a consequence of this partial deconfinement, the algebraic degeneracy of the $ D^{K} \left( \mathds{Z} _{2 \mathsf{n}} \right) $ ground state is neither a maximum nor a minimum because, for instance, all their vacuum states are indexed by $ 1 < \left\vert \ \ker \left( \mathfrak{f} \right) \ \right\vert < K $ values of $ \tilde{\alpha } $.
				\end{itemize}
				
				\subsubsection{Does the degree of degeneracy of the $ D^{K} \left( \mathds{Z} _{N} \right) $ ground state depend on the (sub)set $ \textnormal{Im} \left( \mathfrak{f} \right) $?}
				
					Note that, as there is no way to manipulate the matter fields when $ K = 1 $, the Hamiltonian of any trivial $ D^{1} \left( \mathds{Z} _{N} \right) $ model (i.e., of any $ D^{K} \left( \mathds{Z} _{N} \right) $ model that is classified as $ \left( N , 1 , 0 \right) $) is given by				
					\begin{equation*}
						\left. H_{D^{K} \left( G \right) } \right\vert _{K=1} = - \mathcal{J} ^{\prime }_{A} \sum _{v  \in \mathcal{L} _{2}} A_{v} - \mathcal{J} ^{\prime } _{B} \sum _{f \in \mathcal{L} _{2}} \ B_{f} - \mathcal{J} ^{\prime } _{C} \sum _{\ell \in \mathcal{L} _{2}} \mathds{1} _{f_{1}} \otimes \mathds{1} _{\ell } \otimes \mathds{1} _{f_{2}} \ ,
					\end{equation*}				
					which only reinforces the existence of a correspondence principle between the $ D^{K} \left( \mathds{Z} _{N} \right) $ and $ D \left( \mathds{Z} _{N} \right) $ models because
					\begin{equation*}
						H_{D \left( \mathds{Z} _{N} \right) } - \left. H_{D^{K} \left( G \right) } \right\vert _{K=1} = \left( \mathcal{J} ^{\prime } _{C} N_{\ell } \right) \cdot \mathds{1} _{\mathcal{L} _{2}} \ .
					\end{equation*}
					But, although we have said (somewhere in this paper) that the cardinality of $ \ker \mathfrak{f} $ is relevant for determining the degree $ \tilde{\mathfrak{n}} $ of degeneracy of the $ D^{K} \left( \mathds{Z} _{N} \right) $ ground states, we have not yet presented the formula for this $ \tilde{\mathfrak{n}} $ when $ \mathfrak{f} $ is not a trivial group homomorphism. So, the natural question that we can ask here is: does this formula exist?
					
					In order to understand the answer to this question, it is interesting that we remember, for instance, that we have already managed to determine this formula when we analysed the trivial $ D^{K} \left( \mathds{Z} _{N} \right) $ models. And an interesting aspect of this formula (\ref{formula-trivial}) that we found is that it clearly shows us that, in fact, there is a dual correspondence between the trivial $ D_{M} \left( \mathds{Z} _{N} \right) $ and $ D^{K} \left( \mathds{Z} _{N} \right) $ models. After all, according to what has been said on page \pageref{duality-comment-formula}, all the trivial group homomorphisms (\ref{trivial-via-proposition}) always can be induced by a trivial (sub)group action (\ref{trivial-dual-group-action}) that maximizes $ \bigl\vert \mathfrak{Fix} _{\tilde{\mu } _{f}} \bigr\vert $. But what happens when, for instance, the $ D^{K} \left( \mathds{Z} _{N} \right) $ models can be defined by using non-trivial group homomorphisms?
					
					One of the things that happens is that, since all these non-trivial group homomorphisms are induced by non-trivial (sub)group actions $ \tilde{\mu } _{f} : \textnormal{Im} \left( \mathfrak{f} \right) \times \tilde{S} \rightarrow \tilde{S} $ that define only $ k $-cycles where $ k > 1 $, this allows us to conclude that the dual correspondence, between the non-trivial $ D_{M} \left( \mathds{Z} _{N} \right) $ and $ D^{K} \left( \mathds{Z} _{N} \right) $ models, is not so perfect. Why? Because there are non-trivial $ D_{M} \left( \mathds{Z} _{N} \right) $ models that can be defined, for instance, by using non-trivial group actions that can define $ 1 $-cycles. In plain English, no $ D_{M} \left( \mathds{Z} _{N} \right) $ model, which is defined by using a non-trivial action that defines $ 1 $-cycles, can be interpreted as the perfect dual of any $ D^{K} \left( \mathds{Z} _{N} \right) $ model: this interpretation occurs only when
					\begin{itemize}
						\item $ \textnormal{Im} \left( \mathfrak{f} \right) = \mathds{Z} _{N} $, and
						\item the $ D_{M} \left( \mathds{Z} _{N} \right) $ gauge group action can be expressed as $ \mu _{\mathfrak{f}} \bigl( \mathfrak{f} \bigl( \tilde{\alpha } \bigr) , \gamma \bigr) $ because this induces the co-action homomorphism $ \mathcal{F} $.
					\end{itemize}
					Note that, as a consequence of these conditions, it also becomes clear that the $ D_{M} \left( \mathds{Z} _{N} \right) $ models may be interpreted as the perfect dual of the $ D^{K} \left( \mathds{Z} _{N} \right) $ models when $ S = \tilde{S} $.
					
					Nevertheless, it is also worth noting that, in accordance with the definition of the $ D^{K} \left( \mathds{Z} _{N} \right) $ face and link operators, the elements of $ \textnormal{Im} \left( \mathfrak{f} \right) $ must also act on the elements of $ \mathds{Z} _{N} $. After all, since $ \textnormal{Im} \left( \mathfrak{f} \right) $ is a normal subgroup of $ \mathds{Z} _{N} $, it is not difficult to conclude that $ \tilde{\mu } _{\ell } $ allows us to interpret its $ k $-cycles as elements of the quotient group $ \mathds{Z} _{N} / \textnormal{Im} \left( \mathfrak{f} \right) $ [\citen{james}]. And what does it mean? This means that all the magnetic quasiparticles that are inherited from the $ D \left( \mathds{Z} _{N} \right) $ models are divided into equivalence classes in the $ D^{K} \left( \mathds{Z} _{N} \right) $ ones. Thus, by noting that
					\begin{itemize}
						\item the $ D \left( \mathds{Z} _{N} \right) $ models have ground states that are $ \left\vert \mathds{Z} _{N} \right\vert ^{2 \mathfrak{g}} $-fold degenerated, where $ \mathfrak{g} $ is the genus of $ \mathcal{M} _{2} $ [\citen{pachos}], and
						\item all the magnetic quasiparticles $ \tilde{q} ^{\left( 0 , L , 0 \right) } $, which are only detected by the face operators (\ref{b-dk}), are divided into $ \left\vert \ \mathds{Z} _{N} / \textnormal{Im} \left( \mathfrak{f} \right) \ \right\vert $ equivalence classes,
					\end{itemize}
					it becomes clear that the degree of degeneracy of the $ D^{K} \left( \mathds{Z} _{N} \right) $ ground states is given by
					\begin{equation}
						\tilde{\mathfrak{n}} = \left\vert \ \ker \left( \mathfrak{f} \right) \ \right\vert \cdot \left\vert \ \mathds{Z} _{N} / \textnormal{Im} \left( \mathfrak{f} \right) \ \right\vert ^{2 \mathfrak{g}} \ . \label{gsd}
					\end{equation}
					Note that this result in full agreement with the formula (\ref{formula-trivial}) because, when $ \mathfrak{f} $ is a trivial group homomorphism, $ \mathds{Z} _{N} / \textnormal{Im} \left( \mathfrak{f} \right) = \mathds{Z} _{N} $. 
			
	\section{Final remarks} 
	
		As we present in this paper, it is very clear that we can perform a dualization procedure on the $ D_{M} \left( G \right) $ models as long as $ G $ is a finite Abelian gauge group. After all, although it is well known that a group action $ \mu : G \times S \rightarrow S $ can induce a co-action with the help of a group homomorphism $ \mathfrak{f} : S \rightarrow G $, the facts of
		\begin{itemize}
			\item the $ D_{M} \left( G \right) $ and $ D^{K} \left( G \right) $ models have the same gauge group, and bring the $ D \left( G \right) $ models as special cases,
			\item the commutation relations between the $ D^{K} \left( G \right) $ vertex, face and link operators show that the $ D^{K} \left( G \right) $ models only are exactly solvable when $ \mathsf{Im} \left( \mathfrak{f} \right) \subseteq \mathcal{Z} \left( G \right) $, and
			\item the $ D_{M} \left( G \right) $ and $ D^{K} \left( G \right) $ link operators are duals of each other when $ \tilde{S} $ and $ G $ are two finite Abelian groups
		\end{itemize}
		make it clear that the duality between these $ D_{M} \left( G \right) $ and $ D^{K} \left( G \right) $ models exists only when $ G $ is a finite Abelian gauge group. By the way, even though we wrote the entire Sections \ref{QDMf-construction} and \ref{QDMf-properties} by denoting the set that indexes the matter qudits by $ \tilde{S} $, this duality context requires that $ \tilde{S} $ equals $ S $.
		
		Observe that this last requirement is reinforced by the fact that, while the $ D_{M} \left( \mathds{Z} _{N} \right) $ models may differ from the $ D \left( \mathds{Z} _{N} \right) $ ones when $ M $ and $ N $ are coprime numbers, the $ D^{K} \left( \mathds{Z} _{N} \right) $ models cannot do the same when $ K $ and $ N $ are coprime numbers. After all, as $ K $ and $ N $ index the cyclic groups $ \tilde{S} = \mathds{Z} _{K} $ and $ G = \mathds{Z} _{N} $ respectively, Theorem \ref{proposition} allows us to interpret these $ D^{K} \left( \mathds{Z} _{N} \right) $ models, in these cases where $ K $ and $ N $ are coprime numbers, as analogues of the $ D \left( \mathds{Z} _{N} \right) $ models, but with an algebraically degenerate ground state, because the only group homomorphism $ \mathfrak{f} : \mathds{Z} _{K} \rightarrow \mathds{Z} _{N} $ that exists is the trivial one. And, no doubt, this is another way of saying the same thing that we already said in the penultimate paragraph of the last Section: i.e., this is another way of saying that the $ D_{M} \left( \mathds{Z} _{N} \right) $ models may be interpreted as the perfect dual of the $ D^{K} \left( \mathds{Z} _{N} \right) $ models when $ S = \tilde{S} $.
		
		Given this duality that we were able to recognize between the $ D_{M} \left( \mathds{Z} _{N} \right) $ and $ D \left( \mathds{Z} _{N} \right) $ models when $ G $ is a finite Abelian group, it is quite tempting to conclude that a new class of self-dual lattice gauge models can be defined in terms of an overlap of the $ D_{M} \left( \mathds{Z} _{N} \right) $ and $ D \left( \mathds{Z} _{N} \right) $ ones. That is, by coupling the Abelian $ D \left( G \right) $ models to new qudits, which would be assigned to the vertices and faces of $ \mathcal{L} _{2} $, by using the same gauge group actions and co-action homomorphisms that were presented here respectively. However, although it is indeed possible to define this new class, whose Hamiltonian seems to be better defined as
		\begin{equation*}
			H_{\mathrm{total}} = - \mathcal{J} _{A} \sum _{v  \in \mathcal{L} _{2}} A_{v} - \mathcal{J} ^{\prime } _{B} \sum _{f \in \mathcal{L} _{2}} \ B^{\prime } _{f} - \mathcal{J} _{C} \sum _{\ell \in \mathcal{L} _{2}} C_{\ell } - \mathcal{J} ^{\prime } _{C} \sum _{\ell \in \mathcal{L} _{2}} C^{\prime } _{\ell } \ ,
		\end{equation*}
		its models do not look as nice: as these new lattice gauge models bring the Abelian $ D_{M} \left( G \right) $ and $ D^{K} \left( G \right) $ ones as special cases, these new models depict a situation where all the electric and magnetic quasiparticles inherited from the Abelian $ D \left( G \right) $ models can/may be confined. In other words, this is an important aspect that may not be very good, for instance, from a quantum-computational point of view.
		
		By speaking of these electric and magnetic quasiparticles, it is important to summarize some of the reasons why we have said, at various points in this paper, that they are the same ones that appear in the $ D \left( G \right) $ models. And one of the first reasons has to do with the fact that both the $ D_{M} \left( G \right) $ and $ D^{K} \left( G \right) $ models were defined not only by using the same gauge structure as the $ D \left( G \right) $ ones, but mainly without modifying it: after all, note that
		\begin{itemize}
			\item in the $ D_{M} \left( G \right) $ models, the gauge qudits act on the matter ones and not the other way around, and
			\item as much as the $ D^{K} \left( G \right) $ link operators do transformations on the gauge qudits, these transformations can also be interpreted as gauge transformations because, as $ \bigl[ B^{\prime } _{f,R} , C^{\prime } _{\ell , \Lambda } \bigr] = 0 $, they are completely unable to modify the \textquotedblleft fake holonomy\textquotedblright \hspace*{0.01cm} around the lattice faces.
		\end{itemize}
		And this is precisely what, for instance, explains the fact that all the electric and magnetic quasiparticles, which can be produced by manipulating the gauge qudits in the $ D_{M} \left( G \right) $ and $ D^{K} \left( G \right) $ models, are produced by the operators that have the same expressions in the $ D \left( G \right) $ models. That is, despite the group actions and co-actions homomorphisms, which define the $ D_{M} \left( G \right) $ and $ D^{K} \left( G \right) $ models respectively, make these electric and magnetic quasiparticles capable of fusing with the new quasiparticles that are produced by manipulating the matter qudits, these electric and magnetic quasiparticles are exactly the same as those of the $ D \left( G \right) $ models. By the way, and by remembering that all these quantum-computational models are always defined with the intention of modelling some reality that can be physically implemented, it is also interesting to remember that, even though we know, for instance, that an electron is already capable of interacting with several particles, there is nothing that prevents nature from showing that there are other particles that are also capable of interacting with an electron. And this is precisely one of the other reasons that allows to assert that the $ D_{M} \left( G \right) $ and $ D^{K} \left( G \right) $ electric and magnetic quasiparticles are exactly the same as those of the $ D \left( G \right) $ ones, since it does not make much sense that an electron ceases to be an electron just because someone discovered that it is capable of interacting/fusing with this new particle.
		
		Note that, due to the recognition that the $ D_{M} \left( G \right) $ and $ D^{K} \left( G \right) $ electric and magnetic quasiparticles are exactly the same as those of the $ D \left( G \right) $ ones, it is correct to say that the operators $ W^{\left( J , L , \Lambda \right) } _{\ell } $ and $ \tilde{W} ^{\left( J , L , \Lambda \right) } _{\ell } $ that produce them, in pairs, may define the same ribbon operators (string operators) as in the $ D \left( G \right) $ models. But what is most striking about these quasiparticles is the fact that they are confinable. And in the case of the $ D_{M} \left( G \right) $ models, the fact that the magnetic particles are confinable is interesting for, at least, two reasons. One of them seems to be related, for instance, to the validation of these $ D_{M} \left( G \right) $ models as an excellent generalization of the $ D \left( G \right) $ ones because, as
		\begin{itemize}
			\item the $ D \left( G \right) $ models can be understood in terms of pure lattice gauge theories, and
			\item the Quantum Chromodynamics is precisely the gauge theory whose non-perturbative problems fostered the development of the lattice gauge theories,
		\end{itemize}
		this confinement of magnetic quasiparticles seems to be quite welcome since, for instance, there are some works that already explored the possibility that the confinement of quarks has some magnetic reasons [\citen{wyld}]. Although the confinement of electric quasiparticles in the $ D^{K} \left( G \right) $ models is perhaps not so interesting from the point of view of elementary particle physics, it seems to be very interesting from the point of view of condensed matter physics. After all, as the confinement of these electric and magnetic quasiparticles points, in the latter context, to the possibility of exploring these $ D_{M} \left( G \right) $ and $ D^{K} \left( G \right) $ models to describe superconductors (or, at least, perfect diamagnets) and topological insulators, this deserves to be better evaluated in our next papers.
		
		Lastly, in addition to being important to say that the $ D^{K} \left( \mathds{Z} _{N} \right) $ models can be classified by the group homomorphism that define them, one thing we need to remember is that there is no impediment, a priori, to define new generalizations of these $ D \left( G \right) $ and $ D_{M} \left( G \right) $ models without the artifice of a dualization procedure. Nevertheless, a relevant question that we can ask ourselves because of this possibility is whether, for instance, any of these new generalizations are able to lead us to the same results as the Abelian $ D^{K} \left( G \right) $ models. And a good possibility, which we can explore to answer this question, is the one where $ \mathfrak{f} $ defines a \emph{crossed module} [\citen{loday}]: i.e., the one where $ \mathfrak{f} $ is a group homomorphism that, together with a group action $ \tilde{\mu } : G \times \tilde{S} \rightarrow \tilde{S} $, respects two conditions
		\begin{equation*}
			\mathfrak{f} \left( \tilde{\mu } \left( g , \tilde{\alpha } \right) \right) = g \cdot \mathfrak{f} \left( \tilde{\alpha } \right) \cdot g^{-1} \quad \textnormal{and} \quad \tilde{\mu } \bigl( \mathfrak{f} \left( \tilde{\alpha } \right) , \tilde{\beta } \bigr) = \tilde{\alpha } \ast \tilde{\beta } \ast \tilde{\alpha } ^{-1} \ ,
		\end{equation*}
		where the second one is known as the \emph{Peiffer condition} [\citen{peiffer-paper,mantovani}]. Note that the group homomorphisms, which define the Abelian $ D^{K} \left( G \right) $ models, satisfy these two conditions when the co-action homomorphism (\ref{co-action}) is induced by a trivial gauge group action. And the possible advantage of taking $ \mathfrak{f} $ as the group homomorphism that now defines a crossed module lies in the fact that it seems possible to recover the $ D^{K} \left( G \right) $ models as a special case of the \emph{higher lattice gauge theories} [\citen{baez-1}], which are based on the higher-dimensional category theory [\citen{bucur,cheng,group-category-example}]. By the way, a good example of this can be found in Ref. [\citen{faria-martins}], where a $ 2 $-lattice gauge theory is defined by using a three-dimensional lattice in which we can measure $ 1 $- and $ 2 $-holonomies: after all, while the $ 1 $-holonomy is identified as the same \textquotedblleft fake holonomy\textquotedblright \hspace*{0.01cm} (\ref{fake-holonomy}), which is preserved by the gauge transformations that the operator $ A^{\prime } _{v} $ performs, the $ 2 $-holonomy [\citen{abba-wag}] is preserved by the action of the operator
		\begin{equation*}
			\prod _{\ell \in S_{f}} C^{\prime } _{\ell } \ ,
		\end{equation*}
		which corroborates with the perception that the link operator (\ref{qdmf-edge-operator}) actually performs another kind of gauge transformation. Note that, if $ \mathfrak{f} $ is the group homomorphism that defines a crossed module $ \mathcal{G} = \bigl( G , \tilde{S} ; \mathfrak{f} , \tilde{\mu } \bigr) $, the first and second homotopy groups of this crossed module can be defined as $ \pi _{1} \left( \mathcal{G} \right) = G \ / \ \textnormal{Im} \left( \mathfrak{f} \right) = \mathrm{coker} \left( \mathfrak{f} \right) $ and $ \pi _{2} \left( \mathcal{G} \right) = \ker \left( \mathfrak{f} \right) $ respectively [\citen{ricardo-master}], whose orders define, for instance, the formula (\ref{gsd}). We will also return to this topic in another future work. 
	
	\section*{Acknowledgments}
	
		This work has been partially supported by CAPES (ProEx) and CNPq (grant 162117/2015-9). We thank C. Antonio Filho, A. G. Chalom, R. A. Ferraz, R. Figueiredo and U. A. Maciel Neto and P. Teotonio Sobrinho  for some discussions on subjects concerning this project. Special thanks are also due to: J. Beardsley, M. Jibladze, J. P. McCarthy, P. W. Michor and \textquotedblleft John N.\textquotedblright \hspace*{0.01cm} for the discussion, on the fact that a group action always induces a co-action in the context of Hopf algebras, which we quote in the footnote on page \pageref{discussion}; and the reviewers of this paper, who, although we do not know their names, helped us to make this paper a little more consistent and a lot better to read. M. F. is also deeply grateful to L. Daros Gama and F. Diacenco Xavier for friendly support during part of this work.

\end{document}